\documentclass[aps,preprint,floats,epsf,epsfig,nofootinbib,letter]{revtex4}
\usepackage{epsfig}
\usepackage{subfigure}

\begin{document}
\def\be{\begin{eqnarray}}
\def\en{\end{eqnarray}}
\def\non{\nonumber}
\def\ov{\overline}
\def\la{\langle}
\def\ra{\rangle}
\def\B{{\cal B}}
\def\3bar{{\bf \bar 3}}
\def\6bar{{\bf \bar 6}}
\def\10bar{{\bf \ov{10}}}
\def\pr{{\sl Phys. Rev.}~}
\def\prl{{\sl Phys. Rev. Lett.}~}
\def\pl{{\sl Phys. Lett.}~}
\def\np{{\sl Nucl. Phys.}~}
\def\zp{{\sl Z. Phys.}~}
\def\lsim{ {\ \lower-1.2pt\vbox{\hbox{\rlap{$<$}\lower5pt\vbox{\hbox{$\sim$}
}}}\ } }
\def\gsim{ {\ \lower-1.2pt\vbox{\hbox{\rlap{$>$}\lower5pt\vbox{\hbox{$\sim$}
}}}\ } }

\font\el=cmbx10 scaled \magstep2{\obeylines\hfill Aug. 2019} 

\vskip 1.5 cm

\centerline{\large\bf Color-allowed Bottom Baryon to $s$-wave and $p$-wave Charmed Baryon}
\centerline{\large\bf non-leptonic Decays}

\vskip 1.2 cm

\bigskip
\bigskip
\centerline{\bf Chun-Khiang Chua}
\medskip
\medskip
\centerline{Department of Physics and Center for High Energy Physics}
\centerline{Chung Yuan Christian University}
\centerline{Chung-Li, Taiwan 320, Republic of China}

\bigskip
\bigskip
\centerline{\bf Abstract}
We study color allowed bottom baryon to $s$-wave and $p$-wave charmed baryon non-leptonic decays in this work.
The charmed baryons include spin-1/2 and spin-3/2 states.
Explicitly, we consider $\Lambda_b\to \Lambda^{(*,**)}_c M^-$, $\Xi_b\to\Xi_c^{(**)} M^-$ and $\Omega_b\to\Omega^{(*,**)}_c M^-$ decays with $M=\pi, K, \rho, K^*, a_1, D, D_s,  D^*, D^*_s$, 
$\Lambda^{(*,**)}_c=\Lambda_c, \Lambda_c(2595), \Lambda_c(2625), \Lambda_c(2765), \Lambda_c(2940)$, 
$\Xi_c^{(**)}=\Xi_c, \Xi_c(2790), \Xi_c(2815)$ and 
$\Omega^{(*,**)}_c=\Omega_c, \Omega_c(2770), \Omega_c(3050), \Omega_c(3090), \Omega_c(3120)$.
There are six types of transitions, namely, 
(i) ${\cal B}_b({\bf \bar 3_f},1/2^+)$ to ${\cal B}_c({\bf \bar 3_f},1/2^+)$, 
(ii) ${\cal B}_b({\bf 6_f},1/2^+)$ to ${\cal B}_c({\bf 6_f},1/2^+)$, 
(iii) ${\cal B}_b({\bf 6_f},1/2^+)$ to ${\cal B}_c({\bf 6_f},3/2^+)$,
(iv) ${\cal B}_b({\bf 6_f},1/2^+)$ to ${\cal B}_c({\bf 6_f},3/2^-)$,
(v) ${\cal B}_b({\bf \bar 3_f},1/2^+)$ to ${\cal B}_c({\bf \bar 3_f},1/2^-)$, 
and 
(vi) ${\cal B}_b({\bf \bar 3_f},1/2^+)$ to ${\cal B}_c({\bf \bar 3_f},3/2^-)$
transitions. 
Types (i) to (iii) involve spin 1/2 and 3/2 $s$-wave charmed baryons, 
while types (iv) to (vi) involve spin 1/2 and 3/2 $p$-wave charmed baryons. 
The light diquarks are spectating in these transitions.
The transition form factors are calculated in the light-front quark model approach. 
All of the form factors in the $1/2\to 1/2$ and $1/2 \to 3/2$ transitions are extracted,
and they are found to reasonably satisfy the relations obtained in the heavy quark limit, as we are using heavy but finite $m_b$ and $m_c$.
Using na\"{i}ve factorization, decay rates and up-down asymmetries of the above modes 
are predicted and can be checked experimentally.
We compare our results to data and other theoretical predictions.
The study on these decay modes 
may shed light on the quantum numbers of $\Lambda_c(2765)$, $\Lambda_c(2940)$
$\Omega_c(3050)$, $\Omega_c(3090)$ and $\Omega_c(3120)$.

\bigskip
\small

\pagebreak

\section{Introduction}

There are some experimental progresses in the charmed baryon sector recently.
In year 2017 LHCb discovered $\Lambda_c(2864)$ and five $\Omega_c$ states, namely 
$\Omega_c(3000)^0$, 
$\Omega_c(3050)^0$, 
$\Omega_c(3066)^0$, 
$\Omega_c(3090)^0$ and
$\Omega_c(3120)^0$ 
\cite{Aaij:2017vbw,Aaij:2017nav}.
The first four states among the five newly discovered $\Omega_c$  were confirmed by Belle later~\cite{Yelton:2017qxg}. 
In Table~\ref{tab:spectrumC} the mass spectra, decay widths and quantum numbers of charmed baryons observed up to now are summarized.
Note that 16 out of 40 charmed baryons have unspecified quantum numbers, 
while the quantum numbers of the rests are determined with different levels of certainty~\cite{PDG,PDG1}. 
Those with unspecified quantum numbers include $\Lambda_c(2765)^+$, $\Sigma_c(2800)^{++,+,0}$, $\Xi_c(2930)^0$, $\Xi_c(2970)^{+,0}$, $\Xi_c(3055)^{+}$, $\Xi_c(3080)^{+,0}$, $\Xi_c(3123)^+$
and the above mentioned five $\Omega_c$ states.
Various suggestions on the quantum numbers of the newly discovered $\Omega_c$ states were put forward after the discovery, see for example~\cite{Agaev:2017jyt,Chen:2017sci,Karliner:2017kfm,Wang:2017hej,Padmanath:2017lng,Wang:2017zjw,Chen:2017gnu,Cheng:2017ove}. 
Note that even for the one with specified quantum number in PDG, it is still room for different assignments.
For example, two possible quantum numbers of $\Lambda_c(2940)^+$ were proposed.
LHCb and PDG preferred the $\frac{3}{2}^-$ quantum number, but it is not certain~\cite{Aaij:2017vbw,PDG}, while a $\frac{1}{2}^-$ state was advocated by the authors of ref.~\cite{Cheng:2017ove}.
It is timely, of great interest and importance to identify the quantum numbers of these states and to study their
properties. 

The study of bottom baryon to charmed baryon weak decays  
may shed light on the quantum numbers of some of the charmed baryons.
Up to now only several color allowed $\Lambda_b\to\Lambda_c P$ decay rates were measured.
These include the rates of $\Lambda_b\to\Lambda_c \pi^-$,  $\Lambda_c K^-$, $\Lambda_c D^-$ and $\Lambda_c D^-_s$
decays, which were reported by LHCb several years ago~\cite{Aaij:2014jyk,Aaij:2014lpa,Aaij:2014pha}.
It is not unrealistic to expect that further progress on the experimental side, either from LHCb, from Belle II or from elsewhere, will occur soon.
In \cite{Chua:2018lfa} we studied the color allowed $\Lambda^0_b\to \Lambda^{(*,**)}_c M$, $\Xi_b\to\Xi_c^{(**)} M$ and $\Omega_b\to\Omega^{(*)}_c M$ decays with $M=\pi, K, \rho, K^*, a_1, D, D_s,  D^*, D^*_s$,
and 
$\Lambda^{(*,**)}_c=\Lambda_c, \Lambda_c(2595), \Lambda_c(2765), \Lambda_c(2940)$, $\Xi_c^{(**)}=\Xi_c, \Xi_c(2790)$ and $\Omega^{(*)}_c=\Omega_c, \Omega_c(3090)$, which are spin-1/2 charm baryons.
In this work we will extend the study to $s$-wave and $p$-wave charmed baryons up to spin-3/2 states,
these include $\Lambda^{(*,**)}_c=\Lambda_c, \Lambda_c(2595), \Lambda_c(2625), \Lambda_c(2765), \Lambda_c(2940)$(spin-1/2), 
$\Lambda_c(2940)$(spin-3/2),
$\Xi_c^{(**)}=\Xi_c, \Xi_c(2790), \Xi_c(2815)$ and 
$\Omega^{(*,**)}_c=\Omega_c, \Omega_c(2770), \Omega_c(3050), \Omega_c(3090), \Omega_c(3120)$.

In Table~\ref{tab:quantum number C}
configurations of $s$-wave and $p$-wave charmed baryons are shown.  
We consider the color allowed non-leptonic two body weak decays, where the light quarks are spectating in the processes. 
These transitions are straightforward and easier to study.
There are six types of $\B_b\to\B_c$ transitions to be studied in this work, namely, 
(i) ${\cal B}_b({\bf \bar 3_f},1/2^+)$ to ${\cal B}_c({\bf \bar 3_f},1/2^+)$, 
(ii) ${\cal B}_b({\bf 6_f},1/2^+)$ to ${\cal B}_c({\bf 6_f},1/2^+)$, 
(iii) ${\cal B}_b({\bf 6_f},1/2^+)$ to ${\cal B}_c({\bf 6_f},3/2^+)$,
(iv) ${\cal B}_b({\bf 6_f},1/2^+)$ to ${\cal B}_c({\bf 6_f},3/2^-)$,
(v) ${\cal B}_b({\bf \bar 3_f},1/2^+)$ to ${\cal B}_c({\bf \bar 3_f},1/2^-)$, 
and 
(vi) ${\cal B}_b({\bf \bar 3_f},1/2^+)$ to ${\cal B}_c({\bf \bar 3_f},3/2^-)$
transitions. 
They are summarized in Table~\ref{tab:transition}, where more informations of the decaying particles and the final states are given.
Types (i) to (iii) involve spin 1/2 and 3/2 $s$-wave charmed baryons, 
while types (iv) to (vi) involve spin 1/2 and 3/2 $p$-wave charmed baryons. 
Since there are two possible quantum number assignments for $\Lambda_c(2940)$, namely a radial excited $p$-wave spin-1/2 or a spin-3/2 state,
it will be useful to consider both possibilities and compare the predictions on rates and so on. 
The transition form factors are calculated in the light-front quark model approach.
For some other studies
one is referred to
\cite{
Ebert:2006rp,
Mannel:1992ti, 
Cheng:1996cs,
Ivanov:1997ra,
Ivanov:1997hi,
Fayyazuddin:1998ap,
Mohanta:1998iu,
Giri:1997te,
Shih:1999yh,
Albertus:2004wj, 
Ke:2007tg, 
Ke:2012wa, 
Detmold:2015aaa, 
Gutsche:2018utw,
Zhao:2018zcb, 
Zhu:2018jet,
Ke:2019smy}.
We will compare our results to those obtained in other works.

\begin{table}[t!]
\caption{Mass spectra, widths (in units of MeV) and quantum numbers of
charmed baryons are summarized. 
Experimental values of masses and widths and $J^P$ are taken from the Particle
Data Group (PDG) \cite{PDG,PDG1}. 
Note that
$S_{[qq]}$, $L_k$, $L_K$ and $J_l$ are 
the spin of the diquark $[qq]$, 
the orbital angular momentum between the light quarks,  
the orbital angular momentum of the $Q-[qq]$ system
and the total angular momentum of the light degree of freedom, respectively.
See~\cite{Cheng:2006dk,Cheng:2015naa,Chua:2018lfa} for more details.
}
\label{tab:spectrumC}
\begin{center}
\scriptsize{
\begin{tabular}{|c|cc cc c c c c|c|} \hline \hline
State 
   & $J^P$
   & $n$
   & $(L_K,L_k)$ 
   & $S_{[qq]}^P$
   & $J_\ell^{P_\ell}$ 
   & Mass 
   & Width 
   & Decay modes
   \\   
\hline
$\Lambda_c^+$ 
   & ${1\over 2}^+$ 
   & 1
   & (0,0) 
   & $0^+$
   & $0^+$ 
   & $2286.46\pm0.14$ 
   & 
   & weak  
   \\
 \hline
 $\Lambda_c(2595)^+$ 
   & ${1\over 2}^-$ 
   & 1
   & (1,0) 
   & $0^+$
   & $1^-$ 
   & $2592.25\pm0.28$ 
   & $2.6\pm0.6$ 
   & $\Lambda_c\pi\pi,\Sigma_c\pi$ 
   \\
 \hline
 $\Lambda_c(2625)^+$ 
   & ${3\over 2}^-$ 
   & 1
   & (1,0) 
   & $0^+$
   & $1^-$ 
   & $2628.11\pm0.19$ 
   &$<0.97$ 
   & $\Lambda_c\pi\pi,\Sigma_c\pi$ 
   \\
 \hline
 $\Lambda_c(2765)^+$ 
   & $?^?$ 
   & ?
   & ? 
   & ? 
   & $?$ 
   & $2766.6\pm2.4$ 
   & $50$ 
   & $\Sigma_c\pi,\Lambda_c\pi\pi$ 
   \\
 \hline
  $\Lambda_c(2860)^+$ 
   & $\frac{3}{2}^+$ 
   & 1
   & (2,0)
   & $0^+$
   & $2^+$ 
   & $2856.1^{+2.3}_{-6.0}$ 
   & $68^{+12}_{-22}$ 
   & $\Sigma^{(*)}_c\pi,D^0p, D^+ n$ 
   \\
 \hline
 $\Lambda_c(2880)^+$ 
   & ${5\over 2}^+$ 
   & 1
   & (2,0)
   & $0^+$
   & $2^+$ 
   & $2881.63\pm0.24$ 
   & $5.6^{+0.8}_{-0.6}$
   & $\Sigma_c^{(*)}\pi,\Lambda_c\pi\pi,D^0p$ 
   \\
 \hline
 $\Lambda_c(2940)^+$ 
   & $\frac{3}{2}^-$
   & 2
   & (1,0) 
   & $0^+$
   & $1^-$ 
   & $2939.6^{+1.3}_{-1.5}$ 
   & $20^{+6}_{-5}$ 
   & $\Sigma_c^{(*)}\pi,\Lambda_c\pi\pi,D^0p$ 
   \\ 
 \hline
 $\Sigma_c(2455)^{++}$ 
   & ${1\over 2}^+$ 
   & 1
   & $(0,0)$ 
   & $1^+$
   & $1^+$ 
   & $2453.97\pm0.14$ 
   & $1.89^{+0.09}_{-0.18}$ 
   & $\Lambda_c\pi$ 
   \\
 \hline
 $\Sigma_c(2455)^{+}$ 
   & ${1\over 2}^+$ 
   & 1
   & $(0,0)$ 
   & $1^+$
   & $1^+$     
   & $2452.9\pm0.4$ 
   & $<4.6$ 
   & $\Lambda_c\pi$
   \\
 \hline
 $\Sigma_c(2455)^{0}$ 
   & ${1\over 2}^+$ 
   & 1
   & $(0,0)$ 
   & $1^+$
   & $1^+$ 
   & $2453.75\pm0.14$
   & $1.83^{+0.11}_{-0.19}$ 
   & $\Lambda_c\pi$ 
   \\
 \hline
 $\Sigma_c(2520)^{++}$ 
   & ${3\over 2}^+$ 
   & 1
   & $(0,0)$ 
   & $1^+$
   & $1^+$ 
   & $2518.41^{+0.21}_{-0.19}$
   & $14.78^{+0.30}_{-0.40}$ 
   & $\Lambda_c\pi$
   \\
 \hline
 $\Sigma_c(2520)^{+}$ 
   & ${3\over 2}^+$ 
   & 1
   & $(0,0)$ 
   & $1^+$
   & $1^+$     
   & $2517.5\pm2.3$
   & $<17$ 
   & $\Lambda_c\pi$ 
   \\
 \hline
 $\Sigma_c(2520)^{0}$ 
   & ${3\over 2}^+$ 
   & 1
   & $(0,0)$ 
   & $1^+$
   & $1^+$ 
   & $2518.48\pm0.20$
   & $15.3^{+0.4}_{-0.5}$ 
   & $\Lambda_c\pi$ 
   \\
 \hline
 $\Sigma_c(2800)^{++}$ 
   & $?^?$ 
   & ?
   & ? 
   & ?
   & ? 
   & $2801^{+4}_{-6}$ 
   & $75^{+22}_{-17}$ 
   & $\Lambda_c\pi,\Sigma_c^{(*)}\pi,\Lambda_c\pi\pi$ 
   \\
 \hline
 $\Sigma_c(2800)^{+}$ 
   & $?^?$ 
   & ?
   & ? 
   & ? 
   & ?
   & $2792^{+14}_{-~5}$ 
   & $62^{+60}_{-40}$ 
   & $\Lambda_c\pi,\Sigma_c^{(*)}\pi,\Lambda_c\pi\pi$ 
   \\
 \hline
 $\Sigma_c(2800)^{0}$ 
   & $?^?$ 
   & ?
   & ? 
   & ? 
   & ?
   & $2806^{+5}_{-7}$ 
   & $72^{+22}_{-15}$ 
   & $\Lambda_c\pi,\Sigma_c^{(*)}\pi,\Lambda_c\pi\pi$
   \\
 \hline
 $\Xi_c^+$ 
   & ${1\over 2}^+$ 
   & 1
   & (0,0) 
   & $0^+$
   & $0^+$ 
   & $2467.87\pm 0.30$ 
   & 
   & weak 
   \\ 
 \hline
 $\Xi_c^0$ 
   & ${1\over 2}^+$ 
   & 1
   & (0,0) 
   & $0^+$
   & $0^+$  
   & $2470.87^{+0.28}_{-0.31}$ 
   & 
   & weak 
   \\ 
\hline
 $\Xi'^+_c$ 
   & ${1\over 2}^+$ 
   & 1
   & (0,0) 
   & $1^+$
   & $1^+$ 
   & $2577.4\pm1.2$ 
   & 
   & $\Xi_c\gamma$ 
   \\ 
\hline
 $\Xi'^0_c$ 
   & ${1\over 2}^+$ 
   & 1
   & (0,0) 
   & $1^+$
   & $1^+$  
   & $2578.8\pm0.5$ 
   & 
   & $\Xi_c\gamma$ 
   \\ 
\hline
 $\Xi_c(2645)^+$ 
   & ${3\over 2}^+$ 
   & 1
   & (0,0) 
   & $1^+$
   & $1^+$  
   & $2645.53\pm 0.31$ 
   & $2.14\pm0.19$ 
   & $\Xi_c\pi$ 
   \\
\hline
 $\Xi_c(2645)^0$ 
   & ${3\over 2}^+$ 
   & 1
   & (0,0) 
   & $1^+$
   & $1^+$  
   & $2646.32\pm0.31$ 
   & $2.35\pm0.22$ 
   & $\Xi_c\pi$ 
   \\
 \hline
 $\Xi_c(2790)^+$ 
   & ${1\over 2}^-$ 
   & 1
   & (1,0) 
   & $0^+$
   & $1^-$ 
   & $2792.0\pm0.5$ 
   & $8.9\pm 1.0$ 
   & $\Xi'_c\pi$
   \\
 \hline
 $\Xi_c(2790)^0$ 
   & ${1\over 2}^-$ 
   & 1
   & (1,0) 
   & $0^+$
   & $1^-$  
   & $2792.8\pm1.2$ 
   & $10.0\pm1.1$ 
   & $\Xi'_c\pi$ 
   \\
 \hline
 $\Xi_c(2815)^+$ 
   & ${3\over 2}^-$ 
   & 1
   & (1,0) 
   & $0^+$
   & $1^-$   
   & $2816.67\pm0.31$ 
   & $2.43\pm0.26$ 
   & $\Xi^*_c\pi,\Xi_c\pi\pi,\Xi_c'\pi$ 
   \\
 \hline
 $\Xi_c(2815)^0$ 
   & ${3\over 2}^-$ 
   & 1
   & (1,0) 
   & $0^+$
   & $1^-$   
   & $2820.22\pm0.32$ 
   & $2.54\pm0.25$ 
   & $\Xi^*_c\pi,\Xi_c\pi\pi,\Xi_c'\pi$ 
   \\
 \hline
$\Xi_c(2930)^0$ 
  & $?^?$ 
  & ?
  & ? 
  & ? 
  & $?$ 
  & $2931\pm6$ 
  & $36\pm13$
  & $\Lambda_c \ov K$ 
  \\
\hline
 $\Xi_c(2970)^+$ 
   & $?^?$ 
   & ?
   & ? 
   & ? 
   & $?$ 
   & $2969.4\pm0.8$ 
   & $20.9^{+2.4}_{-3.5}$
   & $\Sigma_c \ov K,\Lambda_c \ov K\pi,\Xi_c\pi\pi$  
   \\
 \hline
 $\Xi_c(2970)^0$ 
   & $?^?$ 
   & ?
   & ? 
   & ? 
   & $?$ 
   & $2967.8\pm0.8$ 
   & $28.1^{+3.4}_{-4.0}$
   & $\Sigma_c \ov K,\Lambda_c \ov K\pi,\Xi_c\pi\pi$
   \\
 \hline
 $\Xi_c(3055)^+$ 
   & $?^?$ 
   & ?
   & ? 
   & ? 
   & $?$ 
   & $3055.9\pm0.4$
   & $7.8\pm1.9$
   & $\Sigma_c \ov K,\Lambda_c \ov K\pi,D\Lambda$  
   \\
 \hline
 $\Xi_c(3080)^+$ 
   & $?^?$ 
   & ?
   & ? 
   & ? 
   & $?$ 
   & $3077.2\pm0.4$ 
   & $3.6\pm1.1$ 
   & $\Sigma_c \ov K,\Lambda_c \ov K\pi,D\Lambda$  
   \\
\hline
 $\Xi_c(3080)^0$ 
   & $?^?$ 
   & ?
   & ? 
   & ? 
   & $?$ 
   & $3079.9\pm1.4$ 
   & $5.6\pm2.2$
   & $\Sigma_c \ov K,\Lambda_c \ov K\pi,D\Lambda$ 
   \\
\hline
$\Xi_c(3123)^+$ 
   & $?^?$ 
   & ?
   & ? 
   & ? 
   & $?$ 
   & $3122.9\pm1.3$ 
   & $4\pm4$
   & $\Sigma_c^* \ov K,\Lambda_c \ov K\pi$ 
   \\
 \hline
 $\Omega_c^0$ 
   & ${1\over 2}^+$ 
   & 1
   & (0,0) 
   & $1^+$
   & $1^+$ 
   & $2695.2\pm1.7$ 
   & 
   & weak 
   \\
 \hline
 $\Omega_c(2770)^0$ 
   & ${3\over 2}^+$ 
   & 1
   & (0,0) 
   & $1^+$
   & $1^+$  
   & $2765.9\pm2.0$ 
   & 
   & $\Omega_c\gamma$ 
   \\
\hline 
 $\Omega_c(3000)^0$ 
   & $?^?$
   & ?
   & ?
   & ?
   & ?
   & $3000.4\pm0.4$ 
   & $4.5\pm0.7$
   & $\Xi_c\bar K$ 
   \\
\hline
 $\Omega_c(3050)^0$ 
   & $?^?$ 
   & ?
   & ?
   & ?
   & ? 
   & $3050.2\pm0.33$ 
   & $<1.2$
   & $\Xi_c\bar K$ 
   \\
\hline
 $\Omega_c(3065)^0$ 
   & $?^?$ 
   & ?
   & ?
   & ?
   & ?
   & $3065.6\pm0.4$ 
   & $3.5\pm0.4$
   & $\Xi_c\bar K$ 
   \\
\hline
 $\Omega_c(3090)^0$ 
   & $?^?$ 
   & ?
   & ? 
   & ? 
   & ?
   & $3090.2\pm0.7$ 
   & $8.7\pm1.3$
   & $\Xi^{(\prime)}_c\bar K$ 
   \\
\hline
 $\Omega_c(3120)^0$ 
   & $?^?$ 
   & ?
   & ? 
   & ? 
   & ?
   & $3119.1\pm1.0$ 
   & $<2.6$
   & $\Xi^{(\prime)}_c\bar K$
   \\
\hline
\hline
\end{tabular}
}
\end{center}
\end{table}

The analysis and the scope of this work is improved and enlarged compared to a previous study~\cite{Chua:2018lfa} in several aspects. 
All of the form factors in the $1/2\to 1/2$ and $1/2 \to 3/2$ transitions are extracted, while $1/2 \to 3/2$ transitions were not considered and only 2/3 of the $1/2\to1/2$ form factors were extracted in \cite{Chua:2018lfa}.
It is useful to note that in the heavy quark (HQ) limit, the $\B_b\to \B_c$ transition matrix elements with $s$-wave and $p$-wave $\B_c$ baryon form factors have simple behavior~\cite{Isgur:1990pm,Isgur:1991wr,Yan:1992gz,Cheng:1996cs,Xu:1993mj,Chow:1994ni}. 
Form factors are usually related in the heavy quark limit.
We will compare the form factors obtained in this work with the relations in HQ limit.
Although some deviations are expected as we are using heavy but finite $m_b$ and $m_c$, 
it is still interesting to see how well the form factors exhibiting the patterns required by heavy quark symmetry (HQS).

The layout of this paper is as following. In Sec. 2 we shall work out the formulas of form factors for various bottom baryon to $s$-wave and $p$-wave charmed baryon transitions in the light-front quark model approach. 
In Sec. 3 the numerical results of $\B_b\to \B_c$ transition form factors, decay rates and up-down asymmetries of $\B_b\to \B_c M$ decays will be presented.
In Sec. 4 we give our conclusions. Appendices~A and B are prepared to give some details of the light-front quark model and the derivations of the vertex functions, while some formulas involving kinematics are collected in Appendix~C.

\begin{table}[t!]
\caption{Configurations of $s$-wave and $p$-wave singlely charmed baryons are shown. 
The angular momenta are defined as $\vec S_{qq}$, 
$\vec S_{[qq]}\equiv\vec L_k+\vec S_{qq}$, $\vec J_l\equiv\vec S_{[qq]}+\vec L_K$ and $\vec J\equiv\vec J_l+\vec S_Q$,
which are the angular momenta of the light quark pair (without the relative orbital angular momentum), the whole diquark system, the light-degree of freedom and the whole baryon, respectively. 
The quantum number assignments of these states are from Table \ref{tab:spectrumC} and \cite{Chua:2018lfa}, while those with $(\dagger)$ are taken from \cite{Cheng:2017ove}. 
States with $L_K=0$ and 1 correspond to $s$-wave and $p$-wave states, respectively.
}
 \label{tab:quantum number C}
{
 \begin{center}
\begin{tabular}{| l c c  c c c c c c|}
\hline
$n$
    & $L_K$
    & $L_k$
    & ${\rm flavor}$
    & $S_{qq}$
    & $S_{[qq]}^P$
    & $J_l^P$
    & $J^P$
    & $\B_c$
    \\
    \hline
1
    & 0   
    & 0
    & ${\bf \bar 3_f}$
    &  $0$
    & $0^+$
    & $0^+$
    & $\frac{1}{2}^+$
    & $\Lambda^+_c$, $\Xi_c^{+,0}$
    \\    
2
    & 0
    & 0
    & ${\bf \bar 3_f}$
    &  $0$
    & $0^+$
    & $0^+$
    & $\frac{1}{2}^+$
    & $\Lambda_c(2765)^+ (\dagger)$ 
    \\        
1
    &0
    & 0
    & ${\bf 6_f}$
    & $1$
    & $1^+$
    & $1^+$
    & $\frac{1}{2}^+$
    & $\Sigma_c(2455)^{++,+,0}$, $\Xi_c^{\prime +,0}$, $\Omega^0_c$
    \\
2
    & 0
    & 0
    & ${\bf 6_f}$
    & $1$
    & $1^+$
    & $1^+$
    & $\frac{1}{2}^+$
    & $\Xi'_c(2970)^{+,0}(\dagger)$, $\Omega_c(3090)^0 (\dagger)$ 
    \\ 
1
    & 0
    & 0
    & ${\bf 6_f}$
    & $1$
    & $1^+$
    & $1^+$
    & $\frac{3}{2}^+$
    & $\Sigma_c(2520)^{++,+,0}$, $\Xi_c(2645)^{+,0}$, $\Omega_c(2770)^0$
    \\
2
    & 0
    & 0
    & ${\bf 6_f}$
    & $1$
    & $1^+$
    & $1^+$
    & $\frac{3}{2}^+$
    & $\Omega_c(3120)^0(\dagger)$
    \\
1
    & 1   
    & 0
    & ${\bf \bar 3_f}$
    &  $0$
    & $0^+$
    & $1^-$
    & $\frac{1}{2}^-$
    & $\Lambda_c(2595)^+$, $\Xi_c(2790)^{+,0}$
    \\
2
    & 1   
    & 0
    & ${\bf \bar 3_f}$
    &  $0$
    & $0^+$
    & $1^-$
    & $\frac{1}{2}^-$
    & $\Lambda_c(2940)^+ (\dagger)$
    \\            
1
    & 1   
    & 0
    & ${\bf \bar 3_f}$
    &  $0$
    & $0^+$
    & $1^-$
    & $\frac{3}{2}^-$
    & $\Lambda_c(2625)^+$, $\Xi_c(2815)^{+,0}$
    \\
2
    & 1   
    & 0
    & ${\bf \bar 3_f}$
    &  $0$
    & $0^+$
    & $1^-$
    & $\frac{3}{2}^-$
    & $\Lambda_c(2940)^+$
    \\
1
    & 1
    & 0
    & ${\bf 6_f}$
    & $1$
    & $1^+$
    & $2^-$
    & $\frac{3}{2}^-$
    & $\Sigma_c(2800)^{++,+,0}(\dagger)$, $\Xi'_c(2930)^{ +,0}(\dagger)$, $\Omega_c(3050)^0(\dagger)$
    \\
1
    & 1
    & 0
    & ${\bf 6_f}$
    & $1$
    & $1^+$
    & $2^-$
    & $\frac{5}{2}^-$
    & $\Omega_c(3066)^0(\dagger)$  
    \\                  
    \hline
\end{tabular}
\end{center}
}
\end{table}

\begin{table}[t!]
\caption{
Various bottom baryon to $s$-wave and $p$-wave charmed baryon transitions are shown. 
There are six transition types.
Types (i)-(iii) involve $s$-wave states, the corresponding transitions are (i) $\B_b({\bf \bar 3_f}, 1/2^+)\to \B_c({\bf \bar 3_f}, 1/2^+)$, (ii) $\B_b({\bf 6_f}, 1/2^+)\to \B_c({\bf 6_f}, 1/2^+)$
and (iii) $\B_b({\bf 6_f}, 1/2^+)\to \B_c({\bf 6_f}, 3/2^+)$ transitions.
Types (iv)-(vi) involve $p$-wave baryons in the final states, the corresponding transitions are
(iv) $\B_b({\bf 6_f}, 1/2^+)\to \B_c({\bf 6_f}, 3/2^-)$,
(v) $\B_b({\bf \bar 3_f}, 1/2^+)\to \B_c({\bf \bar 3_f}, 1/2^-)$ and 
(vi) $\B_b({\bf \bar 3_f}, 1/2^+)\to \B_c({\bf \bar 3_f}, 3/2^-)$ transitions.
Note that types (i), (v) and (vi) transitions involve scalar diquarks, 
while type (ii), (iii) and (iv) transitions involve axial-vector diquarks.
These diquarks are spectating in the transitions.
See Tables~\ref{tab:quantum number C} for the quantum number assignments for charmed baryons.  
Note that the asterisks denote the transitions where the final state charmed baryons are radial excited.
The two possibilities of quantum numbers for $\Lambda_c(2940)$, 
namely, a $\B_c({\bf \bar 3_f}, 1/2^-)$ state~\cite{Cheng:2017ove}
or a $\B_c({\bf \bar 3_f}, 3/2^-)$ state~\cite{Aaij:2017vbw}, will be considered in this work.}
 \label{tab:transition}
{
 \begin{center}
\begin{tabular}{| l c c |}
\hline
Type
    &~~~$(n, L_K, S_{[qq]}^P, J_l^P,J^P)_b\to
    (n', L'_K, S_{ [qq]}^P,J_l^{\prime P}, J^{\prime P})_c$
    & $\B_b\to\B_c$
    \\
    \hline
(i) 
    & $ (1, 0, 0^+,0^+, \frac{1}{2}^+)\to(1, 0, 0^+,0^+, \frac{1}{2}^+)$ 
    & $\Lambda_b^0\to \Lambda^+_c$, $\Xi_b^{0(-)}\to \Xi_c^{+(0)}$
    \\ 
(i)$^*$      
    & $(1, 0, 0^+,0^+, \frac{1}{2}^+)\to (2, 0, 0^+,0^+, \frac{1}{2}^+)$ 
    & $\Lambda_b^0\to\Lambda_c(2765)^+ (\dagger)$ 
    \\  
(ii) 
    & $(1, 0, 1^+,1^+, \frac{1}{2}^+)\to (1, 0, 1^+,1^+, \frac{1}{2}^+)$
    & $\Omega^-_b\to\Omega^0_c$
    \\
(ii)$^*$
    & $(1, 0, 1^+,1^+, \frac{1}{2}^+)\to (2, 0, 1^+,1^+, \frac{1}{2}^+)$
    & $\Omega^-_b\to\Omega_c(3090)^0 (\dagger)$  
    \\             
(iii)
    & $(1, 0, 1^+,1^+, \frac{1}{2}^+)\to (1, 0, 1^+,1^+, \frac{3}{2}^+)$      
    & $\Omega^-_b\to\Omega_c(2770)^0$
    \\
(iii)$^*$    
    & $(1, 0, 1^+,1^+, \frac{1}{2}^+)\to (2, 0, 1^+,1^+, \frac{3}{2}^+)$      
    & $\Omega^-_b\to\Omega_c(3120)^0(\dagger)$
    \\       
(iv)
    & $(1, 0, 1^+,1^+, \frac{1}{2}^+)\to(1, 1, 1^+,2^-, \frac{3}{2}^-)$
    & $\Omega_b^-\to\Omega_c(3050)^0(\dagger)$
    \\
(v) 
    & $(1, 0, 0^+,0^+, \frac{1}{2}^+) \to (1, 1, 0^+,1^-, \frac{1}{2}^-)$   
    & $\Lambda_b^0\to\Lambda_c(2595)^+$, $\Xi^{0(-)}_b\to\Xi_c(2790)^{+(0)}$
    \\
(v)$^*$  
    & $(1, 0, 0^+,0^+, \frac{1}{2}^+)\to(2, 1, 0^+,1^-, \frac{1}{2}^-)$   
    & $\Lambda_b^0\to\Lambda_c(2940)^+ (\dagger)$
    \\  
(vi)
    & $(1, 0, 0^+,0^+, \frac{1}{2}^+) \to (1, 1, 0^+,1^-, \frac{3}{2}^-)$  
    & $\Lambda_b^0\to\Lambda_c(2625)^+$, $\Xi_b^{0(-)}\to\Xi_c(2815)^{+(0)}$
    \\                  
(vi)$^*$    
    & $(1, 0, 0^+,0^+, \frac{1}{2}^+)\to (2, 1, 0^+,1^-, \frac{3}{2}^-)$   
    & $\Lambda_b^0\to\Lambda_c(2940)^+$
    \\                    
    \hline
\end{tabular}
\end{center}
}
\end{table}

\section{Obtaining Form factors in the light-front approach}

\subsection{$\B_b(1/2)\to \B_c(1/2)$ and $\B_b(1/2)\to\B_c(3/2)$ weak transitions}

\begin{figure}[t!]
\centerline{
          \includegraphics[width=0.6\textwidth]  {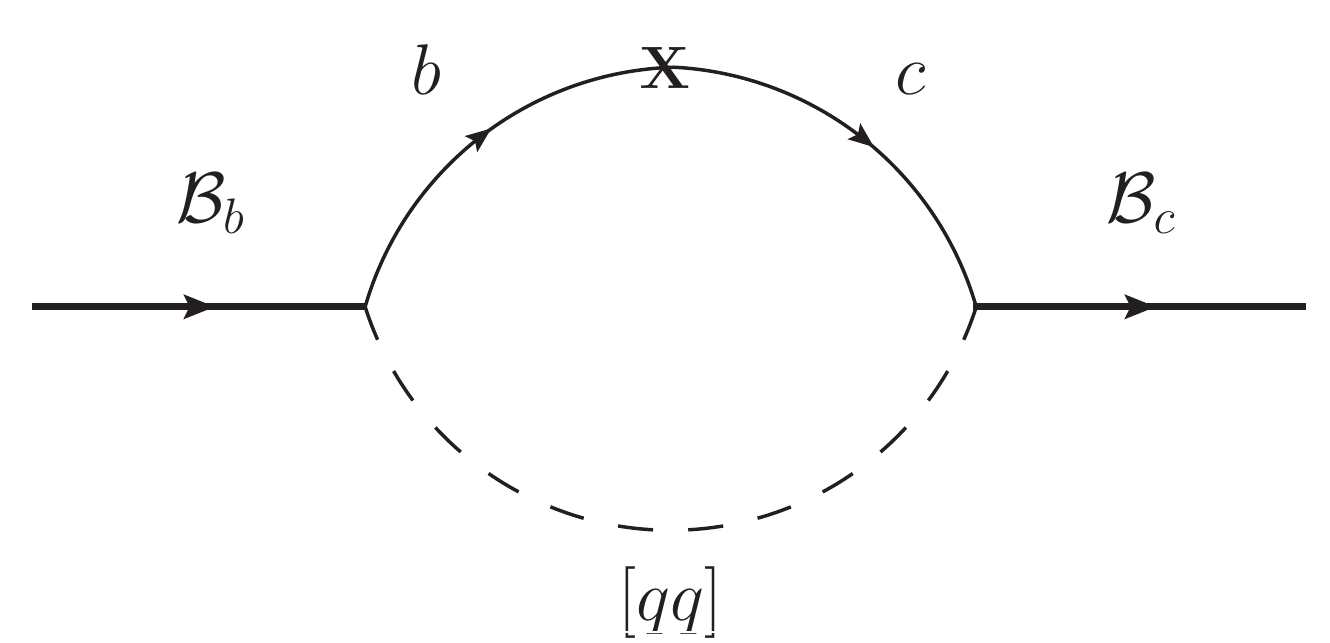}}
\caption{Feynman diagram for a typical $\B_b\to\B_c$
transition, where the scalar or axial-vector diquark $[qq]$
is denoted by a dashed line and the vertex corresponding to the $\bar c\gamma^\mu(1-\gamma_5) b$ current
is denoted by X. The orbital angular momentum of the $Q-[qq]$ system can be $s$-wave or $p$-wave.} \label{fig: BbBc} 
\end{figure}

The Feynman diagram for a typical $\B_b\to\B_c$
transition, is shown in Fig.~\ref{fig: BbBc}. 
For the $\B_{b}(1/2^+)\to \B_{c}(1/2^+)$ transition, 
the matrix elements of $\bar c\gamma_\mu  b$ and $\bar c\gamma_\mu \gamma_5 b$ currents
can be parameterized as
 \be
 \la \B_{c}(P',J'_z)|\bar c\gamma_\mu  b|\B_b(P,J_z)\ra
 =\bar u(P',J'_z)\Big[f^V_1(q^2)\gamma_\mu+i{f^V_2(q^2)\over M+M'} \sigma_{\mu\nu}q^\nu
    +{f^V_3(q^2)\over M+M'}q_\mu\Big] u(P,J_z),
 \non\\
 \la \B_{c}(P',J'_z)|\bar c\gamma_\mu \gamma_5 b|\B_b(P,J_z)\ra
 =\bar u(P',J'_z)\Big[g^A_1(q^2)\gamma_\mu+i{g^A_2(q^2)\over M-M'} \sigma_{\mu\nu}q^\nu
     +{g^A_3(q^2)\over M-M'}q_\mu\Big]\gamma_5 u(P,J_z),
 \label{eq:figi spin1/2}
 \en
with $q\equiv P-P'$. 
We find that it is more convenient to use $g_{2,3}/(M-M')$ instead of $g_{2,3}/(M+M')$ in the above matrix element.
Note that these parametrization are different from those used in \cite{Chua:2018lfa}.
Similarly, for the $\B_{b}(1/2^+)\to \B_{c}(1/2^-)$ transition, we have
  \be
 &&\la \B_{c}(P',J'_z)|\bar c\gamma_\mu  b|\B_b(P,J_z)\ra
 \non\\
 &&\qquad=\bar u(P',J'_z)\Big[g^V_1(q^2)\gamma_\mu+i{g^V_2(q^2)\over M-M'} \sigma_{\mu\nu}q^\nu
    +{g^V_3(q^2)\over M-M'}q_\mu\Big] \gamma_5u(P,J_z),
 \non\\
 &&\la \B_{c}(P',J'_z)|\bar c\gamma_\mu \gamma_5 b|\B_b(P,J_z)\ra
 \non\\
 &&\qquad=\bar u(P',J'_z)\Big[f^A_1(q^2)\gamma_\mu+i{f^A_2(q^2)\over M+M'} \sigma_{\mu\nu}q^\nu
     +{f^A_3(q^2)\over M+M'}q_\mu\Big] u(P,J_z).
 \label{eq:figi1 spin1/2}
 \en
For the $\B_{b}(1/2^+)\to \B_{c}(3/2^+)$ transition, 
the $V^\mu$ and $A^\mu$ matrix elements can be parameterized as
 \be
 \la \B_c(P',J'_z)|\bar c\gamma_\mu  b|\B_b(P,J_z)\ra
 &=&\bar u^\nu(P',J'_z)
 \Big[\bar f^V_1(q^2)g_{\nu\mu}+\frac{\bar f^V_2(q^2)}{M} P_\nu\gamma_\mu+\frac{\bar f^V_3(q^2)}{MM'}P_\nu P'_\mu
 \non\\
 &&\quad+\frac{\bar f^V_4(q^2)}{M^2}P_\nu P_\mu\Big]
 \gamma_5
 u(P,J_z),
 \non\\
\la \B_c(P',J'_z)|\bar c\gamma_\mu \gamma_5 b|\B_b(P,J_z)\ra
 &=&\bar u^\nu(P',J'_z)
  \Big[\bar g^A_1(q^2)g_{\nu\mu}+\frac{\bar g^A_2(q^2)}{M} P_\nu\gamma_\mu+\frac{\bar g^A_3(q^2)}{MM"}P_\nu P'_\mu
\non\\
&&\quad
+\frac{\bar g^A_4(q^2)}{M^2}P_\nu P_\mu\Big]  u (P,J_z),
 \label{eq:figi spin3/2}
 \en
while for the $\B_{b}(1/2^+)\to \B_{c}(3/2^-)$ transition, we have
 \be
 \la \B_c(P',J'_z)|\bar c\gamma_\mu  b|\B_b(P,J_z)\ra
 &=&\bar u^\nu(P',J'_z)
 \Big[\bar g^V_1(q^2)g_{\nu\mu}+\frac{\bar g^V_2(q^2)}{M} P_\nu\gamma_\mu+\frac{\bar g^V_3(q^2)}{MM'}P_\nu P'_\mu
 \non\\
 &&\quad+\frac{\bar g^V_4(q^2)}{M^2}P_\nu P_\mu\Big]
 u(P,J_z),
 \non\\
\la \B_c(P',J'_z)|\bar c\gamma_\mu \gamma_5 b|\B_b(P,J_z)\ra
 &=&\bar u^\nu(P',J'_z)
  \Big[f^A_1(q^2)\bar g_{\nu\mu}+\frac{\bar f^A_2(q^2)}{M} P_\nu\gamma_\mu+\frac{\bar f^A_3(q^2)}{MM"}P_\nu P'_\mu
\non\\
&&\quad
+\frac{\bar f^A_4(q^2)}{M^2}P_\nu P_\mu\Big]  \gamma_5 u (P,J_z).
 \label{eq:figi1 spin3/2}
 \en

Using the light-front quark model, the
general expression for a $\B_b(1/2)\to\B_c(1/2)$ [$\B_b(1/2)\to \B_c(3/2)$] transition matrix element is given by 
\be
&&
     \la \B_{c}(P',J'_z)|\bar c\gamma^\mu (1-\gamma_5) b|\B_{b}(P,J_z)\ra
\non\\
&&\qquad
     =\int \frac{dp^+_2 d^2 p_{2\bot}}{2(2\pi)^3}~\frac{\phi^{\prime*}_{n'L'_K}(\{x'\},\{k'_{\bot}\})\phi_{1 L_K}(\{x\},\{k_{\bot}\})}
                                       {2\sqrt{p_1^+ p_1^{\prime +}(p_1\cdot \bar P+m_1 M_0)(p'_1\cdot\bar P'+m'_1 M'_0)}}
\non\\
&&
     \qquad\qquad\qquad\times~
     \bar u^{(\mu)}(\bar P',J'_z)\bar \Gamma_{(\mu) L'_K S_{[qq]} J'_l} 
     (\not\! p'_1+m'_1)\gamma^\mu(1-\gamma_5)(\not\! p_1+m_1)\Gamma_{L_K S_{[qq]} J_l}  
     u(\bar P,J_z),
\label{eq: B->B'}
\en
where the diquark is spectating in the transition and the kinematics of the constituents are
\begin{eqnarray}
        && p^{(\prime)+}_i=x^{(\prime)}_i P^{(\prime)+},\qquad
           p^{(\prime)}_{i\bot}=x^{(\prime)}_i
           \vec P^{(\prime)}_\bot+\vec k^{(\prime)}_{i\bot},\qquad 1-\sum_{i=1}^2
           x^{(\prime)}_i=\sum_{i=1}^2 \vec k^{(\prime)}_{i\bot}=0,
        \non\\
        && (p_1-p_1^{\prime})^+=q^+,\qquad
         (\vec p_1-\vec p_1^{\,\prime})_\bot=\vec q_\bot,
         \qquad
        p_2=p_2^{\prime},
        \qquad
        p^{(\prime)2}_i=m^{(\prime) 2}_i.
\end{eqnarray}
In Eq.~(\ref{eq: B->B'}) $\bar\Gamma^\mu$ denotes $\gamma_0\Gamma^{\dagger \mu} \gamma_0$ with
the vertex functions $\Gamma^{(\mu)}_{L_K S_{[qq]} J_l}$ given by: 
\begin{eqnarray}
       \Gamma_{s00}(p_1,p_2)&=&1,
\non\\
       \Gamma_{s11}(p_1,p_2,\lambda_2)
       &=&\frac{\gamma_5}{\sqrt3}
          \bigg(\not\!\varepsilon_{LF}^*(p_2,\lambda_2)
        -
        \frac{M_0+m_1+m_2}{\bar P\cdot p_2+m_2 M_0}\varepsilon_{LF}^*(p_2,\lambda_2)\cdot \bar P\bigg),
\non\\
     \Gamma_{p01}(p_1,p_2)&=&\frac{\gamma_5}{2\sqrt3}
     \bigg(\not\! p_1-\not \! p_2-\frac{m_1^2-m_2^2}{M_0} \bigg),
\non\\
    \Gamma^\mu_{p01}(p_1,p_2)&=&-\frac{1}{2}(p_1-p_2)^\mu,    
\non\\
     \Gamma^\mu_{s11}(p_1,p_2,\lambda_2)
     &=&-\bigg(\varepsilon_{LF}^{*\mu}(p_2,\lambda_2)
        -\frac{p^\mu_2}{\bar P\cdot p_2+m_2 M_0}
        \varepsilon_{LF}^*(p_2,\lambda_2)\cdot \bar P\bigg),
\non\\  
     \Gamma^\mu_{p12}(p_1,p_2,\lambda_2)
     &=&-\frac{1}{2\sqrt{10}}\gamma_5
        \bigg[
         \bigg(\varepsilon^{*\mu}_{LF}(p_2,\lambda_2)+(p_1-p_2)^\mu 
         \frac{\varepsilon^{*}_{LF}(p_2,\lambda_2)\cdot\bar P}{\bar P\cdot p_2+M_0m_2}
        \bigg)
         \bigg(\not\! p_1-\not\! p_2- \frac{m_1^2-m_2^2}{M_0}\bigg)
\non\\
&&
         +(p_1-p_2)^\mu\bigg(\not\! \varepsilon^*_{LF}(p_2,\lambda_2) 
                                           -\frac{\varepsilon^*_{LF}(p_2,\lambda_2)\cdot\bar P}{M_0}\bigg)\bigg],           
\label{eq: Gamma}
\end{eqnarray}
for baryon states with a $S_2=0$ or $S_2=1$ diquark. 
Note that the vertex functions $\Gamma_{s00}$, $\Gamma_{s11}$ and $\Gamma_{p01}$
are taken from \cite{Chua:2018lfa}, while $\Gamma^\mu$s are new and the derivations
can be found in Appendices \ref{appendix: wave functions} and \ref{appendix: vertex}.
For the wave function, we have 
\be
\phi_{n L_K}\equiv\sqrt{\frac{d k_{2z}}{d x_2}}\varphi_{n L_K}.
\label{eq: phinLK}
\en
We shall use a Gaussian-like wave function in this work: \cite{Cheng97,CCH,Chua:2018lfa}
\begin{eqnarray} 
\varphi_{n=1, L_K=s}(\vec k,\beta)
    &=&4 \left({\pi\over{\beta^{2}}}\right)^{3\over{4}}
               ~{\rm exp}
               \left(-{k^2_z+k^2_\bot\over{2 \beta^2}}\right),
\non\\               
\varphi_{n=1,L_K=p}(\vec k,\beta)
    &=&\sqrt{2\over{\beta^2}}~\varphi_{n=1}(\vec k,\beta),
\non\\
\varphi_{n=2,L_K=s}(\vec k,\beta)
   &=& \sqrt{\frac{3}{2}}\bigg(1-\frac{2}{3}\frac{\vec k^2}{\beta^2}\bigg) \varphi_{n=1}(\vec k,\beta), 
\non\\   
\varphi_{n=2,L_K=p}(\vec k,\beta)
   &=&\sqrt{\frac{5}{2}}\bigg(1-\frac{2}{5}\frac{\vec k^2}{\beta^2}\bigg) \varphi_{n=1,L_K=p}(\vec k,\beta).
 \label{eq:wavefn}
\end{eqnarray}
One is referred to Appendix A for more details.

We shall extend the formulars in~\cite{Schlumpf, Cheng:2004cc, Chua:2018lfa} to project out all the form
factors from the $\B_b(1/2^+)\to \B_c(1/2^\pm)$ and $\B_b(1/2^+)\to \B_c(3/2^\pm)$ transition matrix elements shown in Eq.~(\ref{eq: B->B'}).
As in \cite{CCH,Schlumpf, Cheng:2004cc, Chua:2018lfa}, we consider the $q^+=0$, $\vec q_\bot\not=\vec 0$ case.
By applying the following identities 
to Eq.~(\ref{eq:figi spin1/2}) for $V^+$, $A^+$, $\vec q_\bot\cdot\vec V$, $\vec q_\bot\cdot\vec A$, $\vec n_\bot\cdot\vec V$ and $\vec n_\bot\cdot\vec A$,  
\be
 \frac{\bar u(P,J_z)\gamma^+ u(P', J'_z)}{2\sqrt{P^+ P^{\prime
 +}}}&=&\delta_{J_z J'_z},
 \qquad
 i\frac{\bar u(P,J_z)\sigma^{+\nu} a_{\bot\nu} u(P', J'_z)}{2\sqrt{P^+
P^{\prime
 +}}}=(\vec\sigma\cdot \vec a_\bot\sigma^3)_{J_z J'_z},
  \non\\
  \frac{\bar u(P,J_z)\gamma^+\gamma_5 u(P', J'_z)}{2\sqrt{P^+ P^{\prime
 +}}}&=&(\sigma^3)_{J_z J'_z},
 \quad
 i\frac{\bar u(P,J_z)\sigma^{+\nu} a_{\bot\nu}\gamma_5 u(P', J'_z)}{2\sqrt{P^+
P^{\prime
 +}}}=(\vec\sigma\cdot \vec a_\bot)_{J_z J'_z},
 \label{eq:spinorprojection spin1/2}
  \en
with $a_\bot=(0,0,\vec a_\bot)$, where $\vec a_\bot$ is an arbitrary 2-$d$ vector,
and the following identities  
to Eq.~(\ref{eq: B->B'}),
  \be
 \sum_{J_z,J'_z} u(\bar P, J_z)\delta_{J_z J'_z}\bar u(\bar P',J'_z)
  &=&\frac{1}{2\sqrt{P^+P^{\prime+}}}(\not\!\bar P+M_0)\gamma^+(\not\!\bar P'+M'_0),
 \non\\
 \sum_{J_z,J'_z} u(\bar P, J_z)(\vec\sigma\cdot\vec a_\bot\sigma^3)_{J_z J'_z}\bar u(\bar P',J'_z)
  &=&\frac{i}{2\sqrt{P^+P^{\prime+}}}(\not\!\bar P+M_0)\sigma^{+\nu}a_{\bot\nu} (\not\!\bar P'+M'_0),
 \non\\
 \sum_{J_z,J'_z} u(\bar P, J_z)(\sigma^3)_{J_z J'_z}\bar u(\bar P',J'_z)
  &=&\frac{1}{2\sqrt{P^+P^{\prime+}}}(\not\!\bar P+M_0)\gamma^+\gamma_5(\not\!\bar P'+M'_0),
 \non\\
 \sum_{J_z,J'_z} u(\bar P, J_z)(\vec\sigma\cdot \vec a_\bot)_{J_z J'_z}\bar u(\bar P',J'_z)
  &=&\frac{i}{2\sqrt{P^+P^{\prime+}}}(\not\!\bar P+M_0)\sigma^{+\nu}a_{\bot\nu}\gamma_5(\not\!\bar P'+M'_0),
  \label{eq:spinorprojectionbar spin1/2}
 \en
we obtain 
\be
f^V_1(q^2) & =&\int \frac{dp^+_2 d^2 p_{2\bot}}{2(2\pi)^3}~
      \frac{\phi^{\prime*}_{n'L'_K}(\{x'\},\{k'_{\bot}\})\phi_{1 L_K}(\{x\},\{k_{\bot}\})}
             {16 P^+P^{\prime+}\sqrt{p_1^{\prime +} p_1^+ (p'_1\cdot\bar P'+m'_1 M'_0)(p_1\cdot \bar P+m_1 M_0)}}
\non\\
&&
     \times~Tr[(\not\!\bar P+M_0)\gamma^+(\not\!\bar P'+M'_0)\bar \Gamma_{L'_K S_{[qq]} J'_l} 
     (\not\! p'_1+m'_1)\gamma^+(\not\! p_1+m_1)\Gamma_{L_K S_{[qq]} J_l}],
\non\\
\frac{f^V_2(q^2) q^2}{M+M'} 
&=&i\int \frac{dp^+_2 d^2 p_{2\bot}}{2(2\pi)^3}~
      \frac{\phi^{\prime*}_{n'L'_K}(\{x'\},\{k'_{\bot}\})\phi_{1 L_K}(\{x\},\{k_{\bot}\})}
             {16 P^+P^{\prime+}\sqrt{p_1^{\prime +} p_1^+ (p'_1\cdot\bar P'+m'_1 M'_0)(p_1\cdot \bar P+m_1 M_0)}}
\non\\
&&
     \times~Tr[(\not\!\bar P+M_0)\sigma^{+\nu} q_{\bot\nu}(\not\!\bar P'+M'_0)\bar \Gamma_{L'_K S_{[qq]} J'_l} 
     (\not\! p'_1+m'_1)\gamma^+(\not\! p_1+m_1)\Gamma_{L_K S_{[qq]} J_l}],
\non\\
f^V_3(q^2)-f^V_1(q^2) & =&\int \frac{dp^+_2 d^2 p_{2\bot}}{2(2\pi)^3}~
      \frac{\phi^{\prime*}_{n'L'_K}(\{x'\},\{k'_{\bot}\})\phi_{1 L_K}(\{x\},\{k_{\bot}\})}
             {8 q^2 \sqrt{P^+P^{\prime +}}\sqrt{p_1^{\prime +} p_1^+ (p'_1\cdot\bar P'+m'_1 M'_0)(p_1\cdot \bar P+m_1 M_0)}}
\non\\
&&
     \times~Tr[(\not\!\bar P+M_0)\gamma^+(\not\!\bar P'+M'_0)\bar \Gamma_{L'_K S_{[qq]} J'_l} 
     (\not\! p'_1+m'_1){\not\! {q}_\bot}(\not\! p_1+m_1)\Gamma_{L_K S_{[qq]} J_l}],
\non\\
f^V_1(q^2)-f^V_2(q^2)  & =&\frac{-i}{M-M'}\int \frac{dp^+_2 d^2 p_{2\bot}}{2(2\pi)^3}~
      \frac{\phi^{\prime*}_{n'L'_K}(\{x'\},\{k'_{\bot}\})\phi_{1 L_K}(\{x\},\{k_{\bot}\})}
             {8 \sqrt{P^+P^{\prime +}}\sqrt{p_1^{\prime +} p_1^+ (p'_1\cdot\bar P'+m'_1 M'_0)(p_1\cdot \bar P+m_1 M_0)}}
\non\\
&&
     \times~Tr[(\not\!\bar P+M_0)\sigma^{+\nu}n_{\bot\nu}(\not\!\bar P'+M'_0)\bar \Gamma_{L'_K S_{[qq]} J'_l} 
     (\not\! p'_1+m'_1){\not\! n}_\bot(\not\! p_1+m_1)\Gamma_{L_K S_{[qq]} J_l}],
\non\\
g^A_1(q^2) & =&\int \frac{dp^+_2 d^2 p_{2\bot}}{2(2\pi)^3}~
      \frac{\phi^{\prime*}_{n'L'_K}(\{x'\},\{k'_{\bot}\})\phi_{1 L_K}(\{x\},\{k_{\bot}\})}
             {16 P^+P^{\prime+}\sqrt{p_1^{\prime +} p_1^+ (p'_1\cdot\bar P'+m'_1 M'_0)(p_1\cdot \bar P+m_1 M_0)}}
\non\\
&&
     \quad\times~Tr[(\not\!\bar P+M_0)\gamma^+\gamma_5(\not\!\bar P'+M'_0)\bar \Gamma_{L'_K S_{[qq]} J'_l} 
     (\not\! p'_1+m'_1)\gamma^+\gamma_5(\not\! p_1+m_1)\Gamma_{L_K S_{[qq]} J_l}],
\non\\
\frac{g^A_2(q^2) q^2}{M-M'} 
&=&-i\int \frac{dp^+_2 d^2 p_{2\bot}}{2(2\pi)^3}~
      \frac{\phi^{\prime*}_{n'L'_K}(\{x'\},\{k'_{\bot}\})\phi_{1 L_K}(\{x\},\{k_{\bot}\})}
             {16 P^+P^{\prime+}\sqrt{p_1^{\prime +} p_1^+ (p'_1\cdot\bar P'+m'_1 M'_0)(p_1\cdot \bar P+m_1 M_0)}}
\non\\
&&
     \times~Tr[(\not\!\bar P+M_0)\sigma^{+\nu}q_{\bot\nu}\gamma_5(\not\!\bar P'+M'_0)\bar \Gamma_{L'_K S_{[qq]} J'_l} 
     (\not\! p'_1+m'_1)\gamma^+\gamma_5(\not\! p_1+m_1)\Gamma_{L_K S_{[qq]} J_l}],
\non\\
g^A_1(q^2)+g^A_3(q^2) & =&-\int \frac{dp^+_2 d^2 p_{2\bot}}{2(2\pi)^3}~
      \frac{\phi^{\prime*}_{n'L'_K}(\{x'\},\{k'_{\bot}\})\phi_{1 L_K}(\{x\},\{k_{\bot}\})}
             {8 q^2 \sqrt{P^+P^{\prime +}}\sqrt{p_1^{\prime +} p_1^+ (p'_1\cdot\bar P'+m'_1 M'_0)(p_1\cdot \bar P+m_1 M_0)}}
\non\\
&&
     \times~Tr[(\not\!\bar P+M_0)\gamma^+\gamma_5(\not\!\bar P'+M'_0)\bar \Gamma_{L'_K S_{[qq]} J'_l} 
     (\not\! p'_1+m'_1){\not\! {q}_\bot}\gamma_5(\not\! p_1+m_1)\Gamma_{L_K S_{[qq]} J_l}],
\non\\
g^A_1(q^2) + g^A_2(q^2) 
&=&\frac{-i}{M+M'}\int \frac{dp^+_2 d^2 p_{2\bot}}{2(2\pi)^3}~
      \frac{\phi^{\prime*}_{n'L'_K}(\{x'\},\{k'_{\bot}\})\phi_{1 L_K}(\{x\},\{k_{\bot}\})}
             {8 \sqrt{P^+P^{\prime +}}\sqrt{p_1^{\prime +} p_1^+ (p'_1\cdot\bar P'+m'_1 M'_0)(p_1\cdot \bar P+m_1 M_0)}}
\non\\
&&
     \times~Tr[(\not\!\bar P+M_0)\sigma^{+\nu}n_{\bot\nu}\gamma_5(\not\!\bar P'+M'_0)\bar \Gamma_{L'_K S_{[qq]} J'_l} 
     (\not\! p'_1+m'_1){\not\! n}_\bot\gamma_5(\not\! p_1+m_1)\Gamma_{L_K S_{[qq]} J_l}],
\non\\
\label{eq: figi (i) and (ii)}
\en
with $q_\bot\equiv (0,0,\vec q_\bot)$, $n_\bot\equiv (0,0,\vec n_\bot)$, $n_\bot^2=-1$ and $\vec n_\bot\cdot \vec q_\bot=0$. 
In ref.~\cite{Chua:2018lfa,Cheng:2004cc,Schlumpf} only $f_{1,2}$ and $g_{1,2}$ can be extracted. Now we can extract all form factors.
Note that the above equations are over constraining in determining $f_i$ and $g_i$. There are consistency relations needed to be satisfied. The equations involving $f_1$, $f_2$ and $f_1-f_2$ need to be consistent and likewise for those involving $g_1$, $g_2$ and $g_1+g_2$.

Similarly, for $\B_b(1/2^+)\to\B_c(1/2^-)$ transition, we have
\be
f^A_1(q^2) & =&\int \frac{dp^+_2 d^2 p_{2\bot}}{2(2\pi)^3}~
      \frac{\phi^{\prime*}_{n'L'_K}(\{x'\},\{k'_{\bot}\})\phi_{1 L_K}(\{x\},\{k_{\bot}\})}
             {16 P^+P^{\prime+}\sqrt{p_1^{\prime +} p_1^+ (p'_1\cdot\bar P'+m'_1 M'_0)(p_1\cdot \bar P+m_1 M_0)}}
\non\\
&&
     \times~Tr[(\not\!\bar P+M_0)\gamma^+(\not\!\bar P'+M'_0)\bar \Gamma_{L'_K S_{[qq]} J'_l} 
     (\not\! p'_1+m'_1)\gamma^+\gamma_5(\not\! p_1+m_1)\Gamma_{L_K S_{[qq]} J_l}],
\non\\
\frac{f^A_2(q^2) q^2}{M+M'} 
&=&i\int \frac{dp^+_2 d^2 p_{2\bot}}{2(2\pi)^3}~
      \frac{\phi^{\prime*}_{n'L'_K}(\{x'\},\{k'_{\bot}\})\phi_{1 L_K}(\{x\},\{k_{\bot}\})}
             {16 P^+P^{\prime+}\sqrt{p_1^{\prime +} p_1^+ (p'_1\cdot\bar P'+m'_1 M'_0)(p_1\cdot \bar P+m_1 M_0)}}
\non\\
&&
     \times~Tr[(\not\!\bar P+M_0)\sigma^{+\nu}q_{\bot\nu}(\not\!\bar P'+M'_0)\bar \Gamma_{L'_K S_{[qq]} J'_l} 
     (\not\! p'_1+m'_1)\gamma^+\gamma_5(\not\! p_1+m_1)\Gamma_{L_K S_{[qq]} J_l}],
\non\\
f^A_3(q^2)-f^A_1(q^2) & =&\int \frac{dp^+_2 d^2 p_{2\bot}}{2(2\pi)^3}~
      \frac{\phi^{\prime*}_{n'L'_K}(\{x'\},\{k'_{\bot}\})\phi_{1 L_K}(\{x\},\{k_{\bot}\})}
             {8 q^2 \sqrt{P^+P^{\prime +}}\sqrt{p_1^{\prime +} p_1^+ (p'_1\cdot\bar P'+m'_1 M'_0)(p_1\cdot \bar P+m_1 M_0)}}
\non\\
&&
     \times~Tr[(\not\!\bar P+M_0)\gamma^+(\not\!\bar P'+M'_0)\bar \Gamma_{L'_K S_{[qq]} J'_l} 
     (\not\! p'_1+m'_1){\not\! {q}_\bot}\gamma_5(\not\! p_1+m_1)\Gamma_{L_K S_{[qq]} J_l}],
\non\\
f^A_1(q^2)-f^A_2(q^2)  & =&\frac{-i}{M-M'}\int \frac{dp^+_2 d^2 p_{2\bot}}{2(2\pi)^3}~
      \frac{\phi^{\prime*}_{n'L'_K}(\{x'\},\{k'_{\bot}\})\phi_{1 L_K}(\{x\},\{k_{\bot}\})}
             {8 \sqrt{P^+P^{\prime +}}\sqrt{p_1^{\prime +} p_1^+ (p'_1\cdot\bar P'+m'_1 M'_0)(p_1\cdot \bar P+m_1 M_0)}}
\non\\
&&
     \times~Tr[(\not\!\bar P+M_0)\sigma^{+\nu}n_{\bot\nu}(\not\!\bar P'+M'_0)\bar \Gamma_{L'_K S_{[qq]} J'_l} 
     (\not\! p'_1+m'_1){\not\! n}_\bot\gamma_5(\not\! p_1+m_1)\Gamma_{L_K S_{[qq]} J_l}],
\non\\
g^V_1(q^2) & =&\int \frac{dp^+_2 d^2 p_{2\bot}}{2(2\pi)^3}~
      \frac{\phi^{\prime*}_{n'L'_K}(\{x'\},\{k'_{\bot}\})\phi_{1 L_K}(\{x\},\{k_{\bot}\})}
             {16 P^+P^{\prime+}\sqrt{p_1^{\prime +} p_1^+ (p'_1\cdot\bar P'+m'_1 M'_0)(p_1\cdot \bar P+m_1 M_0)}}
\non\\
&&
     \times~Tr[(\not\!\bar P+M_0)\gamma^+\gamma_5(\not\!\bar P'+M'_0)\bar \Gamma_{L'_K S_{[qq]} J'_l} 
     (\not\! p'_1+m'_1)\gamma^+(\not\! p_1+m_1)\Gamma_{L_K S_{[qq]} J_l}],
\non\\
\frac{g^V_2(q^2) q^2}{M-M'} 
&=&-i\int \frac{dp^+_2 d^2 p_{2\bot}}{2(2\pi)^3}~
      \frac{\phi^{\prime*}_{n'L'_K}(\{x'\},\{k'_{\bot}\})\phi_{1 L_K}(\{x\},\{k_{\bot}\})}
             {16 P^+P^{\prime+}\sqrt{p_1^{\prime +} p_1^+ (p'_1\cdot\bar P'+m'_1 M'_0)(p_1\cdot \bar P+m_1 M_0)}}
\non\\
&&
     \times~Tr[(\not\!\bar P+M_0)\sigma^{+\nu}q_{\bot\nu}\gamma_5(\not\!\bar P'+M'_0)\bar \Gamma_{L'_K S_{[qq]} J'_l} 
     (\not\! p'_1+m'_1)\gamma^+(\not\! p_1+m_1)\Gamma_{L_K S_{[qq]} J_l}],
\non\\ 
g^V_1(q^2)+g^V_3(q^2) & =&-\int \frac{dp^+_2 d^2 p_{2\bot}}{2(2\pi)^3}~
     \frac{\phi^{\prime*}_{n'L'_K}(\{x'\},\{k'_{\bot}\})\phi_{1 L_K}(\{x\},\{k_{\bot}\})}
             {8 q^2 \sqrt{P^+P^{\prime +}}\sqrt{p_1^{\prime +} p_1^+ (p'_1\cdot\bar P'+m'_1 M'_0)(p_1\cdot \bar P+m_1 M_0)}}
\non\\
&&
     \times~Tr[(\not\!\bar P+M_0)\gamma^+\gamma_5(\not\!\bar P'+M'_0)\bar \Gamma_{L'_K S_{[qq]} J'_l} 
     (\not\! p'_1+m'_1){\not\! {q}_\bot}\gamma_5(\not\! p_1+m_1)\Gamma_{L_K S_{[qq]} J_l}],
\non\\
g^V_1(q^2) + g^V_2(q^2) 
&=&\frac{-i}{M+M'}\int \frac{dp^+_2 d^2 p_{2\bot}}{2(2\pi)^3}~
      \frac{\phi^{\prime*}_{n'L'_K}(\{x'\},\{k'_{\bot}\})\phi_{1 L_K}(\{x\},\{k_{\bot}\})}
             {8 \sqrt{P^+P^{\prime +}}\sqrt{p_1^{\prime +} p_1^+ (p'_1\cdot\bar P'+m'_1 M'_0)(p_1\cdot \bar P+m_1 M_0)}}
\non\\
&&
     \times~Tr[(\not\!\bar P+M_0)\sigma^{+\nu}n_{\bot\nu}\gamma_5(\not\!\bar P'+M'_0)\bar \Gamma_{L'_K S_{[qq]} J'_l} 
     (\not\! p'_1+m'_1){\not\! n}_\bot(\not\! p_1+m_1)\Gamma_{L_K S_{[qq]} J_l}],
\non\\  
\label{eq: figi (v)}
\en
with $q_\bot\equiv (0,0,\vec q_\bot)$, $n_\bot\equiv (0,0,\vec n_\bot)$, $n_\bot^2=-1$ and $\vec n_\bot\cdot \vec q_\bot=0$. 

The following identities are useful in extracting $1/2\to3/2$ transition form factors,
 \be
\delta_{S_z S'_z}      
&=&\sqrt{\frac{3}{2}}\frac{M'}{P^{\prime +}}\frac{\bar u(P',S'_z)\gamma^+ u^+(P, S_z)}
       {2\sqrt{P^+ P^{\prime+}}},
\non\\
(\vec\sigma\cdot \vec a_\bot\sigma^3)_{S'_z S_z}
&=& \sqrt{\frac{3}{2}}\frac{M'}{P^{\prime +}}i\frac{\bar u(P,S_z)\sigma^{+\nu} a_{\bot\nu} u^+(P', S'_z)}
        {2\sqrt{P^+P^{\prime +}}},
\non\\
(\sigma^3)_{S_z S'_z}
&=&\sqrt{\frac{3}{2}}\frac{M'}{P^{\prime +}}\frac{\bar u(P,S_z)\gamma^+\gamma_5 u^+(P', S'_z)}
       {2\sqrt{P^+ P^{\prime +}}},
\non\\
(\vec\sigma\cdot \vec a_\bot)_{S_z S'_z}
&=&
\sqrt{\frac{3}{2}}\frac{M'_0}{P^{\prime +}}i\frac{\bar u(P,S_z)\sigma^{+\nu} a_{\bot\nu}\gamma_5 u^+(P', S'_z)}
       {2\sqrt{P^+ P^{\prime +}}},      
\label{eq:spinorprojection spin3/2}
\en
and~\footnote{Note that since we have $u^+(\bar P',\pm3/2)=u(\bar P',\pm1/2)\varepsilon^+_{LF}(\pm 1)=0$, see Eq.~(\ref{polcom}), we can easily promote the sum in $S'_z=-1/2,+1/2$ to $J'_z=-3/2,-1/2,1/2,3/2$ and, consequently, the convenient full polarization sum formula, Eq. (\ref{eq: Pmunu}) can be used.}
 \be
   \sum_{S_z,S'_z=\pm1/2} u(\bar P, S_z)(\sigma^3)_{S_z S'_z}\bar u^\mu(\bar P',S'_z)
 &=&\frac{\sqrt 3 M_0'}{2\sqrt 2 P^{\prime+}\sqrt{P^+P^{\prime +}}}(\not\!\bar P+M_0)\gamma^+\gamma_5{\cal P}^{+\mu}(\bar P'),
 \non\\
 \sum_{S_z,S'_z=\pm1/2} u(\bar P, S_z)(\sigma_\bot\cdot\vec a_\bot)_{S_z S'_z}\bar u^\mu(\bar P',S'_z)
  &=&\frac{\sqrt 3 M_0' i}{2\sqrt 2 P^{\prime+}\sqrt{P^+P^{\prime +}}}(\not\!\bar P+M_0)\sigma^{+\nu}a_{\bot\nu}\gamma_5{\cal P}^{+\mu}(\bar P'),
  \non\\
 \sum_{S_z,S'_z=\pm1/2} u(\bar P, S_z)\delta_{S_z S'_z}\bar u^\mu(\bar P',S'_z)
  &=&\frac{\sqrt 3 M_0'}{2\sqrt 2 P^{\prime+}\sqrt{P^+P^{\prime +}}}(\not\!\bar P+M_0)\gamma^+{\cal P}^{+\mu}(\bar P'),
 \non\\
 \sum_{S_z,S'_z=\pm1/2} u^\mu(\bar P, S_z)(\vec\sigma_{\bot}\cdot\vec a_\bot \sigma^3 )_{S_z S'_z}\bar u^\mu(\bar P',S'_z)
 &=&\frac{\sqrt 3 M_0' i}{2\sqrt 2 P^{\prime+}\sqrt{P^+P^{\prime +}}}(\not\!\bar P+M_0)\sigma^{+\nu}a_{\bot\nu}{\cal P}^{+\mu(\bar P')},
  \label{eq:spinorprojectionbar spin3/2}
 \en
with (see, for exmple, \cite{Moroi:1995fs})
\be
{\cal P}_{\mu\nu}(\bar P')\equiv\sum_{J'_z=-3/2}^{3/2}u_\mu(\bar P', J'_z) \bar u_\nu(\bar P', J'_z)
=-(\not\!\bar P'+M'_0)
\bigg(G_{\mu\nu}(\bar P')-\frac{1}{3} G_{\mu\sigma}(\bar P') G_{\nu\lambda}(\bar P')\gamma^\sigma\gamma^\lambda\bigg),
\label{eq: Pmunu}
\en
where $G_{\mu\nu}(\bar P')$ is defined as
\be
G_{\mu\nu}(\bar P')\equiv g_{\mu\nu}-\frac{\bar P'_\mu \bar P'_\nu}{M^{\prime 2}_0}.
\en
We apply Eq.~(\ref{eq:spinorprojection spin3/2}) to Eq.~(\ref{eq:figi spin3/2}) for $V^+$, $A^+$, 
$\vec q_\bot\cdot\vec V$, $\vec q_\bot\cdot\vec A$, $\vec n_\bot\cdot\vec V$ and $\vec n_\bot\cdot\vec A$, and apply Eq.~(\ref{eq:spinorprojectionbar spin3/2}) to Eq.~(\ref{eq: B->B'}),
and obtain, for the $\B_b(1/2^+)\to\B_c(3/2^+)$ transition, 
\be
{\cal F}^V_1(q^2) & =&\int \frac{dx_2 d^2 k_{2\bot}}{2 (2\pi)^3}~\frac{3M_0'}{P^{\prime +}}
      \frac{\phi^{\prime*}_{n'L'_K}(\{x'\},\{k'_{\bot}\})\phi_{1 L_K}(\{x\},\{k_{\bot}\})}
             {8 P^+P^{\prime +}\sqrt{[(m_1+x_1 M_0)^2+k_{1\bot}^2][(m'_1+x_1 M'_0)^2+k_{1\bot}^{\prime 2}]}}
\non\\
&&
     \qquad\times~Tr[(\not\!\bar P+M_0)\gamma^+\gamma_5{\cal P}^{+\mu}\bar \Gamma_{\mu L'_K S_{[qq]} J'_l} 
     (\not\! p'_1+m'_1)\gamma^+(\not\! p_1+m_1)\Gamma_{L_K S_{[qq]} J_l}],
\non\\
{\cal F}^V_2(q^2) 
&=&-i \int \frac{dx_2 d^2 k_{2\bot}}{2 (2\pi)^3}~\frac{3M_0'}{P^{\prime +}}
      \frac{\phi^{\prime*}_{n'L'_K}(\{x'\},\{k'_{\bot}\})\phi_{1 L_K}(\{x\},\{k_{\bot}\})}
             {8 P^+P^{\prime +}\sqrt{[(m_1+x_1 M_0)^2+k_{1\bot}^2][(m'_1+x_1 M'_0)^2+k_{1\bot}^{\prime 2}]}}
\non\\
&&
     \qquad\times~Tr[(\not\!\bar P+M_0)\sigma^{+\nu}q_{\bot\nu}\gamma_5{\cal P}^{+\mu}\bar \Gamma_{\mu L'_K S_{[qq]} J'_l} 
     (\not\! p'_1+m'_1)\gamma^+(\not\! p_1+m_1)\Gamma_{L_K S_{[qq]} J_l}],
\non\\
{\cal F}^V_3(q^2) & =&\int \frac{dx_2 d^2 k_{2\bot}}{2 (2\pi)^3}~\frac{3M_0'}{P^{\prime+}}
      \frac{\phi^{\prime*}_{n'L'_K}(\{x'\},\{k'_{\bot}\})\phi_{1 L_K}(\{x\},\{k_{\bot}\})}
             {4 \sqrt{P^+P^{\prime +}}\sqrt{[(m_1+x_1 M_0)^2+k_{1\bot}^2][(m'_1+x_1 M'_0)^2+k_{1\bot}^{\prime 2}]}}
\non\\
&&
     \qquad\times~Tr[(\not\!\bar P+M_0)\gamma^+\gamma_5{\cal P}^{+\mu}\bar \Gamma_{\mu L'_K S_{[qq]} J'_l} 
     (\not\! p'_1+m'_1){\not\! {q}_\bot}(\not\! p_1+m_1)\Gamma_{L_K S_{[qq]} J_l}],
\non\\
{\cal F}^V_4(q^2) 
&=&-i \int \frac{dx_2 d^2 k_{2\bot}}{2 (2\pi)^3}~\frac{3M_0'}{P^{\prime +}}
      \frac{\phi^{\prime*}_{n'L'_K}(\{x'\},\{k'_{\bot}\})\phi_{1 L_K}(\{x\},\{k_{\bot}\})}
             {4 \sqrt{P^+P^{\prime +}}\sqrt{[(m_1+x_1 M_0)^2+k_{1\bot}^2][(m'_1+x_1 M'_0)^2+k_{1\bot}^{\prime 2}]}}
\non\\
&&
     \qquad\times~Tr[(\not\!\bar P+M_0)\sigma^{+\nu}q_{\bot\nu}\gamma_5{\cal P}^{+\mu}\bar \Gamma_{\mu L'_K S_{[qq]} J'_l} 
     (\not\! p'_1+m'_1){\not\!n_\bot}(\not\! p_1+m_1)\Gamma_{L_K S_{[qq]} J_l}],
\non\\
{\cal G}^A_1(q^2) & =&\int \frac{dx_2 d^2 k_{2\bot}}{2 (2\pi)^3}~\frac{3M_0'}{P^{\prime +}}
      \frac{\phi^{\prime*}_{n'L'_K}(\{x'\},\{k'_{\bot}\})\phi_{1 L_K}(\{x\},\{k_{\bot}\})}
             {8 P^+P^{\prime +}\sqrt{[(m_1+x_1 M_0)^2+k_{1\bot}^2][(m'_1+x_1 M'_0)^2+k_{1\bot}^{\prime 2}]}}
\non\\
&&
     \qquad\times~Tr[(\not\!\bar P+M_0)\gamma^+{\cal P}^{+\mu}\bar \Gamma_{\mu L'_K S_{[qq]} J'_l}  
     (\not\! p'_1+m'_1)\gamma^+\gamma_5(\not\! p_1+m_1)\Gamma_{L_K S_{[qq]} J_l}],
\non\\
{\cal G}^A_2(q^2)
&=&i\int \frac{dx_2 d^2 k_{2\bot}}{2 (2\pi)^3}~\frac{3M_0'}{P^{\prime +}}
      \frac{\phi^{\prime*}_{n'L'_K}(\{x'\},\{k'_{\bot}\})\phi_{1 L_K}(\{x\},\{k_{\bot}\})}
             {8 P^+P^{\prime +}\sqrt{[(m_1+x_1 M_0)^2+k_{1\bot}^2][(m'_1+x_1 M'_0)^2+k_{1\bot}^{\prime 2}]}}
\non\\
&&
     \qquad\times~Tr[(\not\!\bar P+M_0)\sigma^{+\nu}q_{\bot\nu}{\cal P}^{+\mu}\bar \Gamma_{\mu L'_K S_{[qq]} J'_l} 
     (\not\! p'_1+m'_1)\gamma^+\gamma_5(\not\! p_1+m_1)\Gamma_{L_K S_{[qq]} J_l}].
\non\\
{\cal G}^A_3(q^2) & =&\int \frac{dx_2 d^2 k_{2\bot}}{2 (2\pi)^3}~\frac{3M_0'}{P^{\prime+}}
      \frac{\phi^{\prime*}_{n'L'_K}(\{x'\},\{k'_{\bot}\})\phi_{1 L_K}(\{x\},\{k_{\bot}\})}
             {4 \sqrt{P^+P^{\prime +}}\sqrt{[(m_1+x_1 M_0)^2+k_{1\bot}^2][(m'_1+x_1 M'_0)^2+k_{1\bot}^{\prime 2}]}}
\non\\
&&
     \qquad\times~Tr[(\not\!\bar P+M_0)\gamma^+{\cal P}^{+\mu}\bar \Gamma_{\mu L'_K S_{[qq]} J'_l} 
     (\not\! p'_1+m'_1){\not\! {q}_\bot}\gamma_5(\not\! p_1+m_1)\Gamma_{L_K S_{[qq]} J_l}],
\non\\
{\cal G}^A_4(q^2)
&=&i\int \frac{dx_2 d^2 k_{2\bot}}{2 (2\pi)^3}~\frac{3M_0'}{P^{\prime +}}
      \frac{\phi^{\prime*}_{n'L'_K}(\{x'\},\{k'_{\bot}\})\phi_{1 L_K}(\{x\},\{k_{\bot}\})}
             {8 P^+P^{\prime +}\sqrt{[(m_1+x_1 M_0)^2+k_{1\bot}^2][(m'_1+x_1 M'_0)^2+k_{1\bot}^{\prime 2}]}}
\non\\
&&
     \qquad\times~Tr[(\not\!\bar P+M_0)\sigma^{+\nu}n_{\bot\nu}{\cal P}^{+\mu}\bar \Gamma_{\mu L'_K S_{[qq]} J'_l} 
     (\not\! p'_1+m'_1){\not\! n_\bot}\gamma_5(\not\! p_1+m_1)\Gamma_{L_K S_{[qq]} J_l}],
\non\\
\label{eq: FG (iii)}
\en
with $q_\bot\equiv (0,0,\vec q_\bot)$, $n_\bot\equiv (0,0,\vec n_\bot)$, $n_\bot^2=-1$, $\vec n_\bot\cdot \vec q_\bot=0$
and
 \be
 {\cal F}_1^V(q^2)
&\equiv&\frac{M'-M}{M'}\bigg\{
\bar f_1^V(q^2)
-\bigg(M+M'+\frac{q^2}{(M'-M)}\bigg) \frac{\bar f_2^V(q^2)}{M}
\non\\
&&+\frac{1}{2}\bigg(M^2-M^{\prime 2}-q^2\frac{2 M'-M}{M'-M}\bigg) \bigg(\frac{\bar f_3^V(q^2)}{MM'}+\frac{\bar f_4^V(q^2)}{M^2} \bigg)\bigg\} ,
 \non\\
 {\cal F}_2^V(q^2)
&\equiv&\frac{q^2}{M'}
\bigg\{\bar f_1^V(q^2)- \frac{M'}{M} \bar f_2^V(q^2)
+\frac{1}{2}\bigg(M^2+M'M-2M^{\prime 2}-q^2\bigg)\bigg(\frac{\bar f_3^V(q^2)}{MM'}+\frac{\bar f_4^V(q^2)}{M^2} \bigg)\bigg\} ,
\non\\
 {\cal F}_3^V(q^2)
&\equiv&\frac{q^2}{M'}
\bigg\{(2M-3M')\bar f_1^V(q^2)+(q^2-(M-2 M')(M+M'))\frac{ \bar f_2^V(q^2)}{M}
\non\\
&&
+[(M-M')^2(M+M')-q^2(M-2M')]\frac{\bar f_3^V(q^2)}{MM'}\bigg\} ,
\non\\
 {\cal F}_4^V(q^2)
&\equiv&
(M'-M)
\bigg\{
 \bar f_1^V(q^2)
+\bigg((M+M')^2-q^2\frac{M+2M'}{(M'-M)}\bigg) \frac{\bar f_2^V(q^2)}{M M'}\bigg\},
\non\\
{\cal G}_1^A(q^2)
&\equiv&\frac{M'+M}{M'}
\bigg\{
\bar g_1^A(q^2)+\bigg(M-M'-\frac{q^2}{(M'+M)}\bigg)\frac{\bar g_2^A(q^2)}{M}
\non\\
&&
+\frac{1}{2}\bigg(M^2-M^{\prime 2}-q^2\frac{2M'+M}{M'+M}\bigg)
 \bigg(\frac{\bar g_3^A(q^2)}{MM'} +\frac{\bar g_4^A(q^2)}{M^2}\bigg)
 \bigg\},
\non\\
{\cal G}_2^A(q^2)
&\equiv&\frac{q^2}{M'}
\bigg\{\bar g_1^A(q^2)- \frac{M'}{M} \bar g_2^A(q^2)
+\frac{1}{2}\bigg(M^2-M'M-2M^{\prime 2}-q^2\bigg) 
\bigg(\frac{\bar g_3^A(q^2)}{MM'} +\frac{\bar g_4^A(q^2)}{M^2}\bigg)\bigg\},
\non\\
 {\cal G}_3^A(q^2)
&\equiv&\frac{q^2}{M'}
\bigg\{-(2M+3M')\bar f_1^V(q^2)+(q^2-(M+2 M')(M-M'))\frac{ \bar f_2^V(q^2)}{M}
\non\\
&&
+[-(M+M')^2(M-M')+q^2(M+2M')]\frac{\bar f_3^V(q^2)}{MM'}\bigg\}.
\non\\
{\cal G}_4^A(q^2)
&\equiv&
(M'+M)
\bigg\{
 \bar g_1^A(q^2)
+\bigg(-(M'-M)^2+q^2\frac{M-2M'}{(M'+M)}\bigg) \frac{\bar g_2^A(q^2)}{M M'}\bigg\}.
\label{eq: FG}
 \en

Similarly, for $\B_b(1/2^+)\to\B_c(3/2^-)$ transition, we have
\be
{\cal F}^A_1(q^2) & =&\int \frac{dx_2 d^2 k_{2\bot}}{2 (2\pi)^3}~\frac{3M_0'}{P^{\prime +}}
      \frac{\phi^{\prime*}_{n'L'_K}(\{x'\},\{k'_{\bot}\})\phi_{1 L_K}(\{x\},\{k_{\bot}\})}
             {8 P^+P^{\prime +}\sqrt{[(m_1+x_1 M_0)^2+k_{1\bot}^2][(m'_1+x_1 M'_0)^2+k_{1\bot}^{\prime 2}]}}
\non\\
&&
     \qquad\times~Tr[(\not\!\bar P+M_0)\gamma^+\gamma_5{\cal P}^{+\mu}\bar \Gamma_{\mu L'_K S_{[qq]} J'_l} 
     (\not\! p'_1+m'_1)\gamma^+\gamma_5(\not\! p_1+m_1)\Gamma_{L_K S_{[qq]} J_l}],
\non\\
{\cal F}^A_2(q^2) 
&=&-i \int \frac{dx_2 d^2 k_{2\bot}}{2 (2\pi)^3}~\frac{3M_0'}{P^{\prime +}}
      \frac{\phi^{\prime*}_{n'L'_K}(\{x'\},\{k'_{\bot}\})\phi_{1 L_K}(\{x\},\{k_{\bot}\})}
             {8 P^+P^{\prime +}\sqrt{[(m_1+x_1 M_0)^2+k_{1\bot}^2][(m'_1+x_1 M'_0)^2+k_{1\bot}^{\prime 2}]}}
\non\\
&&
     \qquad\times~Tr[(\not\!\bar P+M_0)\sigma^{+\nu}q_{\bot\nu}\gamma_5{\cal P}^{+\mu}\bar \Gamma_{\mu L'_K S_{[qq]} J'_l} 
     (\not\! p'_1+m'_1)\gamma^+\gamma_5(\not\! p_1+m_1)\Gamma_{L_K S_{[qq]} J_l}],
\non\\
{\cal F}^A_3(q^2) & =&\int \frac{dx_2 d^2 k_{2\bot}}{2 (2\pi)^3}~\frac{3M_0'}{P^{\prime+}}
      \frac{\phi^{\prime*}_{n'L'_K}(\{x'\},\{k'_{\bot}\})\phi_{1 L_K}(\{x\},\{k_{\bot}\})}
             {4 \sqrt{P^+P^{\prime +}}\sqrt{[(m_1+x_1 M_0)^2+k_{1\bot}^2][(m'_1+x_1 M'_0)^2+k_{1\bot}^{\prime 2}]}}
\non\\
&&
     \qquad\times~Tr[(\not\!\bar P+M_0)\gamma^+\gamma_5{\cal P}^{+\mu}\bar \Gamma_{\mu L'_K S_{[qq]} J'_l} 
     (\not\! p'_1+m'_1){\not\! {q}_\bot}\gamma_5(\not\! p_1+m_1)\Gamma_{L_K S_{[qq]} J_l}],
\non\\
{\cal F}^A_4(q^2) 
&=&-i \int \frac{dx_2 d^2 k_{2\bot}}{2 (2\pi)^3}~\frac{3M_0'}{P^{\prime +}}
      \frac{\phi^{\prime*}_{n'L'_K}(\{x'\},\{k'_{\bot}\})\phi_{1 L_K}(\{x\},\{k_{\bot}\})}
             {4 \sqrt{P^+P^{\prime +}}\sqrt{[(m_1+x_1 M_0)^2+k_{1\bot}^2][(m'_1+x_1 M'_0)^2+k_{1\bot}^{\prime 2}]}}
\non\\
&&
     \qquad\times~Tr[(\not\!\bar P+M_0)\sigma^{+\nu}q_{\bot\nu}\gamma_5{\cal P}^{+\mu}\bar \Gamma_{\mu L'_K S_{[qq]} J'_l} 
     (\not\! p'_1+m'_1){\not\!n_\bot}\gamma_5(\not\! p_1+m_1)\Gamma_{L_K S_{[qq]} J_l}],
\non\\
{\cal G}^V_1(q^2) & =&\int \frac{dx_2 d^2 k_{2\bot}}{2 (2\pi)^3}~\frac{3M_0'}{P^{\prime +}}
      \frac{\phi^{\prime*}_{n'L'_K}(\{x'\},\{k'_{\bot}\})\phi_{1 L_K}(\{x\},\{k_{\bot}\})}
             {8 P^+P^{\prime +}\sqrt{[(m_1+x_1 M_0)^2+k_{1\bot}^2][(m'_1+x_1 M'_0)^2+k_{1\bot}^{\prime 2}]}}
\non\\
&&
     \qquad\times~Tr[(\not\!\bar P+M_0)\gamma^+{\cal P}^{+\mu}\bar \Gamma_{\mu L'_K S_{[qq]} J'_l}  
     (\not\! p'_1+m'_1)\gamma^+(\not\! p_1+m_1)\Gamma_{L_K S_{[qq]} J_l}],
\non\\
{\cal G}^V_2(q^2)
&=&i\int \frac{dx_2 d^2 k_{2\bot}}{2 (2\pi)^3}~\frac{3M_0'}{P^{\prime +}}
      \frac{\phi^{\prime*}_{n'L'_K}(\{x'\},\{k'_{\bot}\})\phi_{1 L_K}(\{x\},\{k_{\bot}\})}
             {8 P^+P^{\prime +}\sqrt{[(m_1+x_1 M_0)^2+k_{1\bot}^2][(m'_1+x_1 M'_0)^2+k_{1\bot}^{\prime 2}]}}
\non\\
&&
     \qquad\times~Tr[(\not\!\bar P+M_0)\sigma^{+\nu}q_{\bot\nu}{\cal P}^{+\mu}\bar \Gamma_{\mu L'_K S_{[qq]} J'_l} 
     (\not\! p'_1+m'_1)\gamma^+(\not\! p_1+m_1)\Gamma_{L_K S_{[qq]} J_l}].
\non\\
{\cal G}^V_3(q^2) & =&\int \frac{dx_2 d^2 k_{2\bot}}{2 (2\pi)^3}~\frac{3M_0'}{P^{\prime+}}
      \frac{\phi^{\prime*}_{n'L'_K}(\{x'\},\{k'_{\bot}\})\phi_{1 L_K}(\{x\},\{k_{\bot}\})}
             {4 \sqrt{P^+P^{\prime +}}\sqrt{[(m_1+x_1 M_0)^2+k_{1\bot}^2][(m'_1+x_1 M'_0)^2+k_{1\bot}^{\prime 2}]}}
\non\\
&&
     \qquad\times~Tr[(\not\!\bar P+M_0)\gamma^+{\cal P}^{+\mu}\bar \Gamma_{\mu L'_K S_{[qq]} J'_l} 
     (\not\! p'_1+m'_1){\not\! {q}_\bot}(\not\! p_1+m_1)\Gamma_{L_K S_{[qq]} J_l}],
\non\\
{\cal G}^V_4(q^2)
&=&i\int \frac{dx_2 d^2 k_{2\bot}}{2 (2\pi)^3}~\frac{3M_0'}{P^{\prime +}}
      \frac{\phi^{\prime*}_{n'L'_K}(\{x'\},\{k'_{\bot}\})\phi_{1 L_K}(\{x\},\{k_{\bot}\})}
             {8 P^+P^{\prime +}\sqrt{[(m_1+x_1 M_0)^2+k_{1\bot}^2][(m'_1+x_1 M'_0)^2+k_{1\bot}^{\prime 2}]}}
\non\\
&&
     \qquad\times~Tr[(\not\!\bar P+M_0)\sigma^{+\nu}n_{\bot\nu}{\cal P}^{+\mu}\bar \Gamma_{\mu L'_K S_{[qq]} J'_l} 
     (\not\! p'_1+m'_1){\not\! n_\bot}(\not\! p_1+m_1)\Gamma_{L_K S_{[qq]} J_l}],
\non\\
\label{eq: FG (iv) and (vi)}
 \en
with $q_\bot\equiv (0,0,\vec q_\bot)$, $n_\bot\equiv (0,0,\vec n_\bot)$, $n_\bot^2=-1$ and $\vec n_\bot\cdot \vec q_\bot=0$.  
The expressions of ${\cal F}^A_i$ and ${\cal G}^V_i$ in terms of $f^A_i$ and $g^V_i$ are similar to those of ${\cal F}^V_i$ and ${\cal G}^A_i$ in Eq.~(\ref{eq: FG}), by with $V$ and $A$ exchanged.   
 
One can solve for $\bar f_i$ and $\bar g_i$ once $F_i$ and $G_i$ are known.
It should be noted that there are relations in the $q^2\to 0$ limit:
\be
\lim_{q^2\to0}\bigg\{{\cal F}^{V,A}_4(q^2)-(M+2 M') {\cal F}^{V,A}_1(q^2)+(M^{\prime 2}-M^2) \frac{{\cal F}^{V,A}_2(q^2)}{q^2}\bigg\}&=&0,
\non\\
\lim_{q^2\to0}\bigg\{{\cal G}^{A,V}_4(q^2)+(M-2 M') {\cal G}^{A,V}_1(q^2)+(M^{\prime 2}-M^2)  \frac{{\cal G}^{A,V}_2(q^2)}{q^2}\bigg\}&=&0.
\label{eq: constraint}
\en 
These are the consistency relations in the $1/2\to 3/2$ case. 
In fact, they ensure the resulting form factors $\bar f_i$ and $\bar g_i$ to be finite in the $q^2\to 0$ limit, i.e. $\lim_{q^2\to 0} q^2 \bar f_i(q^2)=\lim_{q^2\to 0} q^2 \bar g_i(q^2)=0$.~\footnote{To see this we denote equations of ${\cal F}_i$ in Eq. (\ref{eq: FG}) as
\be
\left(
\begin{array}{c}
{\cal F}_1\\
{\cal F}_2/q^2\\
{\cal F}_3/q^2\\
{\cal F}_4
\end{array}
\right)
=A
\left(
\begin{array}{c}
\bar f_1\\
\bar f_2\\
\bar f_3\\
\bar f_4
\end{array}
\right),
\en
where $A$ is a $4\times 4$ matrix with elements correspond to the coefficients of $\bar f_i$ in Eq. (\ref{eq: FG}).
It is well known that $\bar f_i$ can be obtained by acting the inverse matrix $A^{-1}=adj(A)/|A|$ on 
$({\cal F}_1,{\cal F}_2/q^2,{\cal F}_3/q^2,{\cal F}_4)^T$, where $adj(A)$ is the adjugate matrix of $A$. 
For $q^2$ approaches 0 the determinant of $A$, $|A|={\cal O}(q^2)$, approaches 0, which seems to lead to diverging $\bar f_i(0)$ 
as the denominator of $[adj(A)\cdot({\cal F}_1,{\cal F}_2/q^2,{\cal F}_3/q^2,{\cal F}_4)^T]/|A|$ is approching 0, 
but it can be shown that the numerator, $adj(A)\cdot ({\cal F}_1,{\cal F}_2/q^2,{\cal F}_3/q^2,{\cal F}_4)^T$, is proportional to the l.h.s. of first equation in Eq. (\ref{eq: constraint}) for small $q^2$ and also approaches 0. 
Hence, with the help of the first equation in Eq. (\ref{eq: constraint}) the limit of $\bar f_i(q^2)$ for $q^2$ approaching 0 can be obtained by using the L'H\^{o}pital's rule and we find that finite $\bar f_i(0)$ can be obtained in this way. Similar argument holds for the ${\cal G}_i$, $\bar g_i$ case.
} 
Furthermore, we will expect
\be
\lim_{q^2\to 0}\bar f_i(q^2)=\lim_{q^2\to 0} \frac{d}{d q^2} [q^2 \bar f_i (q^2)],
\quad
\lim_{q^2\to 0}\bar g_i(q^2)=\lim_{q^2\to 0} \frac{d}{d q^2} [q^2 \bar g_i (q^2)],
\label{eq: figi q2=0}
\en
to hold.

Most of the traces in Eqs. (\ref{eq: figi (i) and (ii)}), (\ref{eq: figi (v)}), (\ref{eq: FG (iii)}) and (\ref{eq: FG (iv) and (vi)}) are rather complicated to work out.
We find that it is convenient to work in the $\vec P_\bot=\vec 0$ frame and all of the traces can be obtained with the help of the FeynCal program~\cite{Mertig:1990an,Shtabovenko:2016sxi}.
The final expressions of the form factors can be obtained using the kinematics of light-front quantities collected in Appendix~A.
We found that all of the $P^+$ and $P^{\prime +}$ factors in these equations cancel out in the final expressions as they should.

\subsection{Form factors for $\B_b({\bf \bar 3_f}, 1/2^+)\to \B_c({\bf \bar 3_f}, 1/2^+)$ transition [type (i)]}

The $\B_b({\bf \bar 3_f},{1}/{2}^+)\to\B_c({\bf \bar 3_f},{1}/{2}^+)$ transitions involve initial states in
$(n, L_K, S_{[qq]}^P, J_l^P,J^P)_b
 =(1, 0, 0^+,0^+, \frac{1}{2}^+)$ configuration and final states in
$(n', L'_K, S_{[qq]}^P, J_l^{\prime P},J^{\prime P})_c=(n', 0, 0^+,0^+, \frac{1}{2}^+)$ configurations (with $n'$=1,2). 
This type of transition consists of $\Lambda_b^0\to \Lambda^+_c$, $\Xi_b^{0(-)}\to \Xi_c^{+(0)}$ and $\Lambda_b^0\to\Lambda_c(2765)^+$
transitions. 
In these transitions the spectating diquarks are scalar diquarks $[ud]$, $[us]$ and $[ds]$.
To obtain form factors $f^V_i$ and $g^A_i$, we use Eq.~(\ref{eq: figi (i) and (ii)}) with   
$\Gamma_{L_K S_{[qq]} J_l}=\Gamma_{s00}(p_1,p_2)$, 
$\bar\Gamma_{L'_K S_{[qq]} J'_l}=\bar\Gamma_{s00}(p'_1,p_2)$,
which are given in  Eq. (\ref{eq: Gamma}),
and $n'=1,2$.

\subsection{Form factors for $\B_b({\bf 6_f}, 1/2^+)\to \B_c({\bf 6_f}, 1/2^+)$ transition [type (ii)]}

The $\B_b({\bf 6_f}, 1/2^+)\to \B_c({\bf 6_f}, 1/2^+)$ transitions involve initial states in
$(n, L_K, S_{[qq]}^P, J_l^P,J^P)_b
 =(1, 0, 1^+,1^+, \frac{1}{2}^+)$ configuration and final states in
$(n', L'_K, S_{[qq]}^P, J_l^{\prime P},J^{\prime P})_c=(n', 0, 1^+,1^+, \frac{1}{2}^+)$ configurations (with $n'$=1,2). 
This type of transition consists of  
$\Omega^-_b\to\Omega^0_c$ and $\Omega^-_b\to\Omega_c(3090)^0$ 
transitions.
In these transitions the spectating diquarks are axial-vector diquarks $[ss]$. 
To obtain form factors $f^V_i$ and $g^A_i$,  we use Eq.~(\ref{eq: figi (i) and (ii)}) with   
$\Gamma_{L_K S_{[qq]} J_l}=\Gamma_{s11}(p_1,p_2,\lambda_2)$, 
$\bar\Gamma_{L'_K S_{[qq]} J'_l}=\bar\Gamma_{s11}(p'_1,p_2,\lambda_2)$,
which are given in  Eq. (\ref{eq: Gamma}),
and $n'=1,2$.
Note that one needs to sum over the  axial-vector diquark polarization ($\lambda_2$).

\subsection{Form factors for $\B_b({\bf 6_f}, 1/2^+)\to \B_c({\bf 6_f}, 3/2^+)$ transition [type (iii)]}

The $\B_b({\bf 6_f}, 1/2^+)\to \B_c({\bf 6_f}, 3/2^+)$ transitions involve initial states in
$(n, L_K, S_{[qq]}^P, J_l^P,J^P)_b
 =(1, 0, 1^+,1^+, \frac{1}{2}^+)$ configuration and final states in
$(n', L'_K, S_{[qq]}^P, J_l^{\prime P},J^{\prime P})_c=(n', 0, 1^+,1^+, \frac{3}{2}^+)$ configurations (with $n'$=1,2). 
This type of transition consists of 
$\Omega^-_b\to\Omega_c(2770)^0$ and $\Omega^-_b\to\Omega_c(3119)^0$ 
transitions.
In these transitions the spectating diquarks are axial-vector diquarks $[ss]$. 
To obtain form factors $\bar f^V_i$ and $\bar g^A_i$, we use Eq.~(\ref{eq: FG (iii)}) with   
$\Gamma_{L_K S_{[qq]} J_l}=\Gamma_{s11}(p_1,p_2,\lambda_2)$,
$\bar\Gamma^\mu_{L'_K S_{[qq]} J'_l}=\bar\Gamma^\mu_{s11}(p'_1,p_2,\lambda_2)$,
which are given in  Eq. (\ref{eq: Gamma}), and $n'$=1,2.
Note that one needs to sum over the  axial-vector diquark polarization ($\lambda_2$). 

\subsection{Form factors for $\B_b({\bf 6_f}, 1/2^+)\to \B_c({\bf 6_f}, 3/2^-)$ transition [type (iv)]}

The $\B_b({\bf 6_f}, 1/2^+)\to \B_c({\bf 6_f}, 3/2^-)$ transitions involve initial states in
$(n, L_K, S_{[qq]}^P, J_l^P,J^P)_b
 =(1, 0, 1^+,1^+, \frac{1}{2}^+)$ configuration and final states in
$(n', L'_K, S_{[qq]}^P, J_l^{\prime P},J^{\prime P})_c=(1, 1, 1^+,1^+, \frac{3}{2}^+)$ configurations. 
This type of transition consists of 
$\Omega^-_b\to\Omega_c(3050)^0$ 
transition.
We follow ref.~\cite{Cheng:2017ove} to consider $\Omega_c(3050)^0$ as a $p$-wave state. 
In these transitions the spectating diquarks are axial-vector diquarks $[ss]$. 
To obtain form factors $\bar f^A_i$ and $\bar g^V_i$, we use Eq.~(\ref{eq: FG (iv) and (vi)}) with   
$\Gamma_{L_K S_{[qq]} J_l}=\Gamma_{s11}(p_1,p_2,\lambda_2)$,
$\bar\Gamma^\mu_{L'_K S_{[qq]} J'_l}=\bar\Gamma^\mu_{p12}(p'_1,p_2,\lambda_2)$,
which are given in  Eq. (\ref{eq: Gamma}), and $n'$=1.
Note that one needs to sum over the  axial-vector diquark polarization ($\lambda_2$).

\subsection{Form factors for $\B_b({\bf \bar 3_f}, 1/2^+)\to \B_c({\bf \bar 3_f}, 1/2^-)$ transition [type (v)]}

The $\B_b({\bf \bar 3_f},{1}/{2}^+)\to\B_c({\bf \bar 3_f},{1}/{2}^-)$ transitions
involve initial states in
$(n, L_K, S_{[qq]}^P, J_l^P,J^P)_b
 =(1, 0, 0^+,0^+, \frac{1}{2}^+)$ configuration and final states in
$(n', L'_K, S_{[qq]}^P, J_l^{\prime P},J^{\prime P})_c=(n', 1, 0^+,1^-, \frac{1}{2}^-)$ configurations (with $n'$=1,2). 
This type of transition consists of 
$\Lambda_b^0\to\Lambda_c(2595)^+$, $\Xi^{0(-)}_b\to\Xi_c(2790)^{+(0)}$ and $\Lambda_b^0\to\Lambda_c(2940)^+$ 
transitions.
In these transitions the spectating diquarks are scalar diquarks $[ud]$, $[us]$ and $[ds]$.
To obtain form factors $f^A_i$ and $g^V_i$, we use Eq.~(\ref{eq: figi (v)}) with   
$\Gamma_{L_K S_{[qq]} J_l}=\Gamma_{s00}(p_1,p_2)$, 
$\bar\Gamma_{L'_K S_{[qq]} J'_l}=\bar\Gamma_{p10}(p'_1,p_2)$,
which are given in  Eq. (\ref{eq: Gamma}),
and $n'=1,2$.

\subsection{Form factors for $\B_b({\bf \bar 3_f}, 1/2^+)\to \B_c({\bf \bar 3_f}, 3/2^-)$ transition [type (vi)]}

The $\B_b({\bf \bar 3_f},{1}/{2}^+)\to\B_c({\bf \bar 3_f},{3}/{2}^-)$ transitions
involve initial states in
$(n, L_K, S_{[qq]}^P, J_l^P,J^P)_b
 =(1, 0, 0^+,0^+, \frac{1}{2}^+)$ configuration and final states in
$(n', L'_K, S_{[qq]}^P, J_l^{\prime P},J^{\prime P})_c=(n', 1, 0^+,1^-, \frac{3}{2}^-)$ configurations with $n'=1,2$. 
This type of transition consists of 
$\Lambda_b^0\to\Lambda_c(2625)^+$, $\Lambda_c(2940)^+$ and $\Xi^{0(-)}_b\to\Xi_c(2815)^{+(0)}$ 
transitions.
We follow LHCb and PDG \cite{Aaij:2017vbw,PDG} to take $\Lambda_c(2940)$ as a spin-3/2 particle.
To obtain form factors $\bar f^A_i$ and $\bar g^V_i$, we use Eq.~(\ref{eq: FG (iv) and (vi)}) with 
$\Gamma_{L_K S_{[qq]} J_l}=\Gamma_{s00}(p_1,p_2)$, 
$\bar\Gamma^\mu_{L'_K S_{[qq]} J'_l}=\bar\Gamma^\mu_{p01}(p'_1,p_2)$,
which are given in  Eq. (\ref{eq: Gamma}), and $n'=1,2$.

\subsection{Heavy quark limit}

It is useful to note that in the heavy quark (HQ) limit, baryon form factors have simple behavior~\cite{Isgur:1990pm,Isgur:1991wr,Yan:1992gz}. 
The $\B_b\to \B_c$ transition matrix elements with $s$-wave and $p$-wave $\B_c$ can be expressed as~\cite{Isgur:1990pm,Isgur:1991wr,Yan:1992gz,Cheng:1996cs,Xu:1993mj}~\footnote{Note that in the $\B_b(1/2^+)\to \B_c(3/2^\pm)$ transition matrix elements, we apply overall minus signs to match our sign convention, which follow from the sign convention of the Clebsch-Gordon coefficients, see Eq.~(\ref{eq: Psi}).}
\be
\la\B_c(\bar{3}_f,1/2^+)(v')|j^\mu_{V-A}|\B_b(\bar{\bf 3}_f,1/2^+)(v)\ra
&=&\zeta(\omega) \bar u(v') \gamma^\mu(1-\gamma_5) u(v), 
\non\\
\la\B_c(\bar{3}_f,1/2^-)(v')|j^\mu_{V-A}|\B_b(\bar{\bf 3}_f,1/2^+)(v)\ra
&=&\frac{\sigma(\omega)}{\sqrt3} \bar u(v') \gamma_5(\not\! v+v\cdot v') \gamma^\mu (1-\gamma_5)  u(v), 
\non\\
\la\B_c(\bar{3}_f,3/2^-)(v')|j^\mu_{V-A}|\B_b(\bar{\bf 3}_f,1/2^+)(v)\ra
&=&-\sigma(\omega) \bar u_\nu(v') v^\nu \gamma^\mu (1-\gamma_5) u(v), 
\non\\
\la \B_c({\bf 6}_f,1/2^+)(v')|j^\mu_{V-A}|\B_b({\bf 6}_f,1/2^+)(v)\ra
&=&-\frac{1}{3} 
(g^{\rho\sigma}\xi_1-v^\rho v^{\prime \sigma}\xi_2)
\non\\
&&\times
\bar u(v')(\gamma_\rho-v'_\rho)\gamma^\mu(1-\gamma_5)(\gamma_\sigma-v_\sigma) u(v),
\non\\
\la \B_c({\bf 6}_f,3/2^+)(v')|j^\mu_{V-A}|\B_b({\bf 6}_f,1/2^+)(v)\ra
&=&-\frac{1}{\sqrt3} 
(g^{\rho\sigma}\xi_1-v^\rho v^{\prime \sigma}\xi_2)
\non\\
&&\times
\bar u_\rho(v')\gamma^\mu(1-\gamma_5)(\gamma_\sigma+v_\sigma)\gamma_5 u(v),
\non\\
\la \B_c({\bf 6}_f,3/2^-)(v')|j^\mu_{V-A}|\B_b({\bf 6}_f,1/2^+)(v)\ra
&=&-\frac{1}{\sqrt{30}} 
[g^{\rho\sigma}\xi_5+(v-v')^\rho (v-v')^\sigma\xi_6]
\non\\
&&\times
[\bar u\cdot v (\gamma_\rho-v'_\rho)+\bar u_\rho (\not\! v-\omega))]
\non\\
&&\times
\gamma^\mu(1-\gamma_5)(\gamma_\sigma-v_\sigma) u(v),
\en
which imply, 
in the type (i) $\B_b(\bar{\bf 3}_f ,1/2^+)\to\B_c(\bar{\bf 3}_f , 1/2^+)$ transition,
\be
f^V_1(\bar{\bf 3}_f)=g^A_1(\bar{\bf 3}_f)=\zeta(\omega),
\qquad 
f^V_{2,3}(\bar{\bf 3}_f)=g^A_{2,3}(\bar{\bf 3}_f)=0;
\label{eq: HQ type i}
\en
in the type (ii) $\B_b({\bf 6}_f, 1/2^+)\to\B_c({\bf 6}_f, 1/2^+)$ transition,
\be
f^V_1({\bf 6}_f)&=&\frac{1}{3}[(2-\omega) \xi_1+(1-\omega)^2\xi_2]+\frac{1}{3}\frac{M^2+M^{\prime 2}}{M M'}[\xi_1+(1-\omega)\xi_2],
\non\\
f^V_2({\bf 6}_f)&=&\frac{1}{3}\frac{(M+M')^2}{M M'}[\xi_1+(1-\omega)\xi_2],
\quad
f^V_3({\bf 6}_f)=-\frac{1}{3}\frac{M^2-M^{\prime 2}}{M M'}[\xi_1+(1-\omega)\xi_2],
\non\\
g^A_1({\bf 6}_f)&=&\frac{1}{3}[-(2+\omega) \xi_1+(1+\omega)^2\xi_2]+\frac{1}{3}\frac{M^2+M^{\prime2}}{M M'}[\xi_1-(1+\omega)\xi_2],
\non\\
g^A_2({\bf 6}_f)&=&-\frac{1}{3}\frac{(M-M')^2}{M M'}[\xi_1-(1+\omega)\xi_2],
\quad
g^A_3({\bf 6}_f)=\frac{1}{3}\frac{M^2-M^{\prime 2}}{M M'}[\xi_1-(1+\omega)\xi_2];
\label{eq: HQ type ii}
\en
in the type (iii) $\B_b({\bf 6}_f, 1/2^+)\to\B_c({\bf 6}_f, 3/2^+)$ transition,
\be
\bar f^V_1({\bf 6}_f)&=&-\bar g^A_1({\bf 6}_f)=-\frac{2}{\sqrt 3}\xi_1(\omega),
\quad
\bar f^V_3({\bf 6}_f)=-\bar g^A_3({\bf 6}_f)=+\frac{2}{\sqrt3}\xi_2(\omega),
\quad
\bar f^V_4({\bf 6}_f)=\bar g^A_4({\bf 6}_f)=0,
\non\\
\bar f^V_2({\bf 6}_f)&=&-\frac{1}{\sqrt 3}[\xi_1+(1-\omega)\xi_2(\omega)],
\quad
\bar g^A_2({\bf 6}_f)=-\frac{1}{\sqrt 3}[\xi_1-(1+\omega)\xi_2(\omega)];
\label{eq: HQ type iii}
\en
in the type (iv) $\B_b({\bf 6}_f, 1/2^+)\to\B_c({\bf 6}_f, 3/2^-)$ transition,
\be
\bar f^A_1({\bf 6}_f)&=&-\frac{2}{\sqrt {30}}(1+\omega)\xi_5(\omega),
\quad
\bar f^A_2({\bf 6}_f)=-\frac{2}{\sqrt {30}}[(1+\omega)\xi_5(\omega)+(1-\omega^2)\xi_6(\omega)],
\non\\
\bar f^A_3({\bf 6}_f)&=&-\frac{2}{\sqrt {30}}[\xi_5(\omega)-2(1+\omega)\xi_6(\omega)],
\quad
\bar f^A_4({\bf 6}_f)=\frac{4}{\sqrt {30}}[\xi_5(\omega)-(1+\omega)\xi_6(\omega)],
\non\\
\bar g^V_1({\bf 6}_f)&=&-\frac{2}{\sqrt {30}}(1-\omega)\xi_5(\omega),
\quad
\bar g^V_2({\bf 6}_f)=\frac{1}{\sqrt {30}}[(1-2\omega)\xi_5(\omega)-2(1-\omega^2)\xi_6(\omega)],
\non\\
\bar g^V_3({\bf 6}_f)&=&\frac{2}{\sqrt {30}}[\xi_5(\omega)+2(1-\omega)\xi_6(\omega)],
\quad
\bar g^V_4({\bf 6}_f)=\frac{4}{\sqrt {30}}[\xi_5(\omega)+(1-\omega)\xi_6(\omega)];
\label{eq: HQ type iv}
\en
in the type (v) $\B_b(\bar{\bf 3}_f, 1/2^+)\to\B_c(\bar{\bf 3}_f, 1/2^-)$ transition,
\be
f^A_1(\bar{\bf 3}_f)&=&g^V_1(\bar{\bf 3}_f)=\bigg(\omega-\frac{M'}{M}\bigg)\frac{\sigma(\omega)}{\sqrt3},
\non\\
f^A_2(\bar{\bf 3}_f)&=&f^A_3(\bar{\bf 3}_f)=-\frac{M+M'}{M}\frac{\sigma(\omega)}{\sqrt3},
\quad
g^V_2(\bar{\bf 3}_f)=g^V_3(\bar{\bf 3}_f)=-\frac{M-M'}{M}\frac{\sigma(\omega)}{\sqrt3};
\label{eq: HQ type v}
\en
and in the type (vi) $\B_b(\bar{\bf 3}_f, 1/2^+)\to\B_c(\bar{\bf 3}_f, 3/2^-)$ transition,
\be
\bar f^A_2(\bar{\bf 3}_f)=\bar g^V_2(\bar{\bf 3}_f)=\sigma(\omega),
\quad
\bar f^A_{1,3,4}(\bar {\bf 3}_f)=\bar g^V_{1,34}(\bar{\bf 3}_f)=0.
\label{eq: HQ type vi}
\en

For low lying $\B_c$ states, the following normalizations are applied,
\be
\zeta(1)=1, 
\quad
\xi_1(1)=1,
\label{eq: w=1}
\en
and it has been shown that in the large $N_c$ limit, one has~\cite{Chow:1994ni}
\be
\xi_1(\omega)=(1+\omega)\xi_2(\omega)=\zeta(\omega).
\label{eq: large Nc}
\en
These imply very simple and specify relations of form factors at $q^2=q^2_{max}$ (or $\omega=1$), namely
\be
f^V_1(\bar{\bf 3}_f)=g^A_1(\bar{\bf 3}_f)=1,
\qquad 
f^V_{2,3}(\bar{\bf 3}_f)=g^A_{2,3}(\bar{\bf 3}_f)=0,
\label{eq: HQ type i w=1}
\en
in the type (i) $\B_b(\bar{\bf 3}_f ,1/2^+)\to\B_c(\bar{\bf 3}_f , 1/2^+)$ transition;
\be
f^V_1({\bf 6}_f)&=&\frac{1}{3}+\frac{1}{3}\frac{M^2+M^{\prime 2}}{M M'},
\quad
f^V_2({\bf 6}_f)=\frac{1}{3}\frac{(M+M')^2}{M M'},
\quad
f^V_3({\bf 6}_f)=-\frac{1}{3}\frac{M^2-M^{\prime 2}}{M M'},
\non\\
g^A_1({\bf 6}_f)&=&-\frac{1}{3},
\quad
g^A_2({\bf 6}_f)=g^A_3({\bf 6}_f)=0,
\label{eq: HQ type ii w=1}
\en
in the type (ii) $\B_b({\bf 6}_f, 1/2^+)\to\B_c({\bf 6}_f, 1/2^+)$ transition;
and
\be
-\bar f^V_1({\bf 6}_f)
&=&\bar g^A_1({\bf 6}_f)
=-2\bar f^V_2({\bf 6}_f)
=2\bar f^V_3({\bf 6}_f)=-2\bar g^A_3({\bf 6}_f)=\frac{2}{\sqrt 3},
\non\\
\bar g^A_2({\bf 6}_f)&=&\bar f^V_4({\bf 6}_f)=\bar g^A_4({\bf 6}_f)=0,
\label{eq: HQ type iii w=1}
\en
in the type (iii) $\B_b({\bf 6}_f, 1/2^+)\to\B_c({\bf 6}_f, 3/2^+)$ transition.
Furthermore, for other transitions at $q^2=q^2_{\max}$, we have
\be
-\bar f^A_1({\bf 6}_f)
&=&-\bar f^A_2({\bf 6}_f)
=-4\bar g^V_2({\bf 6}_f)
=2\bar g^V_3({\bf 6}_f)
=\bar g^V_4({\bf 6}_f)
=\frac{4}{\sqrt {30}}\xi_5(1),
\non\\
\bar f^A_3({\bf 6}_f)
&=&-\frac{2}{\sqrt {30}}[\xi_5(1)-4\xi_6(1)],
\quad
\bar f^A_4({\bf 6}_f)=\frac{4}{\sqrt {30}}[\xi_5(1)-2\xi_6(1)],
\quad
\bar g^V_1({\bf 6}_f)=0,
\label{eq: HQ type iv w=1}
\en
in the type (iv) $\B_b({\bf 6}_f, 1/2^+)\to\B_c({\bf 6}_f, 3/2^-)$ transition;
\be
f^A_1(\bar{\bf 3}_f)&=&
g^V_1(\bar{\bf 3}_f)
=-g^V_2(\bar{\bf 3}_f)
=-g^V_3(\bar{\bf 3}_f)
=\bigg(\frac{M-M'}{M}\bigg)\frac{\sigma(1)}{\sqrt3},
\non\\
f^A_2(\bar{\bf 3}_f)&=&f^A_3(\bar{\bf 3}_f)=-\frac{M+M'}{M}\frac{\sigma(1)}{\sqrt3},
\label{eq: HQ type v w=1}
\en
in the type (v) $\B_b(\bar{\bf 3}_f, 1/2^+)\to\B_c(\bar{\bf 3}_f, 1/2^-)$ transition;
and
\be
\bar f^A_2(\bar{\bf 3}_f)=\bar g^V_2(\bar{\bf 3}_f)=\sigma(1),
\quad
\bar f^A_{1,3,4}(\bar {\bf 3}_f)=\bar g^V_{1,34}(\bar{\bf 3}_f)=0,
\label{eq: HQ type vi w=1}
\en
in the type (vi) $\B_b(\bar{\bf 3}_f, 1/2^+)\to\B_c(\bar{\bf 3}_f, 3/2^-)$ transition.

Although obtaining these Isgur-Wise functions is beyond the scope of this work, the above relations on form factors can still be useful.
Indeed, we expect our form factors to roughly exhibit the above patterns, since we have large but finite $m_{b,c}$.

\section{Numerical results}

In this section we will present the numerical results of
all the relevant $\B_b\to \B_c$ transition form factors.  
We will give predictions on
the decay rates and up-down asymmetries of various
$\Lambda_b\to \Lambda^{(*,**)}_c M^-$, $\Xi_b\to\Xi_c^{(**)} M^-$ and $\Omega_b\to\Omega^{(*)}_c M^-$ decays
using na\"{i}ve factorization.

\subsection{$\B_b\to \B_c$ form factors}

In Table~\ref{tab:input} we summerize the input parameters $m_{[qq']}$, $m_q$ and $\beta$. 
Note that the constituent quark and diquark masses are close to but smaller than those in ref.~\cite{Ebert:2010af}.
For the diquark masses, we use $m^S_{[ud]}$ for $\Lambda_b$ and $\Lambda^{(*,**)}_c$, 
$m^S_{[us]}$ and $m^S_{[ds]}$ for $\Xi_b$ and $\Xi^{(**)}_c$, 
and $m^A_{[ss]}$ for $\Omega_b$ and $\Omega_c^{(*)}$.
The $\beta$s for states only differ in their radial quantum numbers should be identical.
For example, the $\beta$s of $\Lambda_c$ and of the radial excited state $\Lambda_c(2765)$ are identical,
and the $\beta$s of the low-lying $3/2^-$ state $\Lambda_c(2625)$ and of the radial excited $3/2^-$ state $\Lambda_c(2765)$ are identical.
These input parameters are chosen to satisfy the consistency constraints [see discussions after Eqs. (\ref{eq: figi (i) and (ii)}) and (\ref{eq: constraint})] and to reproduce the $Br(\Lambda_b\to\Lambda_c P)$ data.
In practice it is more convenient to enforce the consistency constraints by using floating $M$ and $M'$, and the input parameters are determined by requiring $M$ and $M'$ to reproduce the physical masses of $\B_b$ and $\B_c$, respectively, within 10\%. 
In fact, in most cases the agreements are better than 10\%, 
but in the case of $\Lambda_c(2940)$ as a radial excited $3/2^-$ state, the corresponding $M'$ is larger than $m_{\Lambda_c(2940)}$ by 25\%.

Using the results in the previous section the form factors of various $\B_b\to\B_c$ transitions can be obtained.
The form factors are calculated in spacelike region, as we are using the $q^+=0$ frame,
we shall follow~\cite{Jaus:1989au,Jaus:1989av,Jaus96,Jaus:1999zv,CCH,CC2004,Chua:2018lfa} to analytically
continue them to the timelike region.
We follow~\cite{Jaus:1989au,Jaus:1989av,Jaus:1999zv,CCH,CC2004} and parameterize the form factors  
in the three-parameter form:
 \be 
 F(q^2)&=&\,\frac{F(0)}{1-a(q^2/M^2)+b(q^2/M^2)^2 }
 \label{eq:FFpara1}
 \en
for $\B_b\to \B_c$ transitions with the parameters $a,b$ expected to be of order
${\cal O}(1)$,
while for some cases, where the corresponding $a$ and $b$ are much larger than 1, we shall use the following form~\cite{Chua:2018lfa,CCH,Cheng:2004cc,CC2004} 
 \be 
 F(q^2)&=&\,\frac{F(0)}{(1-q^2/M^2)[1-a(q^2/M^2)+b(q^2/M^2)^2] },
 \label{eq:FFpara2}
 \en
to reduce the size of $a$ and $b$ and gives better fits. 
As we shall see that there are cases where some of the parameters $a$, $b$ are still larger than ${\cal O}(1)$, 
but usually the corresponding form factors are small and, consequently, 
they do not have much impact on the corresponding $\B_b\to\B_c M$ decay rates.

\begin{table}[t!]
\caption{\label{tab:input} The input
parameters $m^{S}_{[qq']}$,  $m^{A}_{[qq']}$, $m_q$
and $\beta$'s appearing in the Gaussian-type wave function (\ref{eq:wavefn}) (in units of GeV).
The superscript $S$ and $A$ denote scalar and axial vector diquarks, respectively.
}
\begin{ruledtabular}
\begin{tabular}{ccccccc}
           $m^S_{[ud]}$
          & $m^S_{[us],[ds]}$
          & $m^A_{[ss]}$
          & $m_b$
          & $m_c$
          & $\beta(\Lambda_b)$
          & $\beta(\Xi^{0,-}_b)$
          \\
\hline    
           $0.65$
          & $0.86$
          & $1.10$
          & $4.44$
          & $1.42$
          & $0.750$
          & $0.850$
          \\
\hline  
$\beta(\Omega_b)$  
          & $\beta(\Lambda_c)$
          & $\beta[\Lambda_c(2595)]$
          & $\beta[\Lambda_c(2625)]$
          & $\beta[\Lambda_c(2765)]$
          & $\beta[\Lambda_c(2940,\frac{1}{2}^-)]$
          & $\beta[\Lambda_c(2940,\frac{3}{2}^-)]$
           \\
\hline
$0.900$
          & 0.345  
          & 0.350
          & 0.450
          & 0.345
          & 0.350
         & 0.450  
          \\           
\hline 
$\beta(\Xi^{+,0}_c)$
          & $\beta[\Xi^{+,0}_c(2790)]$  
          & $\beta[\Xi^{+,0}_c(2815)]$
          & $\beta(\Omega_c)$  
          & $\beta[\Omega_c(2770)]$
          & $\beta[\Omega_c(3050)]$
          & $\beta[\Omega_c(3090)]$  
         \\
\hline
0.370
         & 0.365 
         & 0.550
         & 0.300
         & 0.370
         & 0.420
         & 0.300  
         \\
\hline
$\beta[\Omega_c(3120)]$
\\
\hline
0.370                 
\end{tabular}
\end{ruledtabular}
\end{table}

\begin{table}[t!]
\caption{\label{tab:fg type i} The transition form factors for various
$\B_b({\bf \bar 3_f},1/2^+)\to\B_c({\bf \bar 3_f},1/2^+)$ transitions [types (i) and (i)$^*$]. 
We employ a three parameter form for these form factors, see Eq.~(\ref{eq:FFpara1}).
}
\footnotesize{
\begin{ruledtabular}
\begin{tabular}{ccccccccccc}
 $\B_b\to\B_c$
          & $F$
          & $F(0)$
          & $F(q^2_{max})$
          & $a$
          & $b$
          & $F$
          & $F(0)$
          & $F(q^2_{max})$
          & $a$
          & $b$
          \\
\hline     
$ \Lambda _b\to\Lambda _c$ 
          & $f^V_1$  
          & $ 0.474_{-0.072}^{+0.069} $ & $ 0.764_{-0.116}^{+0.111} $ & $ 1.426 $ & $ 0.994 $
          & $ g^A_1 $ 
          & $ 0.468_{-0.07}^{+0.067} $ & $ 0.743_{-0.111}^{+0.106} $ & $ 1.394 $ & $ 0.966 $              
          \\      
         & $f^V_2$ 
         & $ -0.153_{-0.029}^{+0.027} $ & $ -0.262_{-0.050}^{+0.046} $ & $ 1.753 $ & $ 1.623 $
         & $ g^A_2 $ 
         & $0.030_{-0.007}^{+0.005} $ & $ 0.053_{-0.012}^{+0.009} $ & $ 1.921 $ & $ 1.963 $ 
         \\
         & $f^V_3$ 
         & $ 0.069_{-0.022}^{+0.021} $ & $ 0.130_{-0.041}^{+0.039} $ & $ 2.068 $ & $ 2.100 $ 
         & $ g^A_3 $ 
         & $-0.070_{-0.009}^{+0.010} $ & $ -0.114_{-0.015}^{+0.016} $ & $ 1.65 $ & $ 1.587 $ 
         \\
        \hline
$ \Xi _b\to\Xi _c^0$  
        & $f^V_1$  
        & $ 0.437_{-0.072}^{+0.070} $ & $ 0.714_{-0.118}^{+0.114} $ & $ 1.676 $ & $ 1.504 $ 
        & $ g^A_1  $ 
        & $ 0.429_{-0.071}^{+0.068} $ & $ 0.693_{-0.115}^{+0.110} $ & $ 1.635 $ & $ 1.452 $
        \\
        & $f^V_2$ 
        & $ -0.175_{-0.035}^{+0.033} $ & $ -0.294_{-0.059}^{+0.056} $ & $ 1.968 $ & $ 2.233 $
        & $ g^A_2  $ 
        & $ 0.034_{-0.007}^{+0.006} $ & $ 0.057_{-0.012}^{+0.010} $ & $ 2.067 $ & $ 2.503 $  
        \\
        & $f^V_3$ 
        & $ 0.081_{-0.025}^{+0.025} $ & $ 0.146_{-0.045}^{+0.045} $ & $ 2.257 $ & $ 2.760 $
        & $ g^A_3 $ 
        & $ -0.078_{-0.010}^{+0.011} $ & $ -0.123_{-0.016}^{+0.017} $ & $ 1.825 $ & $ 2.119 $
        \\           
          \hline
$ \Lambda _b\to\Lambda _c(2765)$ 
        & $f^V_1$ 
        & $ -0.354_{-0.028}^{+0.027} $ & $ -0.494_{-0.039}^{+0.038} $ & $ 1.079 $ & $ -0.063 $ 
        & $ g^A_1  $ 
        & $ -0.341_{-0.026}^{+0.024} $ & $ -0.460_{-0.035}^{+0.032} $ & $ 0.985 $ & $ -0.047 $ 
       \\
       & $f^V_2$ 
       & $ 0.135_{-0.018}^{+0.026} $ & $ 0.246_{-0.034}^{+0.047} $ & $ 1.844 $ & $ 0.346 $
       & $ g^A_2  $ 
       & $ 0.012_{-0.010}^{+0.013} $ & $ 0.014_{-0.012}^{+0.014} $ & $ 1.874 $ & $ 5.804 $  
       \\
       & $f^V_3$ 
       & $ 0.047_{-0.037}^{+0.039} $ & $ 0.057_{-0.045}^{+0.048} $ & $ 1.838 $ & $ 4.433 $
       & $ g^A_3 $ 
       & $ 0.062_{-0.007}^{+0.006} $ & $0.120_{-0.014}^{+0.012} $ & $ 2.014 $ & $ 0.551 $  
       \\ 
\end{tabular}
\end{ruledtabular}
}
\end{table}

\begin{table}[t!]
\caption{\label{tab:fg type ii} The transition form factors for various
$\B_b({\bf 6_f},1/2^+)\to\B_c({\bf 6_f},1/2^+)$ transitions [types (ii) and (ii)$^*$]. 
We employ a three parameter form for these form factors, Eq.~(\ref{eq:FFpara1}),
while for those with asterisks we employ Eq.~(\ref{eq:FFpara2}).
}
\footnotesize{
\begin{ruledtabular}
\begin{tabular}{ccccccccccc}
 $\B_b\to\B_c$
          & $F$
          & $F(0)$
          & $F(q^2_{max})$
          & $a$
          & $b$
          & $F$
          & $F(0)$
          & $F(q^2_{max})$
          & $a$
          & $b$
          \\
\hline     
$\Omega _b\to\Omega _c$ 
          & $ f^V_1{}^*  $ 
          & $ 0.292_{-0.062}^{+0.062} $ & $ 0.605_{-0.128}^{+0.128} $ & $ 2.229 $ & $ 4.051 $
          & $ g^A_1{}^* $ 
          & $ -0.097_{-0.021}^{+0.021} $ & $ -0.194_{-0.042}^{+0.042} $ & $ 1.421 $ & $ 1.723 $ 
          \\
          & $ f^V_2{}^* $ 
          & $ 0.440_{-0.093}^{+0.094} $ & $ 0.951_{-0.201}^{+0.203} $ & $ 2.027 $ & $ 3.082 $ 
          & $ g^A_2{}^*  $ 
          & $ -0.009_{-0.002}^{+0.002} $ & $ -0.018_{-0.004}^{+0.004} $ & $ 1.501 $ & $ 2.114 $ 
          \\
         & $ f^V_3{}^* $ 
         & $ -0.125_{-0.026}^{+0.028} $ & $ -0.246_{-0.051}^{+0.055} $ & $ 1.594 $ & $ 2.387 $ 
         & $ g^A_3{}^* $ 
         & $ 0.015_{-0.003}^{+0.002} $ & $ 0.027_{-0.005}^{+0.004} $ & $ 1.230 $ & $ 1.895 $ 
         \\          
         \hline    
$\Omega _b\to\Omega _c(3090)$ 
          & $ f^V_1  $ 
          & $ -0.291_{-0.043}^{+0.043} $ & $ -0.483_{-0.071}^{+0.071} $ & $2.573 $ & $ 3.826 $
          & $ g^A_1  $ 
          & $ 0.097_{-0.014}^{+0.013} $ & $ 0.151_{-0.022}^{+0.02} $ & $ 1.811 $ & $ 1.319 $
          \\
          & $ f^V_2 $ 
          & $ -0.433_{-0.066}^{+0.065} $ & $ -0.757_{-0.115}^{+0.114} $ & $ 2.467 $ & $ 2.836 $
          & $ g^A_2  $ 
          & $ 0.00_{-0.002}^{+0.002} $ & $ 0.00_{-0.002}^{+0.002} $ & $ 1.044 $ & $ 1.006 $ 
          \\
          & $ f^V_3 $ 
          & $ 0.159_{-0.033}^{+0.033} $ & $ 0.259_{-0.054}^{+0.054} $ & $ 2.278 $ & $ 2.763 $ 
          & $ g^A_3 $ 
          & $ -0.010_{-0.003}^{+0.003} $ & $ -0.017_{-0.005}^{+0.005} $ & $ 1.977 $ & $ 0.932 $ 
          \\                                                       
\end{tabular}
\end{ruledtabular}
}
\end{table}

\begin{table}[t!]
\caption{\label{tab: figi type iii and iv} The transition form factors for various
$\B_b({\bf 6_f},1/2^+)\to\B_c({\bf 6_f},3/2^\pm)$ transitions [types (iii), (iii)$^*$ and (iv)]. 
We employ a three parameter form for these form factors, Eq.~(\ref{eq:FFpara1}),
while for those with asterisks we employ Eq.~(\ref{eq:FFpara2}).}
\footnotesize{
\begin{ruledtabular}
\begin{tabular}{ccccccccccc}
 $\B_b\to\B_c$
          & $F$
          & $F(0)$
          & $F(q^2_{max})$
          & $a$
          & $b$
          & $F$
          & $F(0)$
          & $F(q^2_{max})$
          & $a$
          & $b$
          \\
\hline     
$ \Omega _b\to\Omega _c(2770)$ 
         & $\bar f^V_1{}^*$  
         & $ -0.734_{-0.127}^{+0.126} $ & $ -1.270_{-0.220}^{+0.218} $ & $ 1.138 $ & $ 1.780 $ 
         & $\bar g^A_1{}^*$  
         & $ 0.526_{-0.083}^{+0.079} $ & $ 1.023_{-0.161}^{+0.153} $ & $ 1.217 $ & $1.010 $ 
         \\
         & $\bar f^V_2{}^*$ 
         & $ -0.300_{-0.066}^{+0.063} $ & $ -0.570_{-0.125}^{+0.120} $ & $ 1.633 $ & $ 2.614 $
         & $\bar g^A_2{}^*$  
         & $0.088_{-0.019}^{+0.020} $ & $ 0.230_{-0.049}^{+0.052} $ & $ 2.137 $ & $ 1.960 $ 
         \\
         & $\bar f^V_3{}^*$ 
         & $ 0.340_{-0.067}^{+0.069} $ & $ 0.573_{-0.113}^{+0.117} $ & $ 1.503 $ & $ 3.256 $
         & $\bar g^A_3{}^*$ 
         & $ -0.361_{-0.072}^{+0.069} $ & $ -0.838_{-0.167}^{+0.160} $ & $ 1.983 $ & $ 2.245 $
         \\
         & $\bar f^V_4{}^*$ 
         & $ 0.033_{-0.012}^{+0.016} $ & $ 0.037_{-0.014}^{+0.018} $ & $ 0.108 $ & $ 3.272 $ 
         & $\bar g^A_4{}^*$ 
         & $ 0.021_{-0.013}^{+0.014} $ & $0.031_{-0.019}^{+0.021} $ & $ 1.233 $ & $ 3.716 $ 
         \\
         \hline
$ \Omega _b\to\Omega _c(3120)$ 
         & $\bar f^V_1$  
         & $ 0.572_{-0.053}^{+0.062} $ & $ 0.741_{-0.068}^{+0.080} $ & $ 1.250 $ & $ 1.184 $ 
         & $\bar g^A_1$  
         & $ -0.369_{-0.032}^{+0.028} $ & $ -0.487_{-0.042}^{+0.037} $ & $ 1.162 $ & $ 0.547 $ 
         \\
         & $\bar f^V_2$ 
         & $ 0.270_{-0.036}^{+0.039} $ & $ 0.403_{-0.053}^{+0.058} $ & $ 1.851 $ & $ 1.884 $ 
         & $\bar g^A_2$  
         & $ 0.038_{-0.028}^{+0.029} $ & $0.017_{-0.012}^{+0.013} $ & $ 4.054 $ & $ 40.596 $
         \\
         & $\bar f^V_3$ 
         & $ -0.263_{-0.033}^{+0.028} $ & $ -0.330_{-0.041}^{+0.035} $ & $ 1.445 $ & $ 2.477 $ 
         & $\bar g^A_3$ 
         & $0.163_{-0.022}^{+0.021} $ & $ 0.174_{-0.023}^{+0.022} $ & $ 0.639 $ & $ 1.531 $
         \\
         & $\bar f^V_4$ 
         & $ 0.010_{-0.015}^{+0.016} $ & $ 0.013_{-0.018}^{+0.02} $ & $ 1.011 $ & $ 1.000 $ 
         & $\bar g^A_4$ 
         & $ -0.106_{-0.035}^{+0.034} $ & $ -0.151_{-0.05}^{+0.048} $ & $ 2.565 $ & $ 5.564 $  
         \\         
         \hline      
$\Omega _b\to\Omega _c(3050)$ 
         & $\bar f^A_1$  
         & $ -0.667_{-0.096}^{+0.115} $ & $-1.164_{-0.167}^{+0.200} $ & $ 2.007 $ & $ 1.089 $ 
         & $\bar g^V_1$  
         & $ 0.163_{-0.026}^{+0.024} $ & $ 0.127_{-0.020}^{+0.019} $ & $ -0.463 $ & $2.900 $  
         \\
         & $\bar f^A_2$ 
         & $ -0.376_{-0.075}^{+0.077} $ & $ -0.574_{-0.115}^{+0.118} $ & $ 2.122 $ & $ 2.913 $ 
         & $\bar g^V_2{}^*$  
         & $ -0.147_{-0.055}^{+0.048} $ & $ -0.541_{-0.202}^{+0.176} $ & $ 3.160 $ & $ 2.247 $
         \\
         & $\bar f^A_3$ 
         & $ 0.300_{-0.049}^{+0.042} $ & $ 0.405_{-0.066}^{+0.057} $ & $ 2.034 $ & $ 3.976 $ 
         & $\bar g^V_3$ 
         & $ 0.335_{-0.109}^{+0.120} $ & $0.255_{-0.083}^{+0.092} $ & $ -1.021 $ & $ 0.997 $ 
         \\
         & $\bar f^A_4{}^*$ 
         & $ 0.047_{-0.013}^{+0.018} $ & $ 0.157_{-0.045}^{+0.060} $ & $ 2.785 $ & $ 1.298 $ 
         & $\bar g^V_4{}^*$ 
         & $ 0.426_{-0.110}^{+0.115} $ & $ 0.774_{-0.200}^{+0.208} $ & $ 1.813 $ & $ 2.902 $  
         \\                                                                                   
\end{tabular}
\end{ruledtabular}
}
\end{table}

\begin{figure}[t!]
\centering
\subfigure[]{
 \includegraphics[width=0.44\textwidth]  {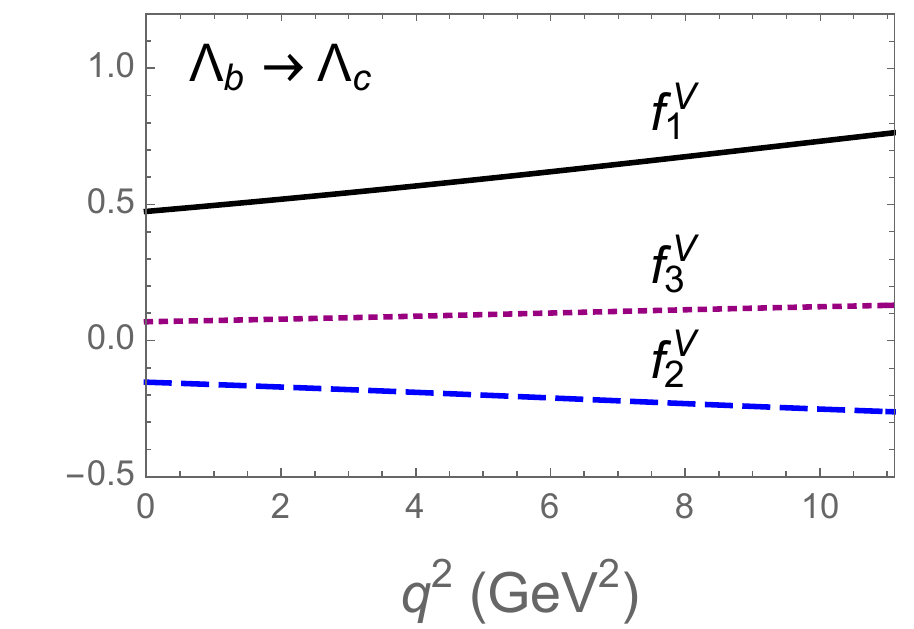}
}
\subfigure[]{
  \includegraphics[width=0.44\textwidth]  {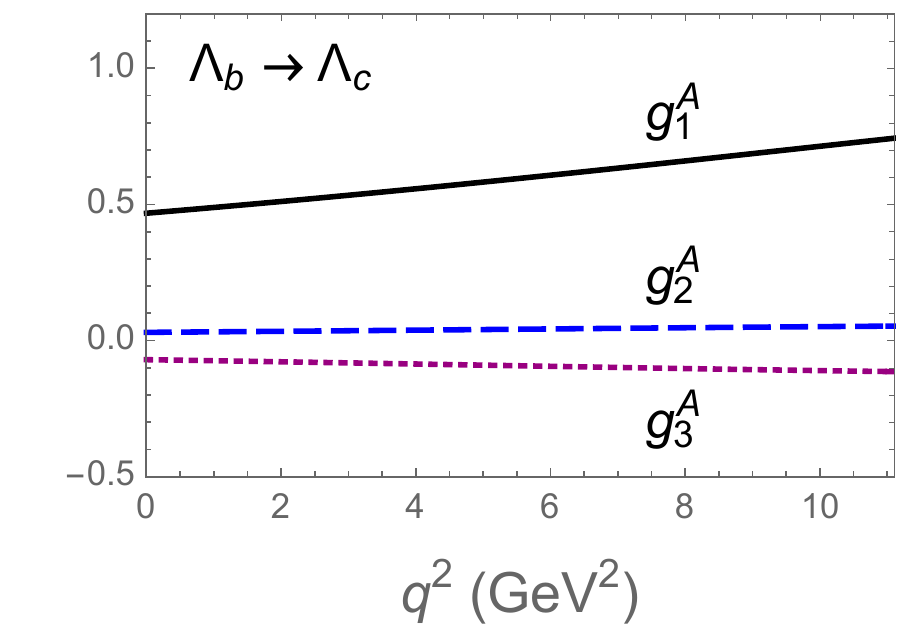}
}
\subfigure[]{
 \includegraphics[width=0.44\textwidth]  {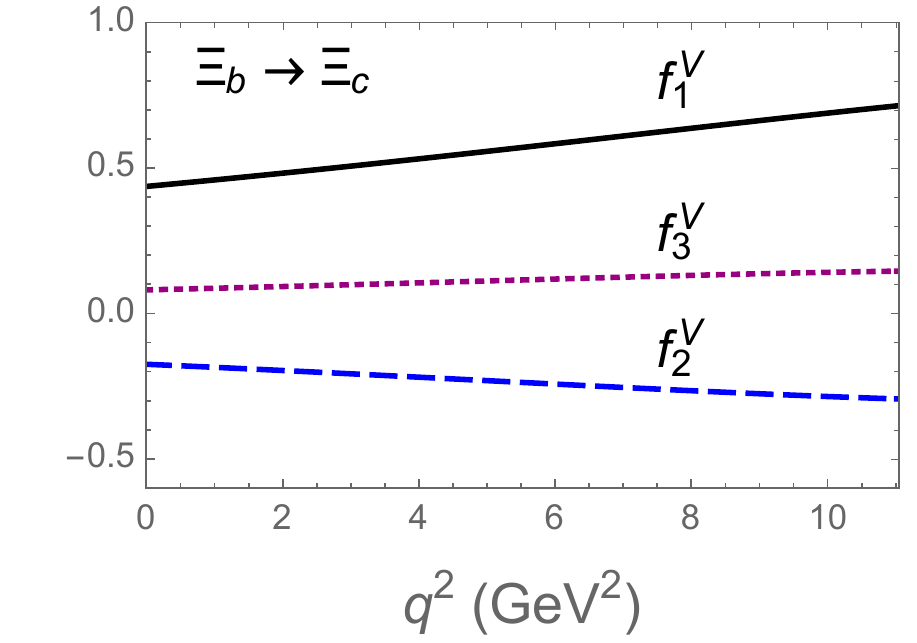}
}
\subfigure[]{
  \includegraphics[width=0.44\textwidth]  {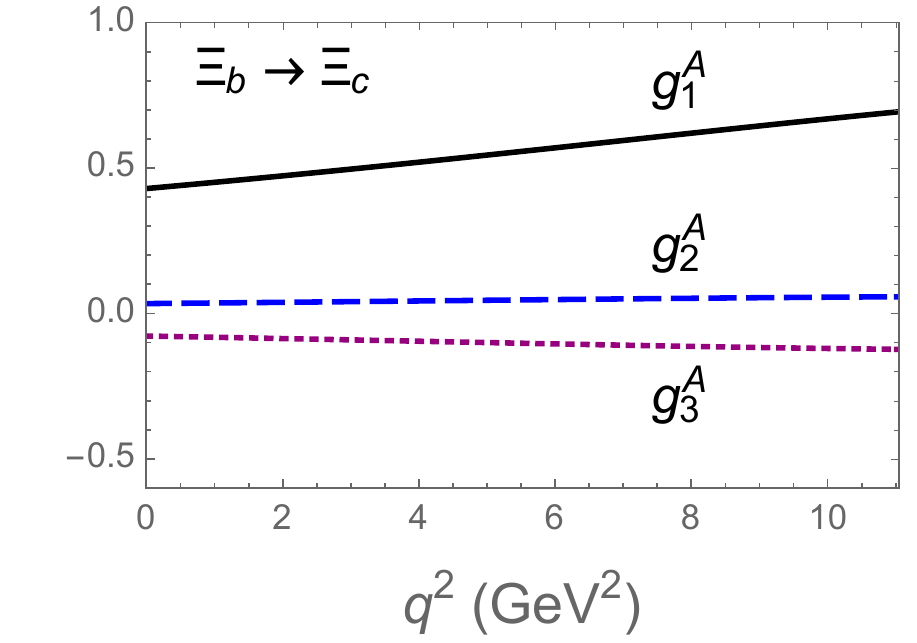}
}
\subfigure[]{
 \includegraphics[width=0.44\textwidth]  {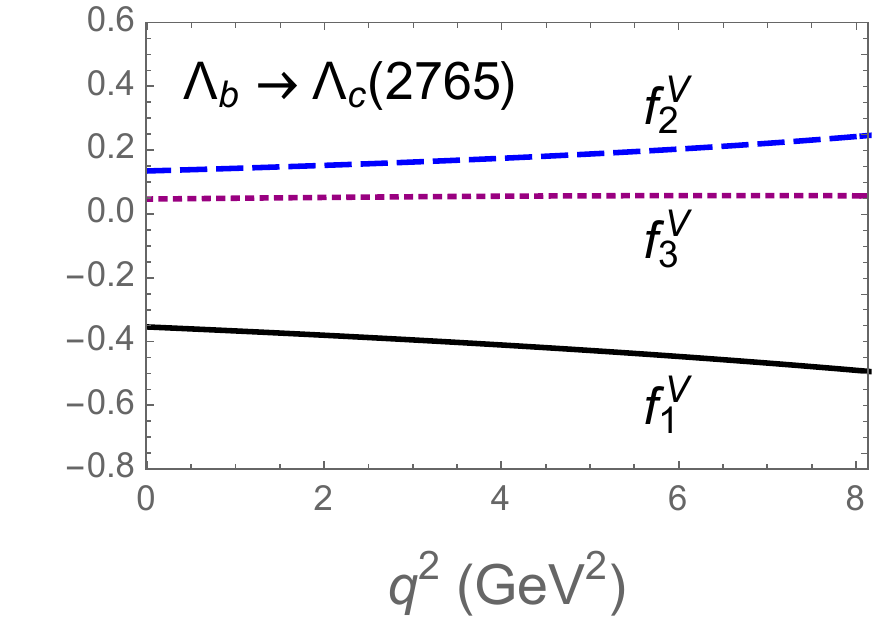}
}
\subfigure[]{
  \includegraphics[width=0.44\textwidth]  {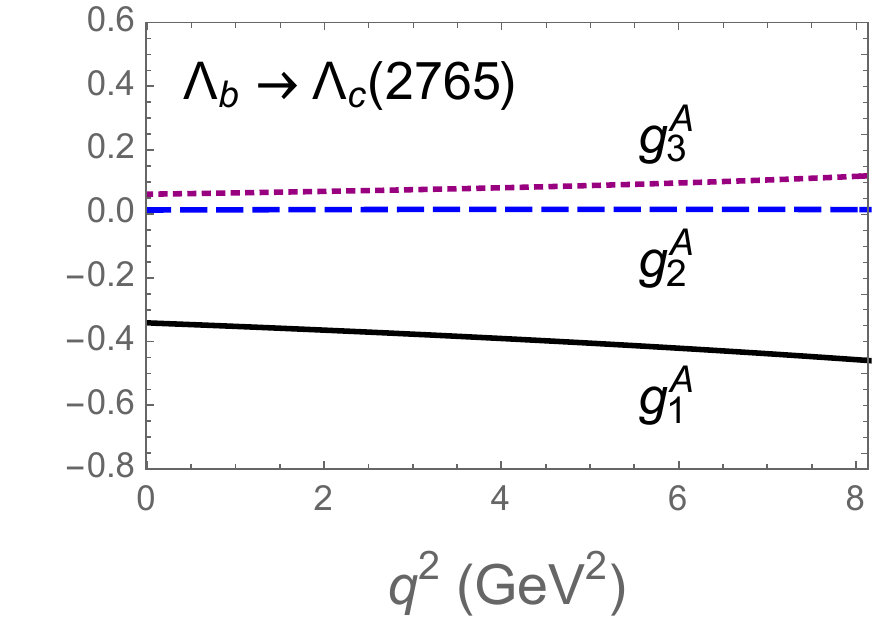}
}
\caption{Form factors $f_{1,2,3}(q^2)$ and $g_{1,2,3}(q^2)$ for 
$\Lambda_b\to\Lambda_c, \Lambda_c(2765)$ and $\Xi_b\to\Xi_c$ transitions.
The transitions are $\B_b({\bf\bar 3_f},1/2^+)\to\B_c({\bf \bar 3_f},1/2^+)$ transitions [types (i) and (i)$^*$].}
\label{fig: figi type i}
\end{figure}

\begin{figure}[t!]
\centering
\subfigure[]{
 \includegraphics[width=0.44\textwidth]  {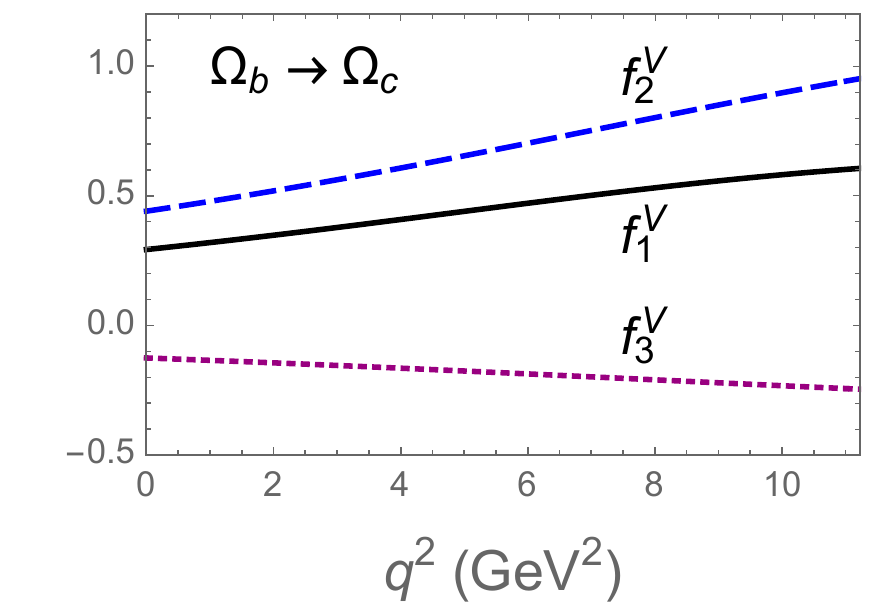}
}
\subfigure[]{
  \includegraphics[width=0.44\textwidth]  {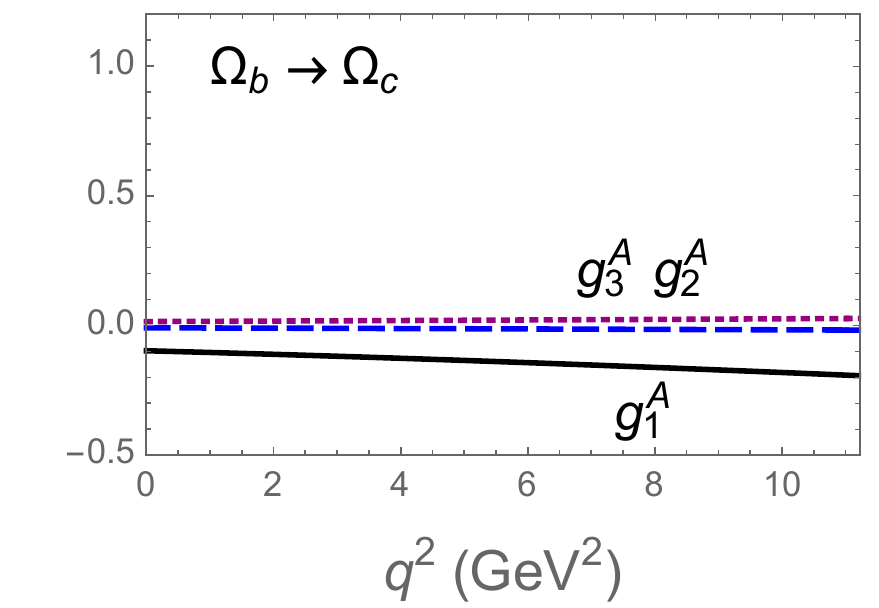}
}
\subfigure[]{
 \includegraphics[width=0.44\textwidth]  {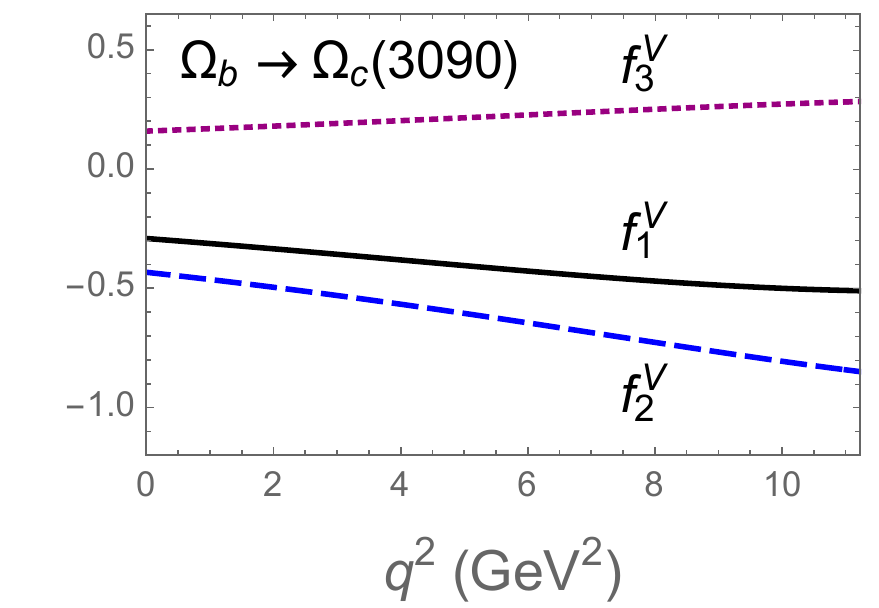}
}
\subfigure[]{
  \includegraphics[width=0.44\textwidth]  {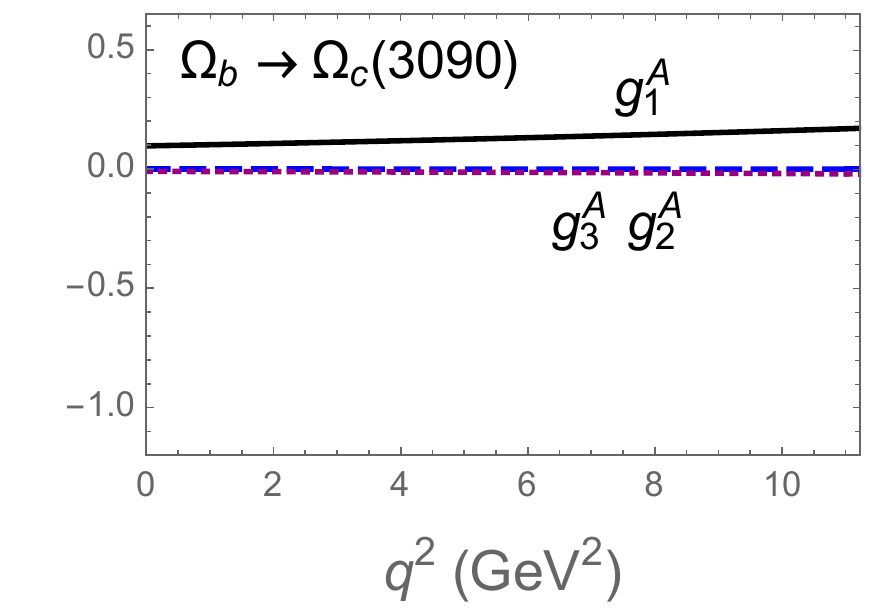}
}
\caption{Form factors $f_{1,2,3}(q^2)$ and $g_{1,2,3}(q^2)$ for
$\Omega_b\to\Omega_c$ and $\Omega_c(3090)$
transitions. The transitions are $\B_b({\bf 6_f},1/2^+)\to\B_c({\bf 6_f},1/2^+)$ transitions [types (ii) and (ii)$^*$].}
\label{fig: figi type ii}
\end{figure}

\begin{figure}[t!]
\centering
\subfigure[]{
 \includegraphics[width=0.44\textwidth]  {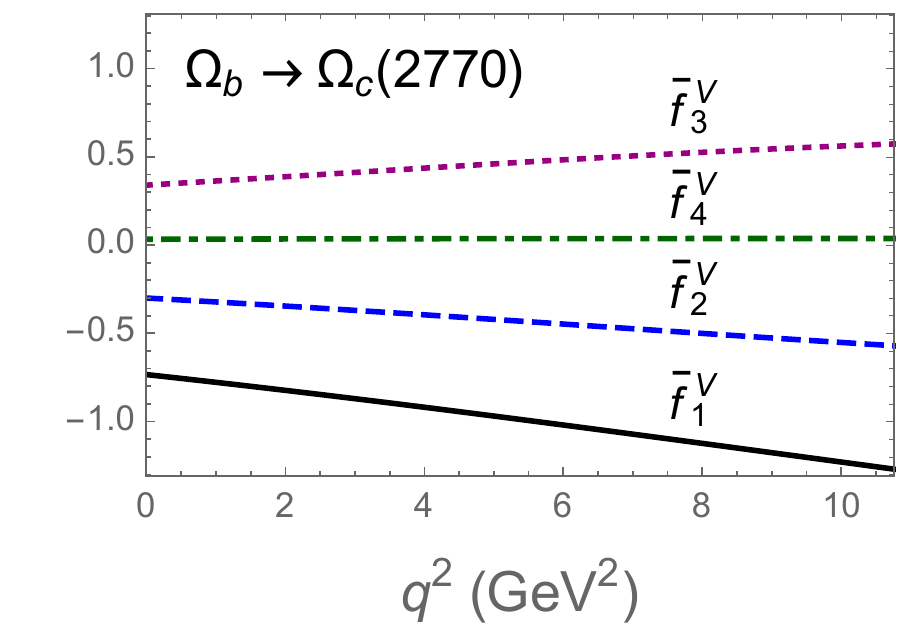}
}
\subfigure[]{
  \includegraphics[width=0.44\textwidth]  {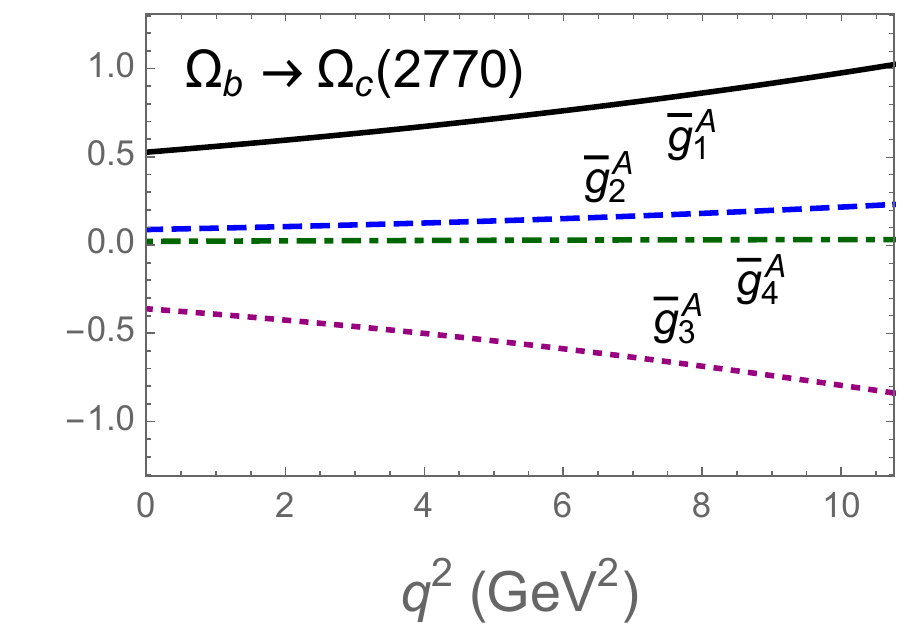}
}
\subfigure[]{
 \includegraphics[width=0.44\textwidth]  {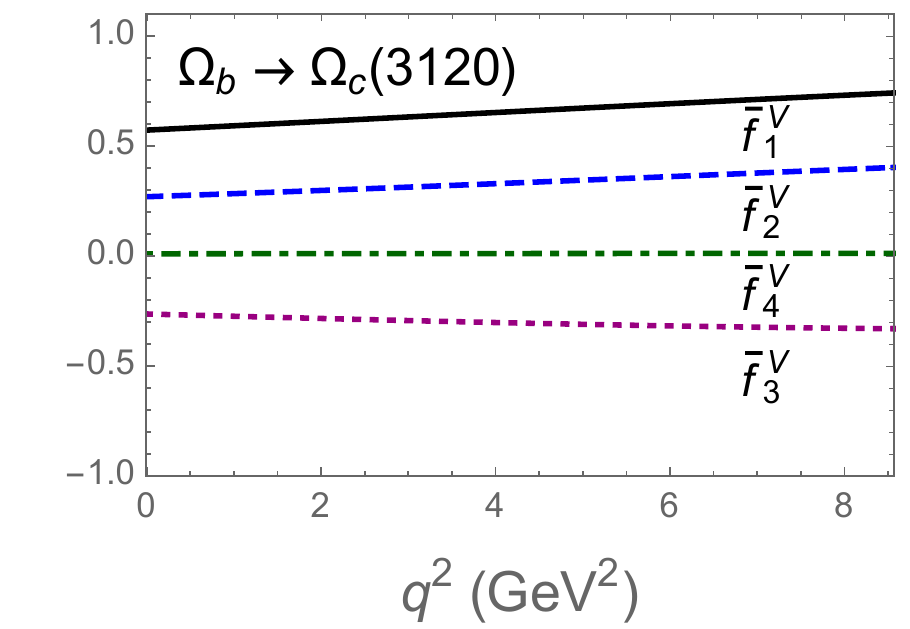}
}
\subfigure[]{
  \includegraphics[width=0.44\textwidth]  {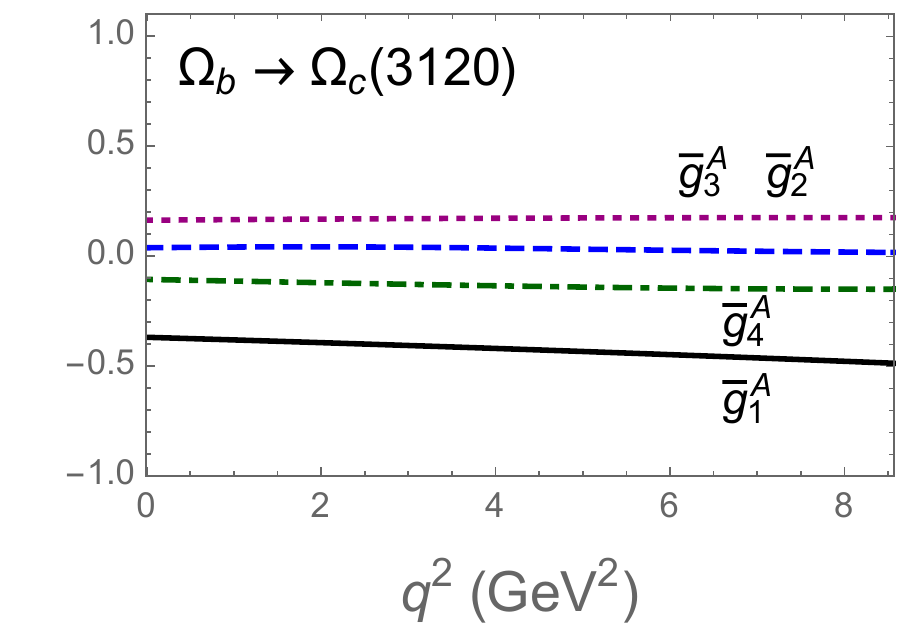}
}
\subfigure[]{
 \includegraphics[width=0.44\textwidth]  {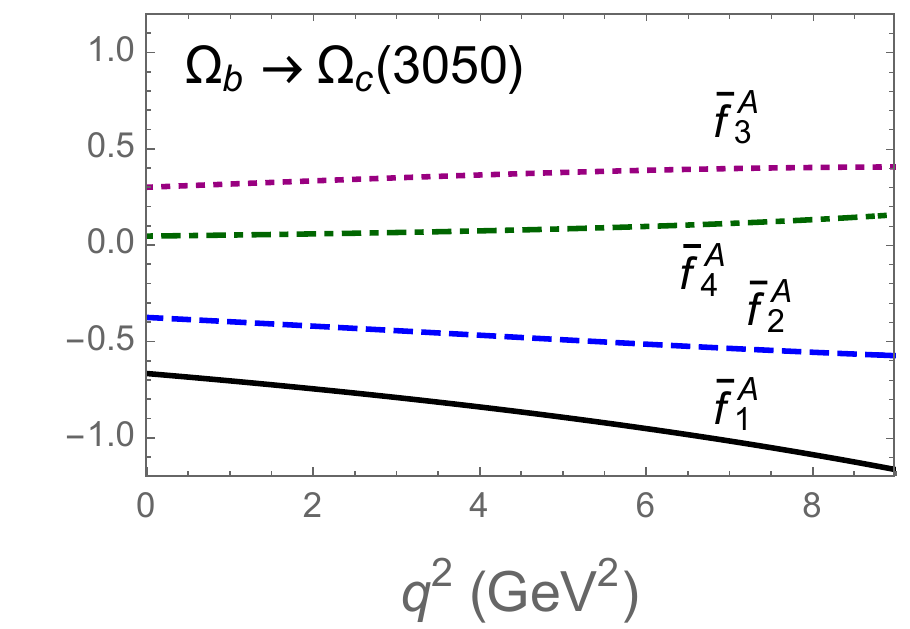}
}
\subfigure[]{
  \includegraphics[width=0.44\textwidth]  {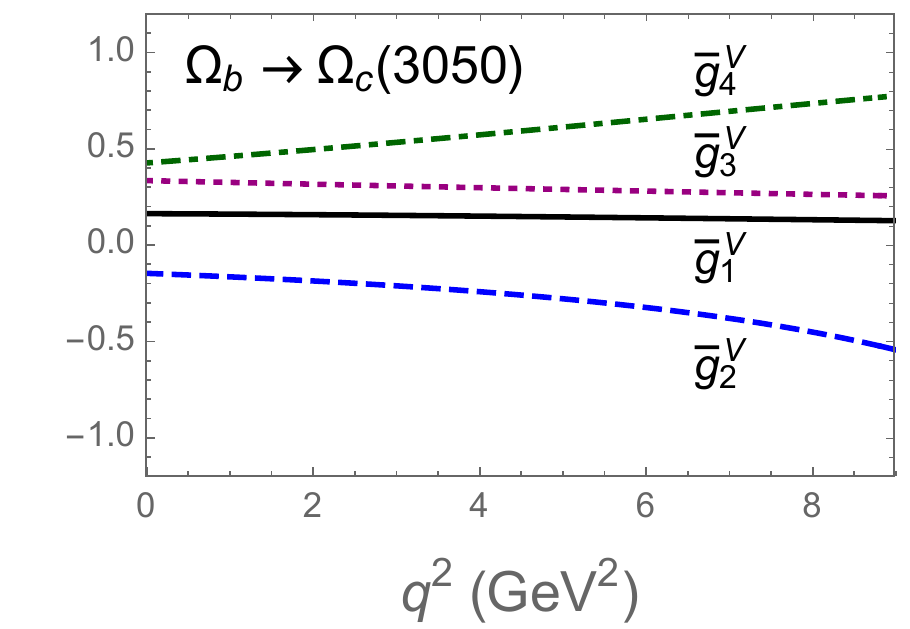}
}
\caption{Form factors $\bar f^{V,A}_i(q^2)$ and $\bar g^{A,V}_i(q^2)$ for
$\Omega_b\to\Omega_c(2770)$, $\Omega_c(3120)$ and $\Omega_c(3050)$
transitions. 
The transitions are $\B_b({\bf 6_f},1/2^+)\to\B_c({\bf 6_f},3/2^+)$ transitions [types (iii) and (iii)$^*$] and 
$\B_b({\bf 6_f},1/2^+)\to\B_c({\bf 6_f},3/2^-)$ transition [type (iv)].}
\label{fig: figi type iii and iv}
\end{figure}

The $\B_b({\bf\bar 3_f},1/2^+)\to\B_c({\bf \bar 3_f},1/2^+)$ transition form factors $f^V_{1,2,3}(q^2)$ and $g^A_{1,2,3}(q^2)$ for $\Lambda_b\to\Lambda_c, \Lambda_c(2765)$ and $\Xi_b\to\Xi_c$ transitions
are given in Table~\ref{tab:fg type i} and they are plotted in Fig.~\ref{fig: figi type i}.
The uncertainties in form factors $F(0)$ are obtained by varying $m_b$, $m_c$, $m_{[qq]}$, $\beta(\B_b)$ and $\beta(\B_c)$ by 10\% separately and combine the uncertainties quadratically.
Note that $\Lambda_c$ and $\Xi_c$ are low lying states, while $\Lambda_c(2765)$ is a radial excited state.
From the table and the figures, we see that $f^V_1\simeq g^A_1$ and they dominate over $f^V_{2,3}$ and $g^A_{2,3}$. 
These are close to the predicted relations of form factors in the heavy quark limit, see Eq. ~(\ref{eq: HQ type i}). 
The values of $f^V_1$ and $g^A_1$ at $q^2_{max}$ in $\Lambda_b\to\Lambda_c$ and $\Xi_b\to\Xi_c$ transitions are smaller than the ones predicted in heavy quark limit with $\zeta(1)=1$, Eq. (\ref{eq: HQ type i w=1}), by roughly $25\%$. 
The reduction can be more or less traced to the mismatch of the overlapping of the wave functions of $\B_b$ and $\B_c$, 
as $\beta(\B_b)\neq \beta(\B_c)$.
Indeed, it is easy to see that, using $\beta(\Lambda_b)$ and $\beta(\Lambda_c)$ given in Table~\ref{tab:input}, 
the overlapping integral of wave functions of $\Lambda_b$ and $\Lambda_c$ is $0.66$, 
which is smaller than the one in the ideal case with $\beta(\Lambda_b)=\beta(\Lambda_c)$.
Things are similar in the $\Xi_b\to\Xi_c$ transition.
In the $\Lambda_b\to \Lambda_c(2765)$ transition 
the sizes of the form factors are smaller than those in $\Lambda_b\to\Lambda_c$ and $\Xi_b\to\Xi_c$ transitions.
and the signs of form factors are flipped.
These changes are closely related to the fact that $\Lambda_c(2765)$ is a radial excited state. 
In fact, in the heavy quark limit, one expects $f^V_1=g^A_1=0$ at $q^2=q^2_{max}$, as the wave functions of the low-lying $\B_b$ state and the radial excited $\B_c$ state are orthogonal~\cite{Isgur:1991wr}. 
This is not borne out as $\beta(\B_b)\neq \beta(\B_c)$, and instead the overlapping integral is $-0.53$.
As the overlapping integral has smaller size and opposite sign compared to those in the low-lying $\B_b$ and $\B_c$ states,   
the reduction in sizes and the flipping of signs in $f_1^V$ and $g_1^A$ in this transition are expected.

The $\B_b({\bf 6_f},1/2^+)\to\B_c({\bf 6_f},1/2^+)$ transition form factors $f^V_{1,2,3}(q^2)$ and $g^A_{1,2,3}(q^2)$ for $\Omega_b\to\Omega_c$ and $\Omega_b\to\Omega_c(3090)$ transitions
are given in Table~\ref{tab:fg type ii} and are plotted in Fig.~\ref{fig: figi type ii}.
Note that $\Omega_c$ is a low lying state, while $\Omega_c(3090)$ is a radial excite state.
Heavy quark symmetry and large $N_c$ QCD predict that in these transitions,
we have $\xi_1(\omega)=(1+\omega)\xi_2(\omega)$, see Eq. ~(\ref{eq: large Nc}), which implies
$g_2^A(q^2)=g_3^A(q^2)=0$. 
We see from the table and the plots that we indeed have $g^A_{2,3}(q^2)\simeq 0$.
In $\Omega_b\to\Omega_c$ transition, Eq. (\ref{eq: w=1})
gives
$\xi_1(1)=2\xi_2(1)=1$, and, consequently, form factors at $q^2_{max}$ take the following values : 
$(f^V_1, f^V_2, f^V_3)=(1.23, 1.56, -0.60)$ and $(g^A_1, g^A_2, g^A_3)=(-0.33,0,0)$, see Eq.~(\ref{eq: HQ type ii w=1}).
From Table~\ref{tab:fg type ii} and Fig.~\ref{fig: figi type ii}(a), (b) we see that the form factors basically exhibit a similar pattern in sizes and signs, 
but the values of $|f^V_i|$ and $|g^A_1|$ are smaller by $40\%$ to $60\%$, while the values of $g_{2,3}^A$ are closer to the HQS predicted values.
The reductions can again be more or less traced to the smaller wave function overlapping integral, which is $0.46$ in this case.
In the $\Omega_b\to\Omega_c(3090)$ transition, we obtain form factors with similar sizes and signs flipped compared to the previous ones. These can also be more or less traced to the overlapping integral of the low lying $\Omega_b$ state and the radial excited $\Omega_c(3090)$ state. Explicitly, the corresponding overlapping integral is $-0.46$,
which is non-vanishing (unlike the one in the ideal case), but has an opposite sign and a similar size compare to the one in the previous case.

\begin{table}[t!]
\caption{\label{tab:fg type v} The transition form factors for various
$\B_b({\bf 3_f},1/2^+)\to\B_c({\bf 3_f},1/2^-)$ transitions [type (v) and (v)$^*$]. 
}
\footnotesize{
\begin{ruledtabular}
\begin{tabular}{ccccccccccc}
 $\B_b\to\B_c$
          & $F$
          & $F(0)$
          & $F(q^2_{max})$
          & $a$
          & $b$
          & $F$
          & $F(0)$
          & $F(q^2_{max})$
          & $a$
          & $b$
          \\
\hline     
$\Lambda _b\to\Lambda _c(2595)$ 
           & $ f^A_1  $ 
           & $ 0.286_{-0.051}^{+0.050} $ & $ 0.338_{-0.060}^{+0.059} $ & $ 0.667 $ & $ 0.483 $ 
           & $ g^V_1  $ 
           & $0.238_{-0.049}^{+0.048} $ & $ 0.232_{-0.048}^{+0.047} $ & $ 0.120 $ & $ 0.722 $  
           \\
          & $ f^A_2 $ 
          & $ -0.313_{-0.081}^{+0.079} $ & $ -0.439_{-0.114}^{+0.111} $ & $ 1.312 $ & $ 1.105 $
          & $ g^V_2 $ 
          & $ -0.080_{-0.022}^{+0.020} $ & $ -0.112_{-0.031}^{+0.028} $ & $ 1.416 $ & $ 1.510 $ 
          \\
          & $ f^A_3 $ 
          & $ -0.299_{-0.083}^{+0.080} $ & $ -0.459_{-0.128}^{+0.123} $ & $ 1.722 $ & $ 1.791 $
          & $ g^V_3 $ 
          & $-0.167_{-0.038}^{+0.039} $ & $ -0.228_{-0.052}^{+0.053} $ & $ 1.286 $ & $ 1.234 $  
          \\          
          \hline
$\Xi _b\to\Xi _c^0(2790)$ 
          & $ f^A_1  $ 
          & $ 0.269_{-0.046}^{+0.056} $ & $ 0.330_{-0.056}^{+0.068} $ & $ 0.895 $ & $ 0.755 $  
          & $ g^V_1 $ 
          & $ 0.221_{-0.046}^{+0.046} $ & $ 0.221_{-0.046}^{+0.046} $ & $ 0.254 $ & $ 0.959 $  
          \\
         & $ f^A_2 $ 
         & $ -0.319_{-0.086}^{+0.083} $ & $ -0.447_{-0.121}^{+0.116} $ & $ 1.487 $ & $ 1.544 $ 
         & $ g^V_2  $ 
         & $ -0.071_{-0.021}^{+0.020} $ & $ -0.099_{-0.029}^{+0.028} $ & $ 1.619 $ & $ 2.187 $  
         \\
        & $ f^A_3 $ 
        & $ -0.289_{-0.084}^{+0.081} $ & $ -0.436_{-0.127}^{+0.122} $ & $ 1.944 $ & $ 2.580 $ 
        & $ g^V_3 $ 
        & $-0.169_{-0.040}^{+0.042} $ & $ -0.228_{-0.054}^{+0.057} $ & $ 1.404 $ & $ 1.663 $
        \\          
        \hline                                                                                       
$ \Lambda _b\to\Lambda _c(2940)$ 
         & $ f^A_1  $ 
         & $ -0.255_{-0.010}^{+0.019} $ & $-0.333_{-0.013}^{+0.025} $ & $ 0.996 $ & $ -0.156 $  
         & $ g^V_1  $ 
         & $ -0.212_{-0.009}^{+0.017} $ & $-0.241_{-0.010}^{+0.019} $ & $ 0.533 $ & $ 0.034 $  
         \\
         & $ f^A_2 $ 
         & $ 0.338_{-0.066}^{+0.064} $ & $ 0.584_{-0.114}^{+0.111} $ & $ 1.997 $ & $ 0.642 $ 
         & $ g^V_2  $ 
         & $ 0.081_{-0.013}^{+0.010} $ & $0.138_{-0.022}^{+0.017} $ & $ 2.029 $ & $ 0.839 $ 
         \\
        & $ f^A_3 $ 
        & $ 0.331_{-0.071}^{+0.071} $ & $ 0.592_{-0.127}^{+0.127} $ & $ 2.176 $ & $ 1.031 $ 
        & $ g^V_3 $ 
        & $ 0.157_{-0.015}^{+0.009} $ & $0.301_{-0.029}^{+0.017} $ & $ 2.324 $ & $ 0.954 $
        \\                
\end{tabular}
\end{ruledtabular}
}
\end{table}

\begin{table}[t!]
\caption{\label{tab:fg type vi} The transition form factors for various
$\B_b({\bf \bar 3_f},1/2^+)\to\B_c({\bf \bar 3_f},3/2^-)$ transitions [type (vi)]. 
Note that the $a$ and $b$ for form factors denoted with asterisks are assumed, the corresponding form factors are small.
}
\footnotesize{
\begin{ruledtabular}
\begin{tabular}{ccccccccccc}
 $\B_b\to\B_c$
          & $F$
          & $F(0)$
          & $F(q^2_{max})$
          & $a$
          & $b$
          & $F$
          & $F(0)$
          & $F(q^2_{max})$
          & $a$
          & $b$
          \\
\hline     
$ \Lambda _b\to\Lambda _c(2625)$ 
         & $\bar f^A_1$ 
         & $ 0.028_{-0.032}^{+0.065} $ & $ 0.035_{-0.040}^{+0.082} $   
         & $ 1 $*
         & $ 1 $*
         & $\bar g^V_1$
         & $ -0.007_{-0.026}^{+0.037} $ & $ -0.009_{-0.033}^{+0.046} $ 
         & $ 1 $*
         & $ 1 $*
         \\
         & $\bar f^A_2$ 
         & $ 0.545_{-0.104}^{+0.111} $ & $ 0.756_{-0.144}^{+0.154} $ & $ 1.310 $ & $ 1.154 $ 
         & $\bar g^V_2$  
         & $ 0.509_{-0.173}^{+0.184} $ & $ 0.737_{-0.251}^{+0.267} $ & $ 1.388 $ & $ 1.043 $
         \\
         & $\bar f^A_3$
         & $ 0.022_{-0.091}^{+0.033} $ & $ 0.027_{-0.114}^{+0.041} $
         & 1*
         & 1*
         & $\bar g^V_3$ 
         & $ 0.088_{-0.043}^{+0.039} $ & $0.115_{-0.056}^{+0.051} $ & $ 2.022 $ & $ 4.221 $ 
         \\
         & $\bar f^A_4$
         & $ -0.005_{-0.068}^{+0.104} $ & $ -0.006_{-0.086}^{+0.131} $ 
         & 1*
         & 1* 
         & $\bar g^V_4$
         & $ 0.004_{-0.053}^{+0.058} $ & $ 0.005_{-0.066}^{+0.072} $ 
         & 1*
         & 1*
         \\
         \hline
$ \Xi _b\to\Xi _c(2815)$ 
         & $\bar f^A_1$ 
         & $ 0.049_{-0.048}^{+0.094} $ & $ 0.033_{-0.033}^{+0.064} $ & $ -1.327 $ & $ 1.830 $ 
         & $\bar g^V_1$*  
         & $ 0.015_{-0.029}^{+0.046} $ & $ 0.018_{-0.036}^{+0.057} $
         & 1* 
         & 1* 
         \\
         & $\bar f^A_2$ 
         & $ 0.675_{-0.122}^{+0.128} $ & $ 0.987_{-0.178}^{+0.187} $ & $ 1.526 $ & $ 1.245 $ 
         & $\bar g^V_2$  
         & $ 0.693_{-0.216}^{+0.247} $ & $1.024_{-0.320}^{+0.365} $ & $ 1.557 $ & $ 1.249 $
         \\
         & $\bar f^A_3$ 
         & $ 0.020_{-0.127}^{+0.045} $ & $ 0.024_{-0.158}^{+0.056} $
         & 1* 
         & 1* 
         & $\bar g^V_3$
         & $ 0.055_{-0.075}^{+0.050} $ & $0.068_{-0.093}^{+0.062} $
         & 1* 
         & 1* 
         \\
         & $\bar f^A_4$
         & $ -0.013_{-0.138}^{+0.173} $ & $ -0.017_{-0.171}^{+0.215} $ 
         & 1* 
         & 1* 
         & $\bar g^V_4$ 
         & $-0.020_{-0.083}^{+0.083} $ & $ -0.024_{-0.103}^{+0.103} $
         & 1*
         & 1*
         \\                            
         \hline
$ \Lambda _b\to\Lambda _c(2940)$
         & $\bar f^A_1$
         & $ -0.036_{-0.047}^{+0.053} $ & $ -0.043_{-0.057}^{+0.065} $ 
         & 1* 
         & 1* 
         & $\bar g^V_1$*
         & $ -0.061_{-0.006}^{+0.013} $ & $-0.074_{-0.007}^{+0.016} $         
         & 1* 
         & 1* 
         \\
         & $\bar f^A_2$ 
         & $ -0.459_{-0.067}^{+0.067} $ & $ -1.049_{-0.153}^{+0.153} $ & $ 2.400 $ & $ -0.326 $ 
         & $\bar g^V_2$  
         & $-0.820_{-0.155}^{+0.192} $ & $ -1.375_{-0.260}^{+0.321} $ & $ 1.957 $ & $ 0.793 $  
         \\
         & $\bar f^A_3{}^*$
         & $ -0.012_{-0.098}^{+0.148} $ & $ -0.004_{-0.033}^{+0.050} $ & $ -8.054 $ & $ 18.712 $ 
         & $\bar g^V_3$ 
         & $ 0.198_{-0.084}^{+0.080} $ & $ 0.358_{-0.152}^{+0.145} $ & $ 2.487 $ & $ 2.307 $
         \\
         & $\bar f^A_4{}^*$ 
         & $ 0.192_{-0.120}^{+0.133} $ & $ 0.139_{-0.087}^{+0.096} $ & $ -1.791 $ & $ 7.419 $ 
         & $\bar g^V_4$ 
         & $ 0.170_{-0.095}^{+0.102} $ & $ 0.183_{-0.102}^{+0.110} $ & $ 0.924 $ & $ 2.713 $ 
     \\         
\end{tabular}
\end{ruledtabular}
}
\end{table}

\begin{figure}[t!]
\centering
\subfigure[]{
 \includegraphics[width=0.44\textwidth]  {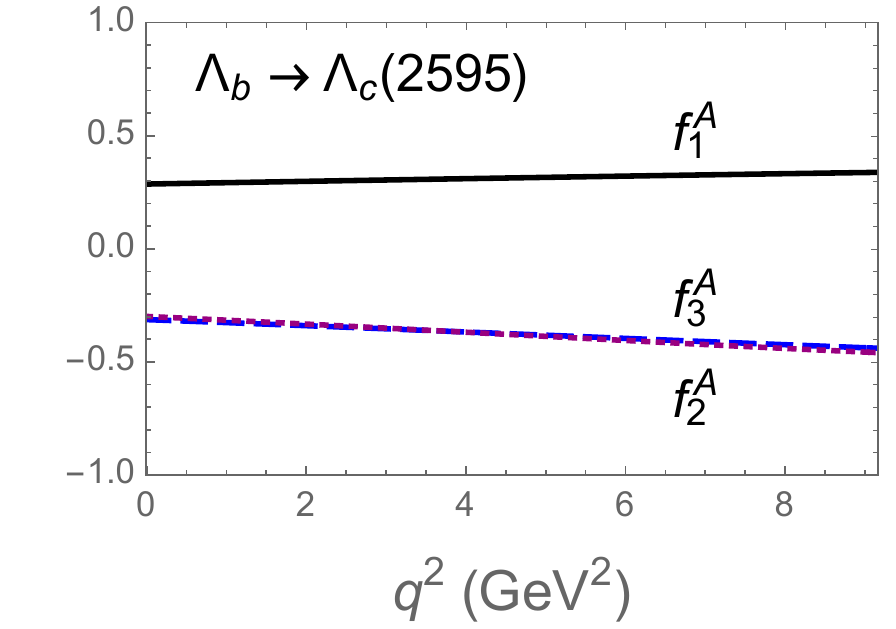}
}
\subfigure[]{
  \includegraphics[width=0.44\textwidth]  {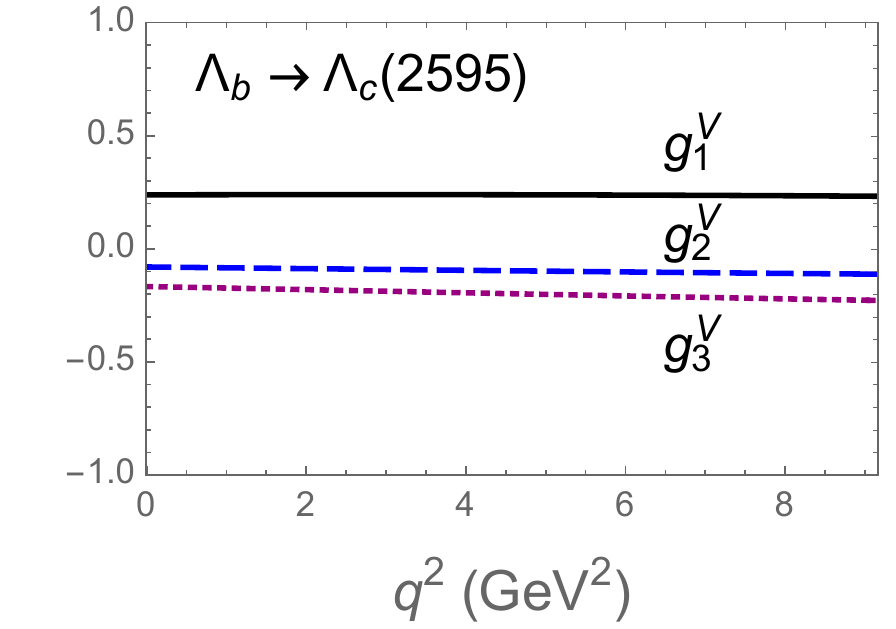}
}
\subfigure[]{
 \includegraphics[width=0.44\textwidth]  {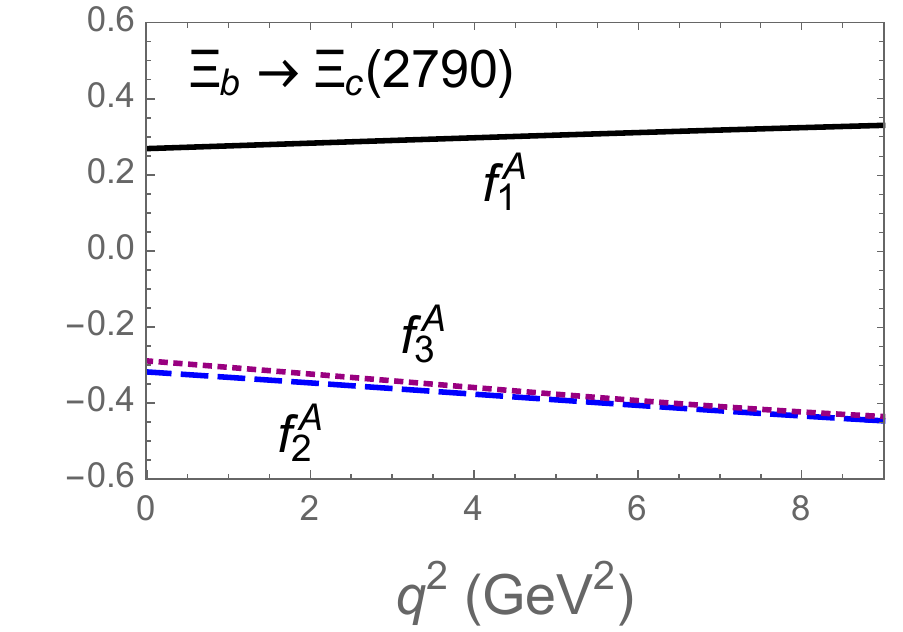}
}
\subfigure[]{
  \includegraphics[width=0.44\textwidth]  {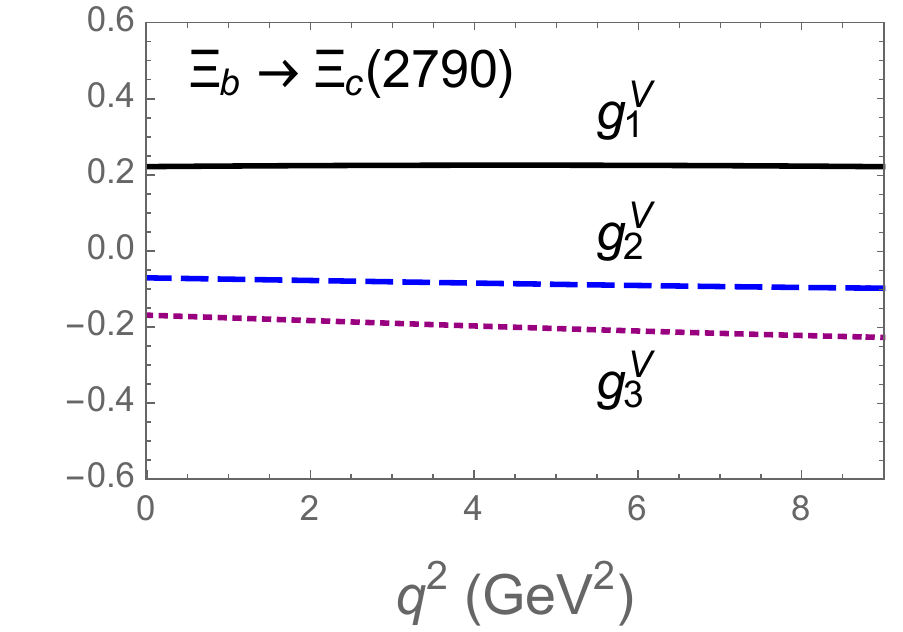}
}
\subfigure[]{
 \includegraphics[width=0.44\textwidth]  {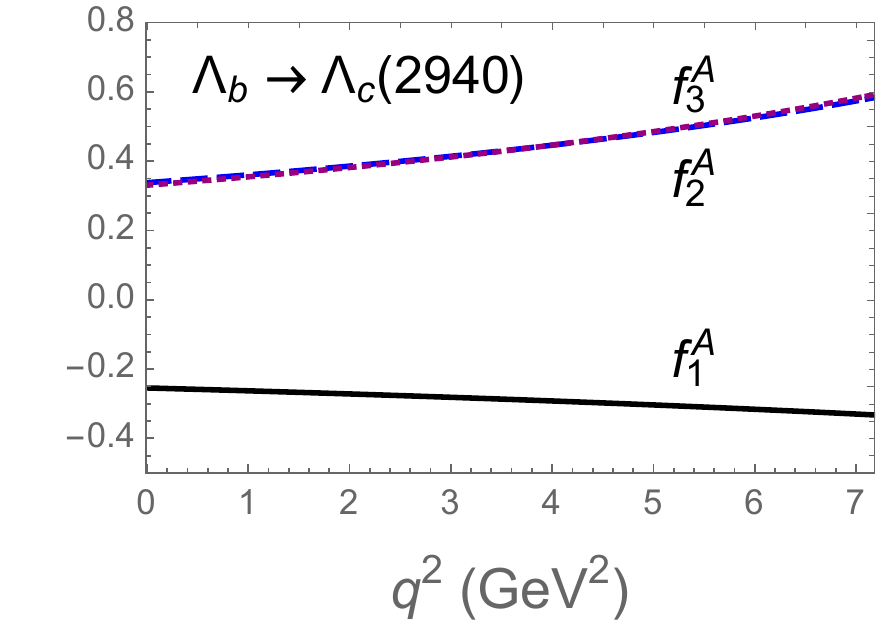}
}
\subfigure[]{
  \includegraphics[width=0.44\textwidth]  {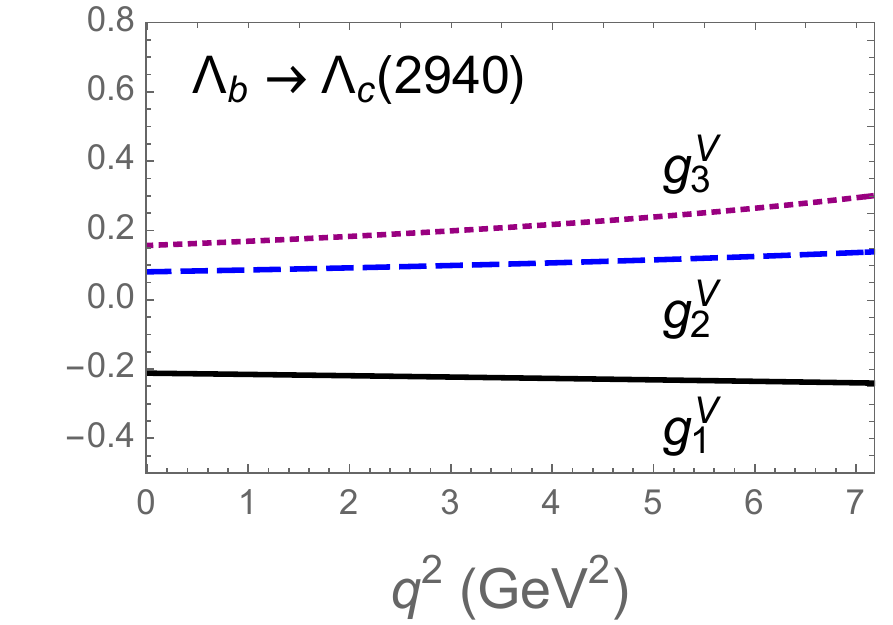}
}
\caption{Form factors $f_{1,2,3}(q^2)$ and $g_{1,2,3}(q^2)$ for
$\Lambda_b\to\Lambda_c(2595)$, $\Lambda_c(2940)$ and $\Xi_b\to\Xi_c(2790)$
transitions. The transitions are $\B_b({\bf\bar 3_f},1/2^+)\to\B_c({\bf\bar 3_f},1/2^-)$ transitions [types (v) and (v)$^*$].}
\label{fig: figi type v}
\end{figure}

\begin{figure}[t!]
\centering
\subfigure[]{
 \includegraphics[width=0.44\textwidth]  {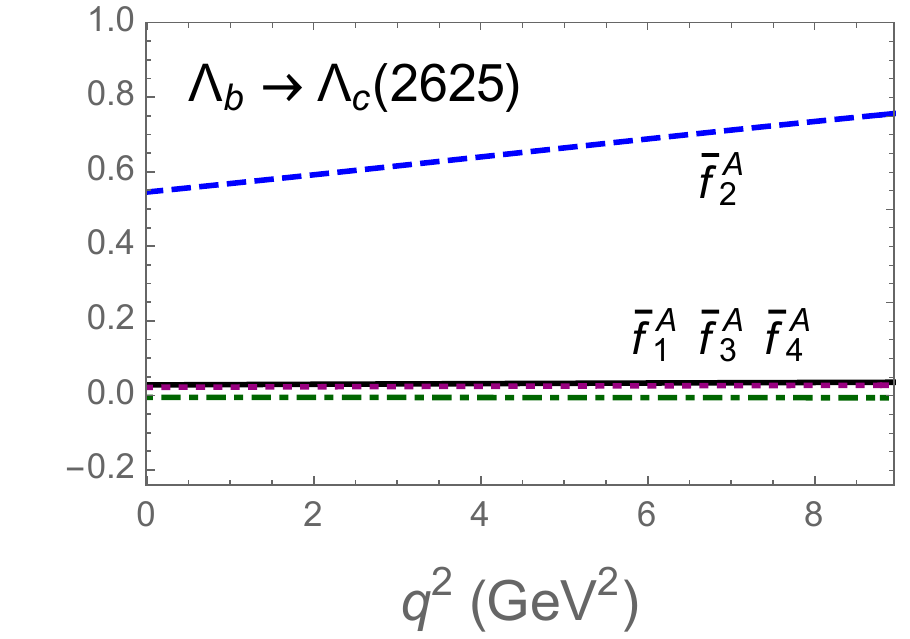}
}
\subfigure[]{
  \includegraphics[width=0.44\textwidth]  {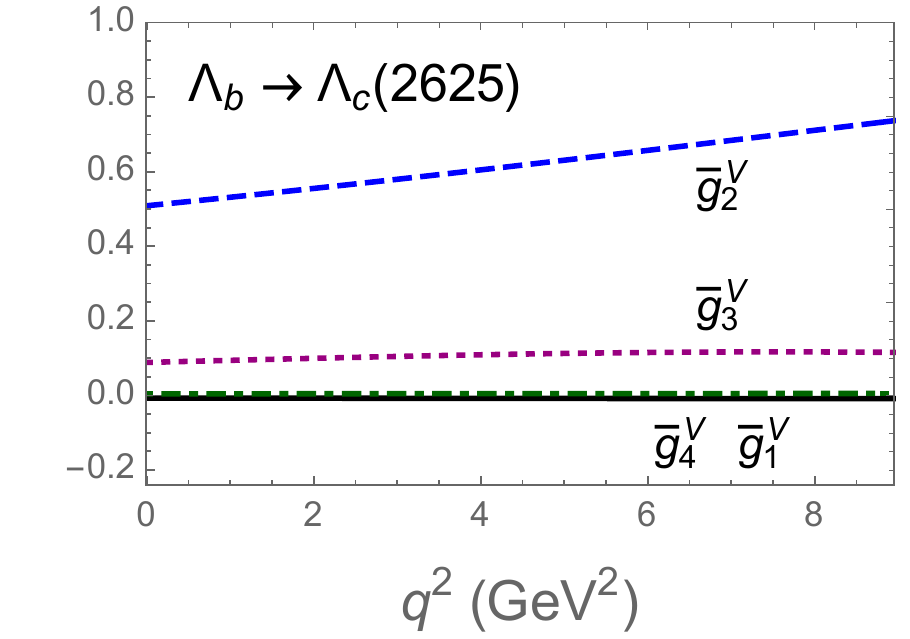}
}
\subfigure[]{
 \includegraphics[width=0.44\textwidth]  {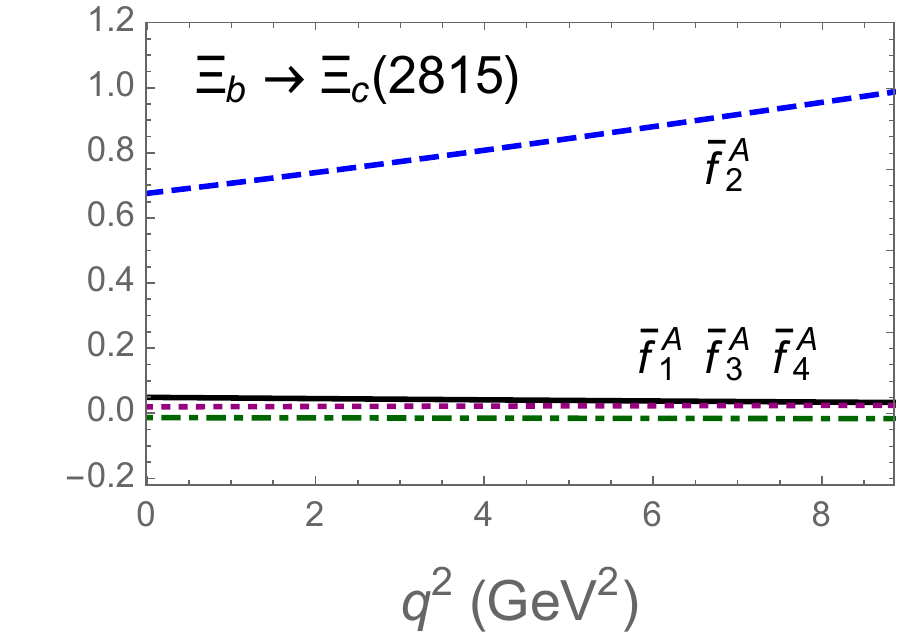}
}
\subfigure[]{
  \includegraphics[width=0.44\textwidth]  {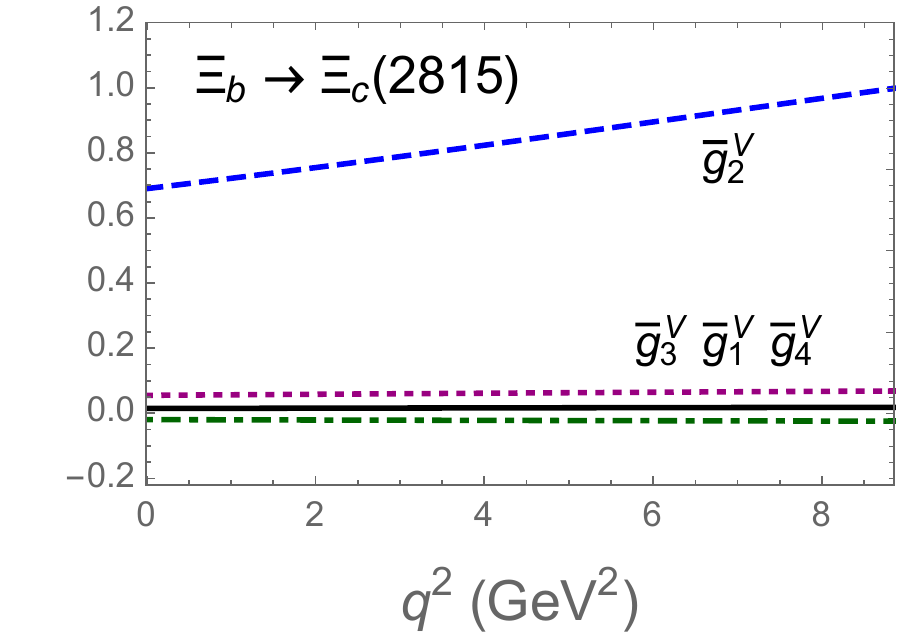}
}
\subfigure[]{
 \includegraphics[width=0.44\textwidth]  {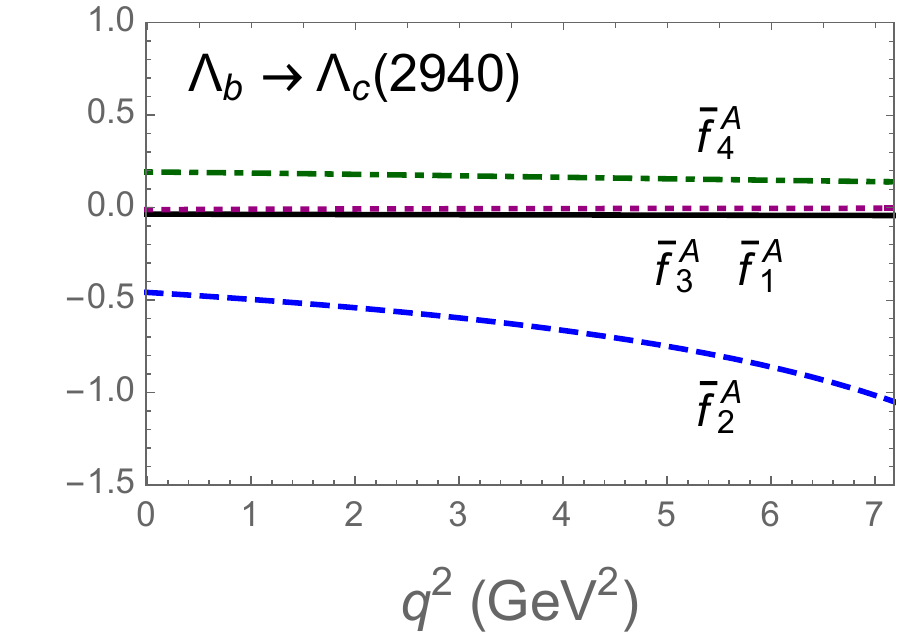}
}
\subfigure[]{
  \includegraphics[width=0.44\textwidth]  {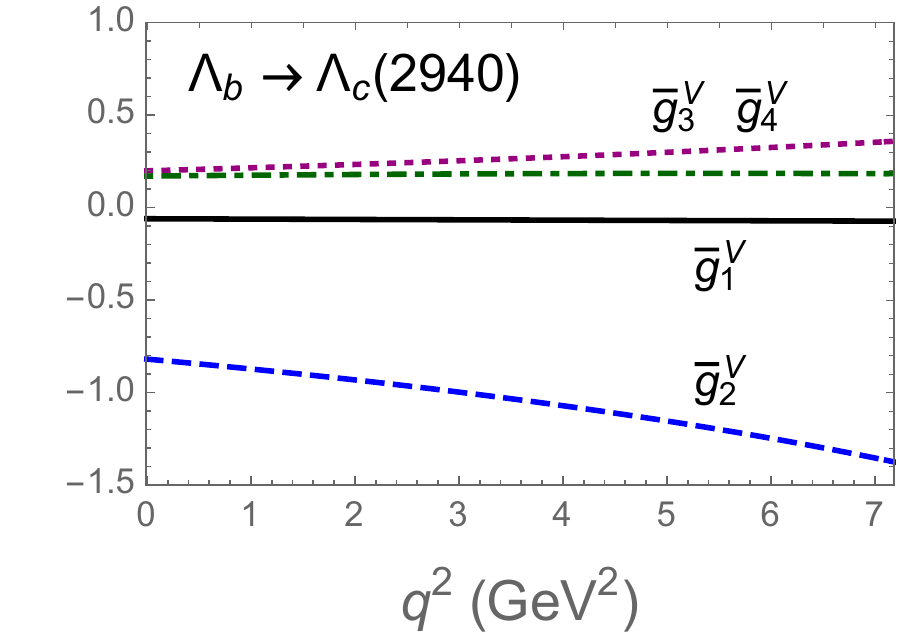}
}
\caption{Form factors $f^A_{i}(q^2)$ and $g^A_{i}(q^2)$ for
$\Lambda_b\to\Lambda_c(2625)$, $\Lambda_c(2940)$ and $\Xi_b\to\Xi_c(2815)$
transitions. The transitions are $\B_b({\bf\bar 3_f},1/2^+)\to\B_c({\bf\bar 3_f},3/2^-)$ transitions [type (vi)].}
\label{fig: figi type vi}
\end{figure}

The $\B_b({\bf 6_f},1/2^+)\to\B_c({\bf 6_f},3/2^+)$ transition form factors are related to the 
$\B_b({\bf 6_f},1/2^+)\to\B_c({\bf 6_f},1/2^+)$ ones in the heavy quark limit, see Eqs. (\ref{eq: HQ type ii}) and (\ref{eq: HQ type iii}). They are governed by the same set of Isgur-Wise functions, namely $\xi_{1,2}$.
In Table~\ref{tab: figi type iii and iv} and Fig.~\ref{fig: figi type iii and iv}, we show the form factors of $\B_b({\bf 6_f},1/2^+)\to\B_c({\bf 6_f},3/2^+)$ transitions. These involve $\Omega_b\to\Omega_c(2770)$ and $\Omega_c(3120)$ transitions, where $\Omega_c(2770)$ is a low lying state, while $\Omega_c(3120)$ a radial excited state. 
In the $\Omega_b\to\Omega_c(2770)$ and $\Omega_b\to \Omega_c(3120)$ transitions, as shown in the table and the figure, 
we have 
$\bar f_1^V(q^2)\simeq -\bar g^A_1(q^2)$, 
$\bar f_3^V(q^2)\simeq -\bar g^A_3(q^2)\simeq -\bar f^V_2(q^2)$ and 
$\bar f_4(q^2),\bar g^A_{2,4}(q^2)$ much smaller than other form factors,
in accordance with the HQS relations using large $N_c$ QCD, see Eqs. (\ref{eq: HQ type iii}) and (\ref{eq: large Nc}). 
The agreement is better in the $\Omega_b\to\Omega_c(2770)$ transition than in the $\Omega_b\to \Omega_c(3120)$ transition.
Heavy quark symmetry and large $N_c$ QCD predict the form factors in the $\Omega_b\to\Omega_c(2770)$ transition have the following values at $q^2_{max}$:
$(\bar f^V_1, \bar f^V_2, \bar f^V_3, \bar f^V_4)=(-1.15,-0.58,0.58,0)$ and 
$(\bar g^A_1, \bar g^A_2, \bar g^A_3, \bar g^A_4)=(1.15,0,-0.58,0)$, see Eq.~(\ref{eq: HQ type iii w=1}).
From Table~\ref{tab: figi type iii and iv}, we see that $\bar f^V_{1,2,3}$ agree with the above predictions within $10\%$, 
$\bar g^A_{1,3}$ within $20\%$ and $45\%$ and $\bar f^V_4, \bar g^A_{4}$ are close to zero, the predicted values, while $\bar g^A_2$ is much smaller than $|\bar g^A_{1,3}|$ and closer to $\bar g^A_4$.
As the overlapping integral is 0.56, the agreements in $\bar f^V_i$ and $\bar g^A_i$ are better than expected.  
In the $\Omega_b\to\Omega_c(3120)$ transition, most of the form factors are smaller in sizes and opposite in signs compared to the corresponding form factors in the previous transition. 
This is understandable as $\Omega_c(3120)$ is a radial excited state, the corresponding overlapping integral is $-0.51$, which has opposite sign compare to the previous one.

In Table~\ref{tab: figi type iii and iv} and Fig.~\ref{fig: figi type iii and iv} we also show the form factors of the $\B_b({\bf 6_f},1/2^+)\to\B_c({\bf 6_f},3/2^-)$ transition, namely the $\Omega_b\to\Omega_c(3050)$ transition, where $\Omega_c(3050)$ is a $p$-wave state.
The signs of $\bar f_{1,2}^V$ and $\bar g^A_{2,3,4}$ at $q^2_{max}$ are in agreement with the HQS predictions, Eq. (\ref{eq: HQ type iv w=1}).
If we na\"ively make use of Eq.~(\ref{eq: HQ type iv w=1}), we obtain $\xi_5(1)\simeq 1.4^{+0.2}_{-0.7}$ and $\xi_6(1)\simeq 0.7$ from these form factors, except from $\bar g^A_2$ we have $\xi_5(1)\simeq 3.0$.

The $\B_b({\bf\bar 3_f},1/2^+)\to\B_c({\bf \bar 3_f},1/2^-)$ transition form factors $f^A_{1,2,3}(q^2)$ and $g^V_{1,2,3}(q^2)$
for $\Lambda_b\to\Lambda_c(2595)$, $\Lambda_c(2940)$ and $\Xi_b\to\Xi_c(2790)$ transitions are given in Table~\ref{tab:fg type v} and they are plotted in Fig.~\ref{fig: figi type v}.
We consider $\Lambda_c(2940)$ as a spin-1/2 $p$-wave radial excited state~\cite{Cheng:2017ove} in this case.
HQS requires $f_1^A(q^2)=g_1^V(q^2)$, $f_2^A(q^2)=f_3^A(q^2)$ and $g^V_2(q^2)=g^V_3(q^2)$, see Eq.~(\ref{eq: HQ type v}).
From the table and the figures, we see that the form factors roughly exhibit this pattern.
By na\"ively compared to the HQ relations at $q^2_{max}$, Eq. (\ref{eq: HQ type v w=1}), we obtain $\sigma(1)\simeq 0.7^{+0.5}_{-0.3}$ in the $\Lambda_b\to\Lambda_c(2595)$ transition and $\sigma(1)\simeq 0.7^{+0.4}_{-0.3}$ in the $\Xi_b\to\Xi_c(2790)$ transition, while for the transition involving radial excite state,  $\Lambda_b\to\Lambda_c(2940)$ transition, we have $\sigma^{(*)}(1)\simeq -0.8^{+0.3}_{-0.4}$. The sign flip in $\Lambda_b\to\Lambda_c(2940)$ transition can again be traced to the overlapping integral, whose value is $-0.53$, while the one in the $\Lambda_b\to\Lambda_c(2595)$ transition is 0.67. 
Roughly speaking these form factors exhibit the pattern predicted by HQS.

We now turn to the last case, the $\B_b(\bar {\bf 3}_f,1/2^+)\to\B_c(\bar {\bf 3}_f,3/2^-)$ transition, which includes
$\Lambda_b\to \Lambda_c(2625)$, $\Xi_b\to \Xi_c(2815)$ and $\Lambda_b\to \Lambda_c(2940)$ transitions.
We consider $\Lambda_c(2940)$ as a spin-3/2 $p$-wave radial excited state~\cite{Aaij:2017vbw} in this case.
The corresponding form factors for these transitions are shown in Table~\ref{tab:fg type vi} and are plotted in Fig.~\ref{fig: figi type vi}. Note that for the form factors denoted with asterisks, the parameters $a$ and $b$ are assumed to be one, while the values of the form factors at $q^2=0$ are obtained by using Eq.~(\ref{eq: figi q2=0}). 
These form factors are small and the assumptions on $a$ and $b$ should not have much impact on decay rates.
HQS requires $\bar f_2^A=\bar g_2^V=\sigma(\omega)$, while all other form factors are vanishing, see Eq. (\ref{eq: HQ type vi}).
We see from the table and the plots that these form factors roughly exhibit the pattern predicted by HQS.
By na\"ively compared to the HQ relations at $q^2_{max}$, Eq. (\ref{eq: HQ type vi w=1}), we obtain $\sigma(1)\simeq 0.7, 0.8$ in the $\Lambda_b\to\Lambda_c(2625)$ transition and $\sigma(1)\simeq 1$ in the $\Xi_b\to\Xi_c(2815)$ transition, while for the transition involving radial excite state,  $\Lambda_b\to\Lambda_c(2940)$ transition, we have $\sigma^{(*)}(1)\simeq -1.0, -1.4$.
These $\sigma(1)$ and $\sigma^{(*)}(1)$ are similar to those obtained in $\B_b({\bf\bar 3_f},1/2^+)\to\B_c({\bf \bar 3_f},1/2^-)$ transitions.

\subsection{$\B_b\to \B_c M$ decay rates and up-down asymmetries}

We will present the results of $\B_b\to \B_c M$ decay rates and up-down asymmetries including $1/2\to 1/2$ and $1/2\to 3/2$ transitions 
in this subsection.
The decay rates and asymmetries for $1/2\to 1/2$ transitions are updated from those in \cite{Chua:2018lfa},
as we are using different input parameters and more form factors.~\footnote{Note that errors in the code in \cite{Chua:2018lfa} are also corrected.}
Under the factorization approximation, the decay amplitudes for
color-allowed $\B_b\to\B_c M^-$ decays are given by
\be
{\cal A}(\B_b\to\B_c M^-)=\frac{G_F}{\sqrt2} V_{cb} V^*_{ij} a_1
 \langle M^-(\bar q_i q_j) |V^\mu-A^\mu|0\rangle \langle \B_c |V_\mu-A_\mu|\B_b\rangle,
\label{eq: Amp}
\en
where $V_{cb}$ and $V_{ij}$ are the Cabibbo-Kobayashi-Maskawa matrix elements
and
$a_1$ is the color-allowed effective Wilson coefficient. 
The effective Wilson coefficient $a_1$ in the na\"ive factorization approximation is given by $c_1+c_2/N_c$ with $N_c=3$, $c_1= 1.081$ and $c_2=-0.190$ at the scale of $\mu=4.2$ GeV~\cite{BBNS}.
The matrix element $\langle \B_c |V_\mu-A_\mu|\B_b\rangle$ is given by Eqs.~(\ref{eq:figi spin1/2}), (\ref{eq:figi1 spin1/2}), (\ref{eq:figi spin3/2}) and (\ref{eq:figi1 spin3/2}),
while $\langle M^-(\bar q_i q_j) |V^\mu-A^\mu|0\rangle$ 
are given by
\be
\langle P |V^\mu-A^\mu|0\rangle=i q^\mu f_P,
\quad
\langle V |V^\mu-A^\mu|0\rangle=m_V f_V \varepsilon^*_V,
\quad
\langle A |V^\mu-A^\mu|0\rangle=-m_A f_A \varepsilon^*_A,
\en
with $f_{P,V,A}$ the corresponding meson decay constants.

\begin{table}[t!]
\caption{\label{tab:rate P} Branching ratios (in the unit of $10^{-3}$) of $\B_b\to \B_c P$ decays. 
Note that the asterisks denote the transitions where the final state charmed baryons are radial excited.
}
\begin{ruledtabular}
\begin{tabular}{rcrrrr}
Transition type          
          & Mode
          & $P=\pi^-$ 
          & $P=K^-$ 
          & $P=D^-$ 
          & $P=D_s^-$ 
          \\
\hline  
(i) $(\frac{1}{2}^+\to\frac{1}{2}^+)$
         & $Br(\Lambda_b\to \Lambda_c P)$ 
         & $4.16_{-1.73}^{+2.43}$ 
         & $0.31_{-0.13}^{+0.18}$ 
         & $0.47_{-0.21}^{+0.30} $ 
         & $11.92_{-5.28}^{+7.69} $ 
         \\
(v) $(\frac{1}{2}^+\to\frac{1}{2}^-)$
         & $Br[\Lambda_b\to \Lambda_c(2595) P]$ 
         & $1.09_{-0.51}^{+0.76}$ 
         & $0.08_{-0.04}^{+0.06}$ 
         & $0.07_{-0.04}^{+0.07} $ 
         & $ 1.72_{-1.01}^{+1.71} $ 
         \\         
(vi) $(\frac{1}{2}^+\to\frac{3}{2}^-)$     
          & $Br[\Lambda_b\to \Lambda_c(2625) P]$   
          & $ 2.40_{-1.82}^{+4.09} $ 
          & $ 0.17_{-0.13}^{+0.30} $ 
          & $ 0.13_{-0.10}^{+0.22} $ 
          & $ 2.88_{-2.16}^{+4.92} $
          \\
(i)$^*$ $(\frac{1}{2}^+\to\frac{1}{2}^+)$
          & $Br[\Lambda_b\to \Lambda_c(2765) P]$ 
          & $ 1.70_{-0.52}^{+0.69} $ 
          & $ 0.13_{-0.04}^{+0.05} $ 
          & $ 0.15_{-0.05}^{+0.07} $ 
          & $ 3.54_{-1.24}^{+1.73} $  
          \\         
(v)$^*$ $(\frac{1}{2}^+\to\frac{1}{2}^-)$
          & $Br[\Lambda_b\to \Lambda_c(2940) P]$ 
          & $ 0.68_{-0.21}^{+0.21} $ 
          & $ 0.05_{-0.02}^{+0.02} $ 
          & $ 0.04_{-0.02}^{+0.02} $ 
          & $ 0.87_{-0.38}^{+0.46} $
          \\        
(vi)$^*$ $(\frac{1}{2}^+\to\frac{3}{2}^-)$
          & $Br[\Lambda_b\to \Lambda_c(2940) P]$  
          & $ 1.00_{-0.83}^{+2.00} $ 
          & $ 0.07_{-0.06}^{+0.14} $ 
          & $ 0.07_{-0.05}^{+0.10} $ 
          & $ 1.69_{-1.23}^{+2.30} $
          \\  
(i) $(\frac{1}{2}^+\to\frac{1}{2}^+)$     
         & $Br(\Xi^-_b\to \Xi^0_c P)$ 
         & $ 3.88_{-1.69}^{+2.43} $ 
         & $ 0.29_{-0.13}^{+0.18} $ 
         & $ 0.45_{-0.21}^{+0.31} $ 
         & $ 11.54_{-5.34}^{+7.98} $ 
         \\       
(i) $(\frac{1}{2}^+\to\frac{1}{2}^+)$     
         & $Br(\Xi^0_b\to \Xi^+_c P)$  
         & $ 3.66_{-1.59}^{+2.29} $ 
         & $ 0.28_{-0.12}^{+0.17} $ 
         & $ 0.43_{-0.20}^{+0.29} $ 
         & $ 10.87_{-5.03}^{+7.51} $ 
         \\        
(v) $(\frac{1}{2}^+\to\frac{1}{2}^-)$
          & $Br[\Xi^-_b\to \Xi^0_c(2790) P]$ 
          & $ 1.03_{-0.48}^{+0.79} $ 
          & $ 0.08_{-0.04}^{+0.06} $ 
          & $ 0.07_{-0.04}^{+0.08} $ 
          & $ 1.70_{-0.99}^{+1.88} $
          \\        
(v) $(\frac{1}{2}^+\to\frac{1}{2}^-)$
          & $Br[\Xi^0_b\to \Xi^-_c(2790) P]$ 
          & $ 0.97_{-0.45}^{+0.74} $ 
          & $ 0.07_{-0.03}^{+0.06} $ 
          & $ 0.07_{-0.04}^{+0.07} $ 
          & $ 1.60_{-0.93}^{+1.76} $
          \\         
(vi) $(\frac{1}{2}^+\to\frac{3}{2}^-)$
          & $Br[\Xi^-_b\to \Xi^0_c(2815) P]$ 
          & $ 3.53_{-2.80}^{+6.46} $ 
          & $ 0.26_{-0.20}^{+0.47} $ 
          & $ 0.20_{-0.15}^{+0.35} $ 
          & $ 4.65_{-3.48}^{+8.08} $ 
          \\     
(vi) $(\frac{1}{2}^+\to\frac{3}{2}^-)$
          & $Br[\Xi^0_b\to \Xi^+_c(2815) P]$  
          & $ 3.32_{-2.63}^{+6.08} $ 
          & $ 0.24_{-0.19}^{+0.44} $ 
          & $ 0.19_{-0.14}^{+0.33} $ 
          & $ 4.34_{-3.25}^{+7.54} $
          \\
(ii) $(\frac{1}{2}^+\to\frac{1}{2}^+)$
          & $Br(\Omega_b\to \Omega_c P)$
          & $ 1.10_{-0.55}^{+0.85} $ 
          & $ 0.08_{-0.04}^{+0.07} $ 
          & $ 0.15_{-0.08}^{+0.14} $ 
          & $ 4.03_{-2.21}^{+3.72} $
          \\         
(iii) $(\frac{1}{2}^+\to\frac{3}{2}^+)$
          & $Br[\Omega_b\to \Omega_c(2770) P]$
          & $ 1.37_{-1.19}^{+3.01} $ 
          & $ 0.11_{-0.09}^{+0.23} $ 
          & $ 0.28_{-0.20}^{+0.38} $ 
          & $ 7.46_{-5.04}^{+9.63} $
          \\
(iv) $(\frac{1}{2}^+\to\frac{3}{2}^-)$ 
          & $Br[\Omega_b\to \Omega_c(3050) P]$  
          & $ 3.40_{-2.25}^{+4.45} $ 
          & $ 0.24_{-0.16}^{+0.31} $ 
          & $ 0.09_{-0.07}^{+0.15} $ 
          & $ 1.78_{-1.38}^{+3.16} $
          \\         
(ii)$^*$ $(\frac{1}{2}^+\to\frac{1}{2}^+)$
          & $Br[\Omega_b\to \Omega_c(3090) P]$ 
          & $ 0.85_{-0.35}^{+0.50} $ 
          & $ 0.06_{-0.03}^{+0.04} $ 
          & $ 0.10_{-0.05}^{+0.07} $ 
          & $ 2.43_{-1.15}^{+1.79} $
          \\                   
(iii)$^*$ $(\frac{1}{2}^+\to\frac{3}{2}^+)$
          & $Br[\Omega_b\to \Omega_c(3120) P]$ 
          & $ 0.96_{-0.52}^{+0.95} $ 
          & $ 0.07_{-0.04}^{+0.07} $ 
          & $ 0.10_{-0.05}^{+0.08} $ 
          & $ 2.37_{-1.10}^{+1.81} $  
\end{tabular}
\end{ruledtabular}
\end{table}

\begin{table}[t!]
\caption{\label{tab:rate V A} Same as Table~\ref{tab:rate P} but for $\B_b\to \B_c V$ and $B_b\to \B_c A$ decays. 
}
\begin{ruledtabular}
\begin{tabular}{lcrrrrr}
Type
          & Mode
          & $M=\rho^-$ 
          & $M=K^{*-}$ 
          & $M=D^{*-}$ 
          & $M=D_s^{*-}$ 
          & $M=a^-_1$ 
          \\
\hline   
(i)
          & $Br(\Lambda_b\to \Lambda_c M)$ 
          & $ 12.28_{-5.11}^{+7.19} $ 
          & $ 0.63_{-0.26}^{+0.37} $ 
          & $ 0.84_{-0.36}^{+0.51} $ 
          & $ 17.49_{-7.48}^{+10.60} $ 
          & $ 11.91_{-4.97}^{+6.98}$           
          \\
(v)
          & $Br[\Lambda_b\to \Lambda_c(2595) M]$  
          & $ 2.99_{-1.44}^{+2.20} $ 
          & $ 0.15_{-0.07}^{+0.11} $ 
          & $ 0.12_{-0.07}^{+0.11} $ 
          & $ 2.28_{-1.29}^{+2.21} $ 
          & $ 2.57_{-1.29}^{+2.01} $
          \\
(vi)
          & $Br[\Lambda_b\to \Lambda_c(2625) M]$  
          & $ 4.38_{-3.17}^{+6.78} $ 
          & $ 0.22_{-0.16}^{+0.33} $ 
          & $ 0.13_{-0.08}^{+0.17} $ 
          & $ 2.41_{-1.52}^{+2.98} $ 
          & $ 3.50_{-2.45}^{+5.11} $
          \\
(i)$^*$
          & $Br[\Lambda_b\to \Lambda_c(2765) M]$ 
          & $ 4.84_{-1.50}^{+2.01} $ 
          & $ 0.25_{-0.08}^{+0.10} $ 
          & $ 0.26_{-0.09}^{+0.12} $ 
          & $ 5.29_{-1.84}^{+2.54} $ 
          & $ 4.45_{-1.42}^{+1.91} $
          \\
(v)$^*$
          & $Br[\Lambda_b\to \Lambda_c(2940) M]$ 
          & $ 1.85_{-0.60}^{+0.63} $ 
          & $ 0.09_{-0.03}^{+0.03} $ 
          & $ 0.06_{-0.03}^{+0.03} $ 
          & $ 1.16_{-0.48}^{+0.62} $ 
          & $ 1.57_{-0.54}^{+0.59} $
          \\
(vi)$^*$
          & $Br[\Lambda_b\to \Lambda_c(2940) M]$  
          & $ 1.93_{-1.43}^{+3.19} $ 
          & $ 0.10_{-0.07}^{+0.15} $ 
          & $ 0.06_{-0.03}^{+0.06} $ 
          & $ 1.11_{-0.62}^{+1.07} $ 
          & $ 1.58_{-1.08}^{+2.26} $ 
          \\ 
(i)
          & $Br(\Xi^-_b\to \Xi^0_c M)$  
          & $ 11.56_{-5.04}^{+7.25} $ 
          & $ 0.60_{-0.26}^{+0.37} $ 
          & $ 0.82_{-0.37}^{+0.53} $ 
          & $ 17.26_{-7.70}^{+11.2} $ 
          & $ 11.37_{-4.97}^{+7.14} $
          \\
(i)
          & $Br(\Xi^0_b\to \Xi^+_c M)$ 
          & $ 10.88_{-4.74}^{+6.83} $ 
          & $ 0.56_{-0.24}^{+0.35} $ 
          & $ 0.77_{-0.35}^{+0.50} $ 
          & $ 16.24_{-7.25}^{+10.54} $ 
          & $ 10.70_{-4.67}^{+6.72}$ 
          \\
(v)
          & $Br[\Xi^-_b\to \Xi^0_c(2790) M]$  
          & $ 2.86_{-1.36}^{+2.28} $ 
          & $ 0.14_{-0.07}^{+0.12} $ 
          & $ 0.12_{-0.06}^{+0.12} $ 
          & $ 2.25_{-1.26}^{+2.33} $ 
          & $ 2.48_{-1.23}^{+2.10} $
          \\
(v)
          & $Br[\Xi^0_b\to \Xi^+_c(2790) M]$  
          & $ 2.69_{-1.28}^{+2.15} $ 
          & $ 0.13_{-0.06}^{+0.11} $ 
          & $ 0.11_{-0.06}^{+0.11} $ 
          & $ 2.11_{-1.19}^{+2.19} $ 
          & $ 2.33_{-1.16}^{+1.98}$
          \\
(vi)
          & $Br[\Xi^-_b\to \Xi^0_c(2815) M]$   
          & $ 6.49_{-4.84}^{+10.58} $ 
          & $ 0.32_{-0.24}^{+0.51} $ 
          & $ 0.20_{-0.13}^{+0.26} $ 
          & $ 3.74_{-2.32}^{+4.58} $ 
          & $ 5.24_{-3.72}^{+7.92}$
          \\
(vi)
          & $Br[\Xi^0_b\to \Xi^+_c(2815) M]$  
          & $ 6.10_{-4.55}^{+9.95} $ 
          & $ 0.30_{-0.22}^{+0.48} $ 
          & $ 0.19_{-0.12}^{+0.24} $ 
          & $ 3.51_{-2.18}^{+4.30} $ 
          & $ 4.92_{-3.50}^{+7.45} $
          \\  
(ii)
          & $Br(\Omega_b\to \Omega_c M)$ 
          & $ 3.07_{-1.53}^{+2.41} $ 
          & $ 0.16_{-0.08}^{+0.12} $ 
          & $ 0.16_{-0.08}^{+0.13} $ 
          & $ 3.18_{-1.61}^{+2.69} $ 
          & $ 2.76_{-1.37}^{+2.20} $
          \\
(iii)
          & $Br[\Omega_b\to \Omega_c(2770) M]$ 
          & $ 2.37_{-1.85}^{+4.68} $ 
          & $ 0.13_{-0.10}^{+0.24} $ 
          & $ 0.28_{-0.16}^{+0.30} $ 
          & $ 6.20_{-3.49}^{+6.19} $ 
          & $ 2.78_{-1.93}^{+4.35} $
          \\
(iv)
          & $Br[\Omega_b\to \Omega_c(3050) M]$  
          & $ 4.09_{-2.71}^{+5.62} $ 
          & $ 0.20_{-0.13}^{+0.27} $ 
          & $ 0.08_{-0.05}^{+0.10} $ 
          & $ 1.43_{-0.88}^{+1.78} $ 
          & $ 2.84_{-1.86}^{+3.88} $
          \\         
(ii)$^*$
          & $Br[\Omega_b\to \Omega_c(3090) M]$  
          & $ 2.29_{-0.94}^{+1.36} $ 
          & $ 0.11_{-0.05}^{+0.07} $ 
          & $ 0.09_{-0.04}^{+0.06} $ 
          & $ 1.69_{-0.71}^{+1.06} $ 
          & $ 1.92_{-0.79}^{+1.15} $
          \\
(iii)$^*$
          & $Br[\Omega_b\to \Omega_c(3120) M]$ 
          & $ 1.50_{-0.76}^{+1.37} $ 
          & $ 0.08_{-0.04}^{+0.07} $ 
          & $ 0.11_{-0.04}^{+0.07} $ 
          & $ 2.37_{-0.88}^{+1.33} $ 
          & $ 1.55_{-0.71}^{+1.21} $
          \\
\end{tabular}
\end{ruledtabular}
\end{table}

\begin{table}[t!]
\caption{\label{tab:rate M compare} Comparisons of data and theoretical results on the branching ratios (in the unit of $10^{-3}$) of $\Lambda_b\to \Lambda_c M$, $\Xi_b\to\Xi_c M$, $\Omega_b\to\Omega_c M$ and $\Omega_b\to\Omega_c(2770) M$ decays. 
}
\begin{ruledtabular}
\begin{tabular}{lccccccccccc}
Mode          
          & Data~\cite{PDG}
          & This work 
          & \cite{Mannel:1992ti} 
          & \cite{Cheng:1996cs}
          & \cite{Ivanov:1997ra,Ivanov:1997hi} 
          & \cite{Giri:1997te}
          & \cite{Fayyazuddin:1998ap}
          & \cite{Mohanta:1998iu}
          & \cite{Zhu:2018jet}
          & \cite{Gutsche:2018utw}
          & \cite{Ke:2019smy}
          \\
\hline  
$\Lambda_b\to \Lambda_c \pi^-$
          & $4.9\pm 0.4$
          & $4.16_{-1.73}^{+2.43}$   
          & $4.6^{+2.0}_{-3.1}$
          & $4.6$
          & $5.62$
          & 3.91 
          & $-$
          & $1.75$
          & $4.96$
          & $-$
          & 5.67
          \\    
$\Lambda_b\to \Lambda_c K^-$      
          & $0.359\pm0.030$
          & $0.31_{-0.13}^{+0.18}$  
          & $-$
          & $-$
          & $-$
          & $-$
          & $-$
          & $0.13$
          & $0.393$
          & $-$
          & 0.46
          \\
$\Lambda_b\to \Lambda_c D^-$
          & $0.46\pm0.06$
          & $0.47_{-0.21}^{+0.30} $ 
          & $-$
          & $-$
          & $-$
          & $-$
          & $-$
          & $0.30$
          & $0.522$
          & $-$
          & 0.76
          \\          
$\Lambda_b\to \Lambda_c D_s^-$
          & $11.0\pm1.0$ 
          & $11.92_{-5.28}^{+7.69} $  
          & $23^{+3}_{-4}$
          & $13.7$
          & $-$
          & $12.91$
          & $22.3$
          & 7.70
          & $12.4$
          & 14.78  
          & 19.94        
          \\  
$\Lambda_b\to \Lambda_c \rho^-$
          & $-$ 
          & $ 12.28_{-5.11}^{+7.19} $   
          & $6.6^{+2.4}_{-4.0}$
          & $12.9$
          & $-$
          & $10.82$
          & $-$
          & $4.91$
          & $8.65$
          & $-$  
          & 16.71        
          \\  
$\Lambda_b\to \Lambda_c K^{*-}$
          & $-$ 
          & $ 0.63_{-0.26}^{+0.37} $     
          & $-$
          & $-$
          & $-$
          & $-$
          & $-$
          & $0.27$
          & $0.441$
          & $-$
          & 0.87
          \\                      
$\Lambda_b\to \Lambda_c D^{*-}$
          & $-$ 
          & $ 0.84_{-0.36}^{+0.51} $ 
          & $-$
          & $-$
          & $-$
          & $-$
          & $-$
          & $0.49$
          & $0.520$
          & $-$
          & 1.38
          \\ 
$\Lambda_b\to \Lambda_c D_s^{*-}$
          & $-$ 
          & $ 17.49_{-7.48}^{+10.60} $ 
          & $17.3^{+2.0}_{-3.0}$
          & $21.8$
          & $-$
          & $19.83$
          & $32.6$
          & 14.14
          & $10.5$
          & $25.16$
          & 30.86
          \\     
$\Lambda_b\to \Lambda_c a_1^-$
          & $-$ 
          & $ 11.91_{-4.97}^{+6.98}$  
          & $-$
          & $-$
          & $-$
          & $-$
          & $-$
          & $5.32$
          & $-$
          & $-$
          & 16.53
          \\
$\Xi^0_b\to \Xi^+_c \pi^-$ 
          & $-$ 
          & $ 3.66_{-1.59}^{+2.29} $
          & $-$
          & $4.9$
          & $7.08$
          & $-$
          & $-$          
          & $-$
          & $-$
          & $-$
          & $-$
          \\    
$\Xi^-_b\to \Xi^0_c \pi^-$ 
          & $-$ 
          & $ 3.88_{-1.69}^{+2.43} $ 
          & $-$
          & $5.2$
          & 10.13
          & $-$
          & $-$
          & $-$
          & $-$
          & $-$
          & $-$
          \\  
$\Xi^0_b\to \Xi^+_c D^-$
          & $-$ 
          & $ 0.43_{-0.20}^{+0.29} $  
          & $-$
          & $-$
          & $-$
          & $-$
          & $-$
          & $-$
          & $-$
          & $0.45$
          & $-$
          \\ 
$\Xi^0_b\to \Xi^+_c D_s^-$
          & $-$ 
          & $ 10.87_{-5.03}^{+7.51} $ 
          & $-$
          & $14.6$
          & $-$
          & $-$
          & $-$
          & $-$
          & $-$
          & $-$
          & $-$
          \\           
$\Xi^0_b\to \Xi^+_c D^{*-}$
          & $-$ 
          & $ 0.77_{-0.35}^{+0.50} $ 
          & $-$
          & $-$
          & $-$
          & $-$
          & $-$
          & $-$
          & $-$
          & $0.95$
          & $-$
          \\  
$\Xi^0_b\to \Xi^+_c D_s^{*-}$
          & $-$ 
          & $ 16.24_{-7.25}^{+10.54} $ 
          & $-$
          & $23.1$
          & $-$
          & $-$
          & $-$
          & $-$
          & $-$
          & $-$
          & $-$
          \\                                         
$\Omega_b\to \Omega_c \pi^-$ 
          & $-$ 
          & $ 1.10_{-0.55}^{+0.85} $ 
          & $-$
          & $4.92$
          & 5.81
          & $-$
          & $-$
          & $-$
          & $-$
          & $1.88$
          & $-$
          \\ 
$\Omega_b\to \Omega_c D_s^-$ 
          & $-$ 
          & $ 4.03_{-2.21}^{+3.72} $ 
          & $-$
          & $17.9$
          & $-$
          & $-$
          & $-$
          & $-$
          & $-$
          & $-$
          & $-$
          \\           
$\Omega_b\to \Omega_c \rho^-$ 
          & $-$ 
          & $ 3.07_{-1.53}^{+2.41} $ 
          & $-$
          & $12.8$
          & $-$
          & $-$
          & $-$
          & $-$
          & $-$
          & $5.43$
          & $-$
          \\
$\Omega_b\to \Omega_c D_s^{*-}$ 
          & $-$ 
          & $ 3.18_{-1.61}^{+2.69} $ 
          & $-$
          & $11.5$
          & $-$
          & $-$
          & $-$
          & $-$
          & $-$
          & $-$
          & $-$
          \\    
$\Omega_b\to \Omega^*_c \pi^-$ 
          & $-$ 
          & $ 1.37_{-1.19}^{+3.01} $
          & $-$
          & $2.69$
          & $-$
          & $-$
          & $-$
          & $-$
          & $-$
          & $1.70$
          & $-$
          \\    
$\Omega_b\to \Omega^*_c D^-$ 
          & $-$ 
          & $ 0.28_{-0.20}^{+0.38} $ 
          & $-$
          & $-$
          & $-$
          & $-$
          & $-$
          & $-$
          & $-$
          & $0.16$
          & $-$
          \\  
$\Omega_b\to \Omega^*_c D_s^-$ 
          & $-$ 
          & $ 7.46_{-5.04}^{+9.63} $ 
          & $-$
          & $3.53$
          & $-$
          & $-$
          & $-$
          & $-$
          & $-$
          & $-$
          & $-$
          \\ 
$\Omega_b\to \Omega^*_c \rho^-$ 
          & $-$ 
          & $ 2.37_{-1.85}^{+4.68} $ 
          & $-$
          & $3.81$
          & $-$
          & $-$
          & $-$
          & $-$
          & $-$
          & $5.58$
          & $-$
          \\ 
$\Omega_b\to \Omega^*_c D^{*-}$ 
          & $-$ 
          & $ 0.28_{-0.16}^{+0.30} $ 
          & $-$
          & $-$
          & $-$
          & $-$
          & $-$
          & $-$
          & $-$
          & $0.58$
          & $-$
          \\   
$\Omega_b\to \Omega^*_c D_s^{*-}$ 
          & $-$ 
          & $ 6.20_{-3.49}^{+6.19} $ 
          & $-$
          & $3.93$
          & $-$
          & $-$
          & $-$
          & $-$
          & $-$
          & $-$
          & $-$
          \\                                                                                                                               
\end{tabular}
\end{ruledtabular}
\end{table}

In type (i) and (ii) transitions 
[$\B_b({\bf\bar 3_f},1/2^+)\to \B_c({\bf\bar 3_f},1/2^+)$ and $\B_b({\bf 6_f},1/2^+)\to \B_c({\bf 6_f},1/2^+)$ transitions],  the decay amplitudes are given by~\cite{Cheng:1996cs}
 \be
 {\cal A}[\B_b\to\B_c(1/2) P]&=&i\bar u'(A+B\gamma_5) u,
 \non\\
  {\cal A}[\B_b\to\B_c(1/2) V]&=&\bar u'\varepsilon^{*\mu}(A_1\gamma_\mu\gamma_5+A_2 P'_\mu\gamma_5
 +B_1\gamma_\mu+B_2 P'_\mu) u,
 \non\\
 {\cal A}[\B_b\to\B_c A]&=&\bar u'\varepsilon^{*\mu}(A'_1\gamma_\mu\gamma_5+A'_2 P'_\mu\gamma_5
 +B'_1\gamma_\mu+B'_2 P'_\mu) u,
 \en
with
\be
A&=&\frac{G_f}{\sqrt2} V_{cb} V^*_{q_1q_2}\, a_1 f_P(M-M') \left( f^V_1(m_P^2)+\frac{m_P^2}{M^2-M^{\prime 2}} f^V_3(m_P^2)\right),
\non\\
B&=&\frac{G_f}{\sqrt2} V_{cb} V^*_{q_1 q_2}\, a_1 f_P (M+M') \left( g^A_1(m_P^2)+\frac{m_P^2}{M^2-M^{\prime 2}} g^A_3(m_P^2)\right),
\non\\
A_1&=&-\frac{G_f}{\sqrt2} V_{cb} V^*_{q_1q_2}\, a_1 f_V m_V
 \left[g^A_1(m_V^2)+g^A_2(m_V^2)\right], 
\non\\
A_2&=&-2\frac{G_f}{\sqrt2} V_{cb} V^*_{q_1q_2}\, a_1 f_V m_V
\frac{g^A_2(m_V^2)}{M-M'},
\non\\
B_1&=&\frac{G_f}{\sqrt2} V_{cb} V^*_{q_1q_2}\, a_1 f_V m_V
\left[f^V_1(m_V^2)-f^V_2(m_V^2)\right],
\non\\
B_2&=&2\frac{G_f}{\sqrt2} V_{cb} V^*_{q_1q_2}\, a_1 f_V m_V\frac{f^V_2(m_V^2)}{M+M'}, 
\non\\
A'_1&=&\frac{G_f}{\sqrt2} V_{cb} V^*_{q_1q_2}\, a_1 f_A m_A
 \left[g^A_1(m_A^2)+g^A_2(m_A^2)\right], 
\non\\
A'_2&=&2\frac{G_f}{\sqrt2} V_{cb} V^*_{q_1q_2}\, a_1 f_A m_A
\frac{g^A_2(m_V^2)}{M-M'},
\non\\
B'_1&=&-\frac{G_f}{\sqrt2} V_{cb} V^*_{q_1q_2}\, a_1 f_A m_A
\left[f^V_1(m_A^2)-f^V_2(m_A^2)\right],
\non\\
B'_2&=&-2\frac{G_f}{\sqrt2} V_{cb} V^*_{q_1q_2}\, a_1 f_A m_A\frac{f^V_2(m_A^2)}{M+M'}.
\label{eq: AB}
\en

For the type (v) transition [$\B_b({\bf\bar 3_f},1/2^+)\to \B_c({\bf\bar 3_f},1/2^-)$ transition], one simply replaces $f_i^V$ and $g_i^A$ in the above equations by $-f^A_i$ and $-g^V_i$, respectively.

In $\B_b({\bf 6_f},1/2^+)\to \B_c({\bf 6_f},3/2^+)$ transitions [type  (iii) transitions], 
the $\B_b\to\B_c P$ and $\B_b\to\B_c V(A)$ decay amplitudes are given by~\cite{Cheng:1996cs}
 \be
 {\cal A}[\B_b\to\B_c(3/2) P]&=&i q_\mu\bar u^{\prime \mu}(P')(C+D\gamma_5) u(P),
 \non\\
  {\cal A}[\B_b\to\B_c(3/2) V]&=&\varepsilon^{*\mu} \bar u^{\prime \nu}(P')
  [g_{\nu\mu}(C_1+C_2 \gamma_5)+q_\nu\gamma_\mu(C_2+D_2\gamma_5)  
   +q_\nu P_\mu (C_3+D_3\gamma_5)]  u(P),
 \non\\
 {\cal A}[\B_b\to\B_c(3/2) A]&=&\varepsilon^{*\mu} \bar u^{\prime \nu}(P')
  [g_{\nu\mu}(C'_1+C'_2 \gamma_5)+q_\nu\gamma_\mu(C'_2+D'_2\gamma_5)  
   +q_\nu P_\mu (C'_3+D'_3\gamma_5)]  u(P),
   \non\\
 \en
with
\be
C&=&-\frac{G_f}{\sqrt2} V_{cb} V^*_{q_1q_2}\, a_1 f_P 
\bigg[g^A_1(m_P^2)+(M-M')\frac{g^A_2(m_P^2)}{M}
\non\\
&&+\frac{1}{2}(M^2-M^{\prime 2}-m^2_P)\bigg(\frac{g^A_3(m^2_P)}{MM'}+ \frac{g^A_4(m^2_P)}{M^2}\bigg)
-m^2_P\frac{g^A_3(m^2_P)}{MM'}\bigg],
\non\\
D&=&\frac{G_f}{\sqrt2} V_{cb} V^*_{q_1 q_2}\, a_1 f_P 
\bigg[f^V_1(m_P^2)-(M+M')\frac{f^V_2(m_P^2)}{M}
\non\\
&&+\frac{1}{2}(M^2-M^{\prime 2}-m^2_P)\bigg(\frac{f^V_3(m^2_P)}{MM'}+ \frac{f^V_4(m^2_P)}{M^2}\bigg)
-m^2_P \frac{f^V_3(m^2_P)}{MM'}\bigg],
\non\\
C^{(\prime)}_1&=&\mp\frac{G_f}{\sqrt2} V_{cb} V^*_{q_1q_2}\, a_1 m_{V(A)} f_{V(A)}  \bar g^A_1(m_{V(A)}^2),
\qquad
D^{(\prime)}_1=\pm\frac{G_f}{\sqrt2} V_{cb} V^*_{q_1q_2}\, a_1 m_{V(A)} f_{V(A)}  \bar f^V_1(m_{V(A)}^2),
\non\\
C^{(\prime)}_2&=&\mp\frac{G_f}{\sqrt2} V_{cb} V^*_{q_1q_2}\, a_1 m_{V(A)} f_{V(A)}  \frac{\bar g^A_2(m_{V(A)}^2)}{M},
\qquad
D^{(\prime)}_2=\pm\frac{G_f}{\sqrt2} V_{cb} V^*_{q_1q_2}\, a_1 m_{V(A)} f_{V(A)}  \frac{\bar f^V_2(m_{V(A)}^2)}{M},
\non\\
C^{(\prime)}_3&=&\mp\frac{G_f}{\sqrt2} V_{cb} V^*_{q_1q_2}\, a_1 m_{V(A)} f_{V(A)} 
\bigg(\frac{\bar g^A_3(m^2_{V(A)})}{MM'}+ \frac{\bar g^A_4(m^2_{V(A)})}{M^2}\bigg),
\non\\
D^{(\prime)}_3&=&\pm\frac{G_f}{\sqrt2} V_{cb} V^*_{q_1q_2}\, a_1 m_{V(A)} f_{V(A)} 
\bigg(\frac{\bar f^V_3(m^2_{V(A)})}{MM'}+ \frac{\bar f^V_4(m^2_{V(A)})}{M^2}\bigg).
\label{eq: CD}
\en
For the $\B_b({\bf 6_f},1/2^+)\to \B_c({\bf 6_f},3/2^-)$ and $\B_b({\bf\bar 3_f},1/2^+)\to \B_c({\bf\bar 3_f},3/2^-)$ transitions 
[type (iv) and (vi) transitions], one simply replaces $\bar f_i^V$ and $\bar g_i^A$ in the above equations by $-\bar f^A_i$ and $-\bar g^V_i$, respectively.
The formulas of decay rates and the up-down asymmetries are collected in Appendix~\ref{appendix: kinematics}.

\begin{table}[t!]
\caption{\label{tab:alpha P} The predicted up-down asymmetries (in the unit of $\%$) of $\B_b\to \B_c P$ decays. 
Note that the asterisks denote the transitions where the final state charmed baryons are radial excited.
}
\begin{ruledtabular}
\begin{tabular}{rcrrrr}
Type
          & Mode
          & $P=\pi^-$ 
          & $P=K^-$ 
          & $P=D^-$ 
          & $P=D_s^-$ 
          \\
\hline  
(i) $(\frac{1}{2}^+\to\frac{1}{2}^+)$
          & $\alpha(\Lambda_b\to \Lambda_c P)$  
          & $ -99.99_{-0.00}^{+4.70} $ 
          & $ -99.97_{-0.01}^{+5.02} $ 
          & $-99.45_{-0.55}^{+7.94} $ 
          & $-99.19_{-0.81}^{+8.59} $
          \\
(v) $(\frac{1}{2}^+\to\frac{1}{2}^-)$
         & $\alpha[\Lambda_b\to \Lambda_c(2595) P]$ 
         & $ -98.33_{-1.67}^{+12.89} $ 
         & $ -98.12_{-1.88}^{+13.51} $ 
         & $ -88.05_{-11.95}^{+26.57} $ 
         & $-86.49_{-13.51}^{+27.83} $
         \\
(vi) $(\frac{1}{2}^+\to\frac{3}{2}^-)$     
          & $\alpha[\Lambda_b\to \Lambda_c(2625) P]$ 
          & $ -97.76_{-2.24}^{+39.39} $ 
          & $ -97.64_{-2.36}^{+39.37} $ 
          & $-97.44_{-2.56}^{+37.86} $ 
          & $-97.07_{-2.93}^{+38.04} $ 
          \\
(i)$^*$ $(\frac{1}{2}^+\to\frac{1}{2}^+)$
          & $\alpha[\Lambda_b\to \Lambda_c(2765) P]$  
          & $ -99.93_{-0.02}^{+1.65} $ 
          & $ -99.87_{-0.11}^{+1.87} $ 
          & $-98.03_{-1.97}^{+5.04} $ 
          & $-97.23_{-2.70}^{+5.70} $ 
          \\
(v)$^*$ $(\frac{1}{2}^+\to\frac{1}{2}^-)$
          & $\alpha[\Lambda_b\to \Lambda_c(2940) P]$ 
          & $ -98.32_{-1.47}^{+2.86} $ 
          & $ -98.10_{-1.64}^{+3.11} $ 
          & $ -86.24_{-9.26}^{+11.11} $ 
          & $-84.04_{-10.66}^{+12.19} $
          \\
(vi)$^*$ $(\frac{1}{2}^+\to\frac{3}{2}^-)$
          & $\alpha[\Lambda_b\to \Lambda_c(2940) P]$ 
          & $ -99.41_{-0.59}^{+65.88} $ 
          & $ -99.06_{-0.94}^{+61.14} $ 
          & $-89.25_{-10.75}^{+31.59} $ 
          & $-86.81_{-13.19}^{+30.72} $
          \\                    
(i) $(\frac{1}{2}^+\to\frac{1}{2}^+)$     
         & $\alpha(\Xi^-_b\to \Xi^0_c P)$  
         & $ -99.98_{-0.00}^{+5.73} $ 
         & $ -99.96_{-0.00}^{+6.10} $ 
         & $-99.29_{-0.71}^{+9.61} $ 
         & $-98.99_{-1.01}^{+10.34} $
         \\
(i) $(\frac{1}{2}^+\to\frac{1}{2}^+)$     
         & $\alpha(\Xi^0_b\to \Xi^+_c P)$ 
         & $ -99.98_{-0.00}^{+5.73} $ 
         & $ -99.96_{-0.00}^{+6.10} $ 
         & $-99.29_{-0.71}^{+9.61} $ 
         & $-98.99_{-1.01}^{+10.34} $
         \\
(v) $(\frac{1}{2}^+\to\frac{1}{2}^-)$
          & $\alpha[\Xi^-_b\to \Xi^0_c(2790) P]$  
          & $ -98.13_{-1.87}^{+14.56} $ 
          & $ -97.88_{-2.11}^{+15.26} $ 
          & $-86.62_{-13.38}^{+28.66} $ 
          & $-84.85_{-15.15}^{+29.84} $ 
          \\
(v) $(\frac{1}{2}^+\to\frac{1}{2}^-)$
          & $\alpha[\Xi^0_b\to \Xi^-_c(2790) P]$  
          & $ -98.13_{-1.87}^{+14.56} $ 
          & $ -97.88_{-2.11}^{+15.26} $ 
          & $-86.60_{-13.40}^{+28.67} $ 
          & $-84.83_{-15.16}^{+29.85} $
          \\
(vi) $(\frac{1}{2}^+\to\frac{3}{2}^-)$
          & $\alpha[\Xi^-_b\to \Xi^0_c(2815) P]$  
          & $ -97.63_{-2.37}^{+42.32} $ 
          & $ -97.48_{-2.52}^{+42.09} $ 
          & $-96.48_{-3.52}^{+38.46} $ 
          & $-95.89_{-4.11}^{+38.40} $
          \\
(vi) $(\frac{1}{2}^+\to\frac{3}{2}^-)$
          & $\alpha[\Xi^0_b\to \Xi^+_c(2815) P]$ 
          & $ -97.70_{-2.30}^{+42.27} $ 
          & $ -97.56_{-2.44}^{+42.03} $ 
          & $ -96.71_{-3.29}^{+38.31} $ 
          & $-96.16_{-3.84}^{+38.26} $
          \\
(ii) $(\frac{1}{2}^+\to\frac{1}{2}^+)$
          & $\alpha(\Omega_b\to \Omega_c P)$ 
          & $ 59.94_{-18.76}^{+21.34} $ 
          & $ 59.39_{-18.70}^{+21.45} $ 
          & $ 56.04_{-19.29}^{+23.79} $ 
          & $ 55.16_{-19.18}^{+23.98} $
          \\
(iii) $(\frac{1}{2}^+\to\frac{3}{2}^+)$
          & $\alpha[\Omega_b\to \Omega_c(2770) P]$ 
          & $ 2.60_{-102.23}^{+97.40} $ 
          & $ 1.17_{-100.15}^{+98.43} $ 
          & $-11.02_{-59.25}^{+55.88} $ 
          & $-11.70_{-55.10}^{+50.63} $
          \\
(iv) $(\frac{1}{2}^+\to\frac{3}{2}^-)$ 
          & $\alpha[\Omega_b\to \Omega_c(3050) P]$ 
          & $ 18.07_{-41.45}^{+52.52} $ 
          & $ 17.73_{-42.50}^{+53.31} $ 
          & $ 9.16_{-71.89}^{+75.06} $ 
          & $ 7.09_{-76.03}^{+78.45} $
          \\
(ii)$^*$ $(\frac{1}{2}^+\to\frac{1}{2}^+)$
          & $\alpha[\Omega_b\to \Omega_c(3090) P]$  
          & $ 59.75_{-13.17}^{+14.13} $ 
          & $ 59.15_{-13.16}^{+14.21} $ 
          & $ 54.01_{-13.95}^{+16.49} $ 
          & $ 52.73_{-13.86}^{+16.61} $
          \\
(iii)$^*$ $(\frac{1}{2}^+\to\frac{3}{2}^+)$
          & $\alpha[\Omega_b\to \Omega_c(3120) P]$ 
          & $ 4.58_{-41.22}^{+42.35} $ 
          & $ 3.81_{-40.26}^{+41.17} $ 
          & $-3.74_{-24.05}^{+22.18} $ 
          & $-4.20_{-22.52}^{+20.50} $ 
          \\ 
\end{tabular}
\end{ruledtabular}
\end{table}

\begin{table}[t!]
\caption{\label{tab:alpha V A} Same as Table~\ref{tab:alpha P} but for $B_b\to \B_c V$ and $B_b\to \B_c A$ decays. 
}
\begin{ruledtabular}
\begin{tabular}{lcrrrrr}
Type
          & Mode
          & $M=\rho^-$ 
          & $M=K^{*-}$ 
          & $M=D^{*-}$ 
          & $M=D_s^{*-}$ 
          & $M=a^-_1$ 
          \\
\hline   
(i)
          & $\alpha(\Lambda_b\to \Lambda_c M)$  
          & $ -86.96_{-0.87}^{+5.60} $ 
          & $ -82.96_{-1.11}^{+6.02} $ 
          & $ -36.85_{-4.88}^{+7.08} $ 
          & $ -32.69_{-5.05}^{+6.82} $
          & $ -70.00_{-2.30}^{+7.00} $ 
          \\
(v)
          & $\alpha[\Lambda_b\to \Lambda_c(2595) M]$ 
          & $ -85.6_{-3.85}^{+12.41} $ 
          & $ -81.65_{-4.46}^{+12.21} $ 
          & $-33.47_{-11.40}^{+15.76} $ 
          & $-28.68_{-11.88}^{+15.97} $
          & $ -68.66_{-6.19}^{+11.32} $
          \\
(vi)
          & $\alpha[\Lambda_b\to \Lambda_c(2625) M]$  
          & $ -91.48_{-3.89}^{+42.04} $ 
          & $ -89.01_{-4.91}^{+41.55} $ 
          & $-53.92_{-18.71}^{+34.18} $ 
          & $-49.80_{-20.08}^{+33.31} $ 
          & $ -80.52_{-8.58}^{+39.81} $
          \\
(i)$^*$
          & $\alpha[\Lambda_b\to \Lambda_c(2765) M]$  
          & $ -86.15_{-1.04}^{+2.71} $ 
          & $ -81.89_{-1.47}^{+3.24} $ 
          & $-31.54_{-4.91}^{+5.31} $ 
          & $-26.87_{-4.82}^{+5.07} $ 
          & $ -67.99_{-2.94}^{+4.60} $
          \\
(v)$^*$
          & $\alpha[\Lambda_b\to \Lambda_c(2940) M]$  
          & $ -84.01_{-3.07}^{+4.44} $ 
          & $ -79.56_{-3.51}^{+4.84} $ 
          & $-24.77_{-9.27}^{+10.24} $ 
          & $-19.04_{-9.71}^{+10.34} $ 
          & $ -64.97_{-4.97}^{+6.29} $
          \\
(vi)$^*$
          & $\alpha[\Lambda_b\to \Lambda_c(2940) M]$ 
          & $ -87.25_{-7.21}^{+66.15} $ 
          & $ -84.14_{-8.45}^{+63.92} $ 
          & $-48.02_{-18.07}^{+36.43} $ 
          & $-45.28_{-18.13}^{+33.21} $
          & $ -73.90_{-12.00}^{+56.57} $
          \\
(i)
          & $\alpha(\Xi^-_b\to \Xi^0_c M)$ 
          & $ -86.61_{-1.09}^{+6.57} $ 
          & $ -82.52_{-1.38}^{+7.04} $ 
          & $-36.00_{-5.51}^{+8.09} $ 
          & $-31.85_{-5.73}^{+7.78} $
          & $ -69.34_{-2.69}^{+8.13} $
          \\
(i)
          & $\alpha(\Xi^0_b\to \Xi^+_c M)$ 
          & $ -86.61_{-1.09}^{+6.57} $ 
          & $ -82.52_{-1.38}^{+7.04} $ 
          & $-35.98_{-5.51}^{+8.08} $ 
          & $-31.84_{-5.72}^{+7.78} $
          & $ -69.33_{-2.69}^{+8.13} $
          \\
(v)
          & $\alpha[\Xi^-_b\to \Xi^0_c(2790) M]$ 
          & $ -85.14_{-4.14}^{+13.71} $ 
          & $ -81.11_{-4.77}^{+13.40} $ 
          & $-32.40_{-11.43}^{+16.14} $ 
          & $-27.57_{-11.88}^{+16.30} $
          & $ -67.92_{-6.55}^{+12.18} $
          \\
(v)
          & $\alpha[\Xi^0_b\to \Xi^+_c(2790) M]$ 
          & $ -85.13_{-4.15}^{+13.71} $ 
          & $ -81.09_{-4.78}^{+13.40} $ 
          & $-32.35_{-11.43}^{+16.14} $ 
          & $-27.52_{-11.88}^{+16.30} $
          & $ -67.90_{-6.55}^{+12.17} $ 
          \\
(vi)
          & $\alpha[\Xi^-_b\to \Xi^0_c(2815) M]$ 
          & $ -91.55_{-4.08}^{+46.65} $ 
          & $ -89.08_{-5.17}^{+45.9} $ 
          & $ -55.07_{-18.5}^{+35.28} $ 
          & $-51.32_{-19.77}^{+34.12} $ 
          & $ -80.63_{-8.94}^{+43.30} $
          \\
(vi)
          & $\alpha[\Xi^0_b\to \Xi^+_c(2815) M]$  
          & $ -91.48_{-4.15}^{+46.84} $ 
          & $ -89.00_{-5.27}^{+46.09} $ 
          & $ -54.83_{-18.69}^{+35.37} $ 
          & $-51.06_{-20.08}^{+34.18} $ 
          & $ -80.52_{-9.07}^{+43.48} $ 
          \\ 
(ii)
          & $\alpha(\Omega_b\to \Omega_c M)$ 
          & $ 61.63_{-19.83}^{+22.03} $ 
          & $ 62.20_{-20.19}^{+22.22} $ 
          & $ 72.15_{-25.51}^{+22.11} $ & $ 73.53_{-26.05}^{+21.54} $
          & $ 64.27_{-21.45}^{+22.74} $ 
          \\
(iii)
          & $\alpha[\Omega_b\to \Omega_c(2770) M]$  
          & $ -5.22_{-88.99}^{+82.03} $ 
          & $ -7.51_{-85.12}^{+78.52} $ 
          & $ -22.24_{-51.42}^{+45.94} $ & $-22.62_{-48.63}^{+42.82} $
          & $ -13.64_{-74.18}^{+68.97} $
          \\
(iv)
          & $\alpha[\Omega_b\to \Omega_c(3050) M]$ 
          & $ 24.95_{-59.79}^{+60.39} $ 
          & $ 24.72_{-60.40}^{+59.86} $ 
          & $ 13.22_{-62.37}^{+49.96} $ & $ 10.39_{-61.11}^{+49.66} $
          & $ 23.53_{-62.28}^{+57.68} $ 
          \\
(ii)$^*$
          & $\alpha[\Omega_b\to \Omega_c(3090) M]$  
          & $ 61.51_{-13.97}^{+14.81} $ 
          & $ 62.10_{-14.24}^{+15.02} $ 
          & $ 73.1_{-17.68}^{+15.51} $ & $ 74.63_{-17.81}^{+14.95} $ 
          & $ 64.31_{-15.17}^{+15.63} $ 
          \\
(iii)$^*$
          & $\alpha[\Omega_b\to \Omega_c(3120) M]$ 
          & $ 1.15_{-48.55}^{+44.49} $ 
          & $ -0.31_{-46.29}^{+41.58} $ 
          & $ -11.06_{-24.74}^{+21.35} $ & $-11.43_{-22.92}^{+19.79} $ 
          & $ -4.44_{-39.57}^{+33.80} $
          \\
\end{tabular}
\end{ruledtabular}
\end{table}

\begin{table}[t!]
\caption{\label{tab:alpha M compare} Various theoretical results on the up-down asymmetries ($\alpha$ in the unit of \%) of $\Lambda_b\to \Lambda_c M$, $\Xi_b\to\Xi_c M$, $\Omega_b\to\Omega_c M$ and $\Omega_b\to\Omega_c(2770) M$ decays are compared. 
}
\begin{ruledtabular}
\begin{tabular}{lccccccccc}
Mode          
          & This work
          & \cite{Mannel:1992ti} 
          & \cite{Cheng:1996cs}
          & \cite{Ivanov:1997ra,Ivanov:1997hi} 
          & \cite{Fayyazuddin:1998ap}
          & \cite{Mohanta:1998iu}
          & \cite{Zhu:2018jet}
          & \cite{Gutsche:2018utw}
          & \cite{Ke:2019smy}
          \\
\hline  
$\Lambda_b\to \Lambda_c \pi^-$
          & $ -99.99_{-0.00}^{+4.70} $    
          & $-100$
          & $-99$
          & $-99$
          & $-$
          & $-99.9$
          & $-99.8$
          & $-$
          & $-100$
          \\    
$\Lambda_b\to \Lambda_c K^-$      
          & $ -99.97_{-0.01}^{+5.02} $
          & $-$
          & $-$
          & $-$
          & $-$
          & $-100$
          & $-100$
          & $-$
          & $-100$
          \\
$\Lambda_b\to \Lambda_c D^-$
          & $-99.45_{-0.55}^{+7.94} $   
          & $-$
          & $-$
          & $-$
          & $-$
          & $-98.7$
          & $-99.9$
          & $-98.9$
          & $-98.3$
          \\          
$\Lambda_b\to \Lambda_c D_s^-$
          & $-99.19_{-0.81}^{+8.59} $ 
          & $-99.1$
          & $-99$
          & $-$
          & $-98$
          & $-98.4$
          & $-100$
          & $-98.6$
          & $-97.8$          
          \\  
$\Lambda_b\to \Lambda_c \rho^-$
          & $ -86.96_{-0.87}^{+5.60} $             
          & $-90.3$
          & $-88$
          & $-$
          & $-$
          & $-89.8$
          & $-88.8$
          & $-$
          & $-87.5$         
          \\  
$\Lambda_b\to \Lambda_c K^{*-}$
          & $ -82.96_{-1.11}^{+6.02} $ 
          & $-$
          & $-$
          & $-$
          & $-$
          & $-86.5$
          & $-85.9$
          & $-$
          & $-83.6$
          \\                      
$\Lambda_b\to \Lambda_c D^{*-}$
          & $-36.85_{-4.88}^{+7.08} $
          & $-$
          & $-$
          & $-$
          & $-$
          & $-45.9$
          & $-47.8$
          & $-$
          & $-37.1$
          \\ 
$\Lambda_b\to \Lambda_c D_s^{*-}$
          & $ -32.69_{-5.05}^{+6.82} $ 
          & $-43.7$
          & $-36$
          & $-$
          & $-40$
          & $-41.9$
          & $-43.9$
          & $-36.4$
          & $-32.7$
          \\     
$\Lambda_b\to \Lambda_c a_1^-$
          & $ -70.00_{-2.30}^{+7.00} $  
          & $-$
          & $-$
          & $-$
          & $-$
          & $-75.8$
          & $-$
          & $-$
          & $-70.9$
          \\
$\Xi^0_b\to \Xi^+_c \pi^-$ 
          & $ -99.98_{-0.00}^{+5.73} $  
          & $-$
          & $-100$
          & $-100$
          & $-$
          & $-$
          & $-$
          & $-$
          & $-$
          \\      
$\Xi^-_b\to \Xi^0_c \pi^-$ 
          & $ -99.98_{-0.00}^{+5.73} $  
          & $-$
          & $-100$
          & $-97$
          & $-$
          & $-$
          & $-$
          & $-$
          & $-$
          \\             
$\Xi^0_b\to \Xi^+_c D_s^-$ 
          & $-98.99_{-1.01}^{+10.34} $  
          & $-$
          & $-99$
          & $-$
          & $-$        
          & $-$
          & $-$
          & $-$
          & $-$
          \\                      
$\Xi^0_b\to \Xi^-_c D_s^{*-}$
          & $ -31.84_{-5.72}^{+7.78} $   
          & $-$
          & $-36$
          & $-$
          & $-$
          & $-$
          & $-$
          & $-$
          & $-$
          \\             
$\Omega_b\to \Omega_c \pi^-$ 
          & $ 59.94_{-18.76}^{+21.34} $   
          & $-$
          & $51$
          & $60$
          & $-$
          & $-$
          & $-$
          & $-$
          & $-$
          \\ 
$\Omega_b\to \Omega_c D_s^-$ 
          & $ 55.16_{-19.18}^{+23.98} $ 
          & $-$
          & $42$
          & $-$
          & $-$
          & $-$
          & $-$
          & $-$
          & $-$
          \\           
$\Omega_b\to \Omega_c \rho^-$ 
          & $ 61.63_{-19.83}^{+22.03} $  
          & $-$
          & $53$
          & $-$
          & $-$
          & $-$
          & $-$
          & $-$
          & $-$
          \\
$\Omega_b\to \Omega_c D_s^{*-}$ 
          & $ 73.53_{-26.05}^{+21.54} $  
          & $-$
          & $64$
          & $-$
          & $-$
          & $-$
          & $-$
          & $-$
          & $-$
          \\ 
$\Omega_b\to \Omega_c(2770) \pi^-$ 
          & $ 2.60_{-102.23}^{+97.40} $  
          & $-$
          & $-38$
          & $-$
          & $-$
          & $-$
          & $-$
          & $-$
          & $-$
          \\    
$\Omega_b\to \Omega_c(2770) D_s^-$ 
          & $-11.70_{-55.10}^{+50.63} $ 
          & $-$
          & $-22$
          & $-$
          & $-$
          & $-$
          & $-$
          & $-$
          & $-$
          \\ 
$\Omega_b\to \Omega_c(2770) \rho^-$ 
          & $ -5.22_{-88.99}^{+82.03} $
          & $-$
          & $-75$
          & $-$
          & $-$
          & $-$
          & $-$
          & $-$
          & $-$
          \\ 
$\Omega_b\to \Omega_c(2770) D_s^{*-}$ 
          & $ -22.62_{-48.63}^{+42.82} $  
          & $-$
          & $-31$
          & $-$
          & $-$
          & $-$
          & $-$
          & $-$
          & $-$
          \\                                                                                                         
\end{tabular}
\end{ruledtabular}
\end{table}

In our numerical study masses and life-times of all hadron are taken from PDG~\cite{PDG},
while the Cabibbo-Kobayashi-Maskawa (CKM) matrix elements are taken from the latest fitting results of the CKM fitter group~\cite{ckmfitter}.
We use \cite{PDG2}
\be
f_\pi=130.2,\, 
f_K=155.6,\,
f_D=211.9,\,
f_{D_s}=249.0,\,  
\en
and \cite{CCH}
\be
f_\rho=216,\, 
f_{K^*}=210,\, 
f_{D^*}=220,\, 
f_{D^*_s}=230,\, 
f_{K^*}=-203,\,
\en
for the values (in unit of MeV) of decay constants of pseudoscalars, vectors and the axial-vector mesons.

The decay rates are calculated using the na\"{i}ve factorization approach.
We assign 10\% uncertainty in the effective Wilson coefficient $a_1$ for estimations and uncertainties in form factors shown in the previous subsection will be used.
Some studies using QCD factorization on $B$ meson and bottom baryon decays indicated that the effective Wilson coefficients $a_1$ in those decays are close to the one in na\"{i}ve factorization~\cite{Beneke:2000ry,Zhu:2018jet}
The authors in ref.~\cite{Beneke:2000ry} obtained 
the $|a_1(DP)|$ agrees with the na\"{i}ve factorization value  
within few \% indicating that for color allowed modes na\"{i}ve factorization is a good approximation.~\footnote{They obtained $|a_1(\bar B\to DP)|=1.055^{+0.019}_{-0.017}-(0.013^{+0.011}_{-0.006})\alpha_1^P$ with $\alpha_1^\pi=0$ and $|\alpha_1^K|<1$. This is to be compared with the na\"{i}ve factorization value, $a^{\rm LO}_1=1.025$.}
A recent study of applying QCD factorization to $\Lambda_b$ decays also shown similar conclusion \cite{Zhu:2018jet}.
However, it has been shown that non-factorizable contributions to $\B_b\to\B_c P$ non-leptonic decay amplitudes can be as large as 30\% of the factorized ones~\cite{Ivanov:1997ra, Ivanov:1997hi}.
Since a precise estimation of non-factorization contributions is beyond the scope of the present work, 
we should stick to the na\"ive factorization approximation.

Note that in the cases of $\B_b\to D_{(s)}^{(*)} \B_c$ decays, penguin terms from $b\to d(s) \bar c c$ decays can contribute to the amplitudes. 
The most dominant penguin contributions to rates are the so-called strong penguin contributions, with effective Wilson coefficients 
$a_4=c_4+c_3/3=-0.03$ and $a_6=c_6+c_5/3=-0.04$ at $\mu=4.2$ GeV~\cite{BBNS}. 
The $a_1 \la M|(V-A)|0\ra\times\la \B_c|(V-A)|\B_b\ra$ term in Eq.~(\ref{eq: Amp}) needs the following additional terms: 
$a_4 \la M|(V-A)|0\ra\times\la \B_c|(V-A)|\B_b\ra$ and 
$a_6 (-2)\la M|(S+P)|0\ra\times\la\B_c|(S-P)|\B_b\ra$.~\footnote{See for example, Eq. (4) in \cite{BBNS} for a similar expression.} 
Note that we have neglected the sub-leading $V_{ub}V_{ud(s)}^*$ terms in the above expression. 
For $M=D^*_{(s)}$, we have $\la M|(S+P)|0\ra=0$ and the resulting $a_6$ contributions are vanishing, 
and the $\B_b\to \B_c D^*_{(s)}$ decay amplitudes can be obtained with the following replacement
\be
a_1\to a_1+a_4,
\en
in the corresponding amplitudes in Eqs. (\ref{eq: AB}) and (\ref{eq: CD}).
The penguin contributions are destructively interfere with the tree contributions in these modes.
For $\B_b\to D_{(s)} \B_c$ decays, through equations of motion, 
we have
\be
-2\la D_{(s)}(q)|(S+P)|0\ra\la |\B_c|(S-P)|\B_b\ra
&=&\frac{2 m^2_{D_{(s)}}}{(m_c+m_{q(s)})(m_b-m_c)}\la D_{(s)}(q)|V-A|0\ra\la\B_c|V|\B_b\ra
\non\\
&&-\frac{2 m^2_{D_{(s)}}}{(m_c+m_{q(s)})(m_b+m_c)}\la D_{(s)}(q)|V-A|0\ra\la\B_c|(-A)|\B_b\ra,
\non\\
\en
where the quark masses are current quark masses with $m_b=4.2$ GeV, $m_c=1.3$ GeV, $m_s=0.08$ GeV and $m_q=0.003$ at $\mu=4.2$ GeV.
Therefore, the $\B_b(1/2^+)\to\B_c(1/2^+,3/2^+) D_{(s)}$ decay amplitudes can be obtained with the following replacements:
\be
a_1&\to& a_1+a_4+\frac{2 m^2_{D_{(s)}}}{(m_c+m_{q(s)})(m_b-m_c)} a_6,
\non\\
a_1&\to& a_1+a_4-\frac{2 m^2_{D_{(s)}}}{(m_c+m_{q(s)})(m_b+m_c)} a_6,
\en
where the first (second) one is applicable for the $A$ ($B$) term
in Eq. (\ref{eq: AB}) and the $D$ $(C)$ term in Eq. (\ref{eq: CD}), while
for the $\B_b(1/2^+)\to\B_c(1/2^-,3/2^-) D_{(s)}$ amplitudes, one needs to switch the above replacements.
We also apply 10\% uncertainties on $a_4$ and $a_6$.

In this work we do not consider the effects of final state interactions (FSI). 
In fact, from the studies of final state interactions in $B$ decays, one will expect the effects of FSI be more prominent in color suppressed modes, where various sub-leading contributions including FSI can compete with each other and be surfaced, see example~\cite{FSI}.
Whereas, for color allowed modes, the leading contributions dominate over sub-leading contributions, and, consequently,
we expect that FSI has little effect on the decay rates and asymmetries of color allowed modes.

The predicted branching ratios for $\B_b\to\B_c P$ and $\B_b\to \B_c V, \B_c A$ decays, 
are summarized in Tables~\ref{tab:rate P} and \ref{tab:rate V A}, respectively
In Tables~\ref{tab:rate M compare}, we compare our results on $\Lambda_b\to \Lambda_c M$, 
$\Xi_b\to\Xi_c M$, $\Omega_b\to\Omega_c M$ and $\Omega_b\to\Omega_c(2770) M$ decay rates to data~\cite{PDG} and the results of other theoretical studies. 
Overall speaking, our results agree reasonably well with data and with most of the results obtained in other works
\cite{Mannel:1992ti,
Cheng:1996cs,
Ivanov:1997ra,Ivanov:1997hi,
Giri:1997te,
Fayyazuddin:1998ap,
Mohanta:1998iu,
Zhu:2018jet,
Gutsche:2018utw,
Ke:2019smy}.
It is interesting that the penguin contributions slightly reduce the $\Lambda_b \to \Lambda_c D$ and $\Lambda D_s$ rates resulting a better agreement with data.
Indeed, when the penguin contributions are included, the central values of the $\Lambda_b \to \Lambda_c D$ and $\Lambda D_s$ branching ratios (in the unit of $10^{-3}$) are reduced from $0.53$ and $13.50$ to $0.47$ and $11.92$, respectively, 
which are closer to the experimental values, $0.46\pm0.06$ and $11.0\pm1.0$, respectively.

In Tables~\ref{tab:alpha P} and \ref{tab:alpha V A}, we show the prediction of up-down asymmetries for $\B_b\to\B_c M$ decays with $M=P,V,A$. 
From the tables we see that most of the signs of $\alpha$ are negative, except those in $\Omega_b\to\Omega_c M$, 
$\Omega_b\to\Omega_c(3050) M$ and $\Omega_b\to\Omega_c(3090) M$ decays.
In Tables~\ref{tab:alpha M compare}, we compare our results on the up-down asymmetries of $\Lambda_b\to \Lambda_c M$, $\Xi_b\to\Xi_c M$, $\Omega_b\to\Omega_c M$ and $\Omega_b\to\Omega_c(2770) M$ decays to other results
\cite{Mannel:1992ti,
Cheng:1996cs,
Ivanov:1997ra,Ivanov:1997hi, 
Fayyazuddin:1998ap,
Mohanta:1998iu,
Zhu:2018jet,
Gutsche:2018utw,
Ke:2019smy}. 
Our results agree well within errors with almost all of the results obtained in other works.

It will be interesting to comparing $\Lambda_b\to\Lambda_c(2940) M$ decays with two different assignments of the configurations of $\Lambda_c(2940)$, a radial excite $p$-wave spin-1/2 or spin-3/2 particle. 
From Tables~\ref{tab:rate P} and \ref{tab:rate V A}, we have
 $Br[\Lambda_b\to\Lambda_c(2940,3/2^-) P]\simeq (1.5\sim2)\times Br[\Lambda_b\to\Lambda_c(2940,1/2^-) P]$,
 $Br[\Lambda_b\to\Lambda_c(2940,3/2^-) V]\simeq Br[\Lambda_b\to\Lambda_c(2940,1/2^-) V]$
 and $Br[\Lambda_b\to\Lambda_c(2940,3/2^-) A]\simeq Br[\Lambda_b\to\Lambda_c(2940,1/2^-) A]$.
The asymmetries in $\Lambda_b\to\Lambda_c(2940,1/2^-) M$ and $\Lambda_b\to\Lambda_c(2940,3/2^-) M$ decays are similar in most cases, 
but have larger deviations in the cases of heavy vector mesons.
In $\Lambda_b\to\Lambda_c(2940) D^{*-}$ and $\Lambda_c(2940) D^{*-}_{s}$ decays, the predictions based on the spin-1/2 configuration, give $-25\%$ and $-19\%$, respectively, while the ones based on the spin-3/2 configurations are $-48\%$ and $-45\%$, respectively. 

It will be useful to understand why the $\Lambda_b\to\Lambda_c(2940,3/2^-) P$ is greater than the $\Lambda_b\to\Lambda_c(2940,1/2^-) P$ rate.
Using Eqs.~(\ref{eq: C1}) and (\ref{eq: C4}), we have
\be
\frac{\Gamma[\Lambda_b\to \Lambda_c(2940,\frac{3}{2}) P]}{\Gamma[\Lambda_b\to \Lambda_c(2940,\frac{1}{2}) P]}
&=&
\frac{2m^2_{\Lambda_b}}{3 m^2_{\Lambda_c(2940)}}
\non\\
&&\times
\frac{[(m_{\Lambda_b}+m_{\Lambda_c(2940)})^2-m_P^2]|p_c C|^2+[(m_{\Lambda_b}-m_{\Lambda_c(2940)})^2-m_P^2]| p_c D|^2}
{[(m_{\Lambda_b}+m_{\Lambda_c(2940)})^2-m_P^2]|A|^2+[(m_{\Lambda_b}-m_{\Lambda_c(2940)})^2-m_P^2]{M^2}|B|^2}.
\non\\
\en
The first factor in the r.h.s. of the above equation is an enhancement factor, while the second factor is expected to be close to unity as the form factors shown in Tables VII and IX for $\Lambda_b\to \Lambda_c(2940,1/2^-)$ and  $\Lambda_b\to \Lambda_c(2940,3/2^-)$ transitions are of similar sizes.
For example, in $\Lambda_b\to \Lambda_c(2940,3/2^-)\pi^-$ and  $\Lambda_b\to \Lambda_c(2940,1/2^-)\pi^-$ decays, 
we have $(A, B)=(-3.09,-8.20)\times 10^{-8}$ and $p_c(C, D)=(-2.33,-6.69)\times 10^{-8}$, 
and the ratio of the decay rates are given by
\be
\frac{\Gamma[\Lambda_b\to \Lambda_c(2940,\frac{3}{2}^-) \pi^-]}{\Gamma[\Lambda_b\to \Lambda_c(2940,\frac{1}{2}^-) \pi^-]}
=
2.44
\times
0.61
=1.49,
\en
and, hence, the $\Lambda_b\to \Lambda_c(2940,3/2^-)\pi^-$ rate is greater than the $\Lambda_b\to \Lambda_c(2940,1/2^-)\pi^-$ rate by about 50\%.
The enhancements in other $\Lambda_b\to \Lambda_c(2940,3/2^-)P$ decay rates can be understood similarly.

The predictions on rates and asymmetries presented in Tables~\ref{tab:rate P}, \ref{tab:rate V A}, \ref{tab:alpha P} and \ref{tab:alpha V A} can be verified experimentally.
These information may shed light on the quantum numbers of $\Lambda_c(2765)$, $\Lambda_c(2940)$
$\Omega_c(3050)$, $\Omega_c(3090)$ and $\Omega_c(3120)$.

\section{Conclusions}

In this work, we study color allowed $\B_b\to\B_c M$ decays with $\B_b=\Lambda_b, \Xi_b,\Omega_b$, $M=\pi, K, \rho, K^*, a_1, D, D_s, D^*, D^*_s$ and $s$-wave and $p$-wave charmed baryons, $\B_c$, including 
$\Lambda^{(*,**)}_c=\Lambda_c, \Lambda_c(2595), \Lambda_c(2625), \Lambda_c(2765), \Lambda_c(2940)$, 
$\Xi_c^{(**)}=\Xi_c, \Xi_c(2815), \Xi_c(2790)$ and 
$\Omega^{(*,**)}_c=\Omega_c, \Omega_c(2770), \Omega_c(3050), \Omega_c(3090), \Omega_c(3120)$. 
There are six types of transitions, namely 
(i) ${\cal B}_b({\bf \bar 3_f},1/2^+)$ to ${\cal B}_c({\bf \bar 3_f},1/2^+)$ transition, 
(ii) ${\cal B}_b({\bf 6_f},1/2^+)$ to ${\cal B}_c({\bf 6_f},1/2^+)$ transition, 
(iii) ${\cal B}_b({\bf 6_f},1/2^+)$ to ${\cal B}_c({\bf 6_f},3/2^+)$ transition,
(iv) ${\cal B}_b({\bf 6_f},1/2^+)$ to ${\cal B}_c({\bf 6_f},3/2^-)$ transition,
(v) ${\cal B}_b({\bf \bar 3_f},1/2^+)$ to ${\cal B}_c({\bf \bar 3_f},1/2^-)$ transition, 
and 
(vi) ${\cal B}_b({\bf \bar 3_f},1/2^+)$ to ${\cal B}_c({\bf \bar 3_f},3/2^-)$
transition. 
Types (i) to (iii) involve spin 1/2 and 3/2 $s$-wave charmed baryons, 
while types (iv) to (vi) involve spin 1/2 and 3/2 $p$-wave charmed baryons. 
We have scalar or axial vector light diquarks in the baryons. The light diquarks are spectating in these transitions.
The bottom baryon to $s$-wave and $p$-wave charmed baryon form factors are calculated in the light-front quark model approach. 
The analysis and the scope of this work is improved and enlarged compared to a previous study~\cite{Chua:2018lfa} in several aspects. 
All of the form factors in the $1/2\to 1/2$ and $1/2 \to 3/2$ transitions are extracted,
while we only have $f_{1,2}$ and $g_{1,2}$ for $1/2\to 1/2$ transition in~\cite{Chua:2018lfa}. 
Some consistency constraints are found and are imposed.
We find that the form factors can reasonably satisfy the relations obtained in the heavy quark limit. 
In fact, we do not expect them to satisfy the relations exactly as we are using heavy but finite $m_b$ and $m_c$.

Using na\"{i}ve factorization decay rates and up-down asymmetries for various $\Lambda_b\to \Lambda^{(*,**)}_c M^-$, $\Xi_b\to\Xi_c^{(**)} M^-$ and $\Omega_b\to\Omega^{(*,**)}_c M^-$ decays are predicted and can be checked experimentally. 
We find that 
most of the signs of $\alpha$ are negative, except those in $\Omega_b\to\Omega_c M$, 
$\Omega_b\to\Omega_c(3050) M$ and $\Omega_b\to\Omega_c(3090) M$ decays.
We compare our results of rates and up-down asymmetries of $\Lambda_b\to \Lambda_c M$, $\Xi_b\to\Xi_c M$, 
$\Omega_b\to\Omega_c M$ and $\Omega_b\to\Omega_c(2770) M$ decays to existing data and other theoretical results. 
Our predictions agree well with data and with most of the results of other works.

The study on these decay modes 
may shed light on the quantum numbers of $\Lambda_c(2765)$, $\Lambda_c(2940)$
$\Omega_c(3050)$, $\Omega_c(3090)$ and $\Omega_c(3120)$,
as the decays depend on bottom baryon to charmed baryon form factors, 
which are sensitive to the configurations of the final state charmed baryons. 
For example, there are two possible quantum numbers for $\Lambda_c(2940)$, 
it can either be a $\B_c({\bf \bar 3_f}, 1/2^-)$ state~\cite{Cheng:2017ove}
or a $\B_c({\bf \bar 3_f}, 3/2^-)$ one~\cite{Aaij:2017vbw}. 
Both possibilities are considered.
Comparing predictions on $\Lambda_b\to\Lambda_c(2940) M$ rates, we have
 $Br[\Lambda_b\to\Lambda_c(2940,3/2^-) P]\simeq (1.5\sim 2)\times Br[\Lambda_b\to\Lambda_c(2940,1/2^-) P]$,
 $Br[\Lambda_b\to\Lambda_c(2940,3/2^-) V]\simeq Br[\Lambda_b\to\Lambda_c(2940,1/2^-) V]$
 and $Br[\Lambda_b\to\Lambda_c(2940,3/2^-) A]\simeq Br[\Lambda_b\to\Lambda_c(2940,1/2^-) A]$.
The asymmetries in $\Lambda_b\to\Lambda_c(2940,1/2^-) M$ and $\Lambda_b\to\Lambda_c(2940,3/2^-) M$ decays have larger deviations in the cases of heavy vector mesons.

\section{Acknowledgments}
The author likes to thank Hai-Yang Cheng for discussion. 
This research was supported in part by the Ministry of Science and Technology of R.O.C. under Grant
No. 106-2112-M-033-004-MY3.

\appendix

\section{Wave functions}\label{appendix: wave functions}

In the light-front quark model the baryon state, which consists of 
a heavy quark $Q=b, c$ and a scalar diquark $[qq]$ or an axial-vector diquark $[qq]$,
can be expressed as (see, \cite{Cheng97,CCH,Chua:2018lfa})
\begin{eqnarray}
        |\B_Q(P,J,J_z)\rangle
                =\int &&\{d^3p_1\}\{d^3p_2\} 2(2\pi)^3 \delta^3(
                \tilde P -\tilde p_1-\tilde p_2)~\nonumber\\
        &&\times \sum_{\lambda_1,\lambda_2,\alpha,\beta,\gamma,b,c}
                \Psi^{JJ_z}_{nL_KS_{[qq]}J_l}(\tilde p_1,\tilde p_2,\lambda_1,\lambda_2)~
                C_{\alpha\beta\gamma} F^{bc}
        \non\\
        &&\times ~
             \Big|Q^\alpha(p_1,\lambda_1) [q_b^\beta q_c^\gamma](p_2,\lambda_2) \Big\rangle,
 \label{lfmbs}
\end{eqnarray}
where $S_{[qq]}$, $L_K$ and $J_l$ denote
the spin of the diquark, 
the orbital angular momentum of the $Q-[qq]$ system,
and the total angular momentum of the light degree of freedom, respectively.
In the above equation, 
$n$,
$(\alpha,\beta,\gamma)$, $(b,c)$
$\lambda_i$
and
$p_{1,2}$
are the quantum number of the wave-function,
color indices,
flavor indices, helicity and 
the on-mass-shell light-front momenta, respectively.
The following notations are used
\begin{equation}
        \tilde p=(p^+, \vec p_\bot)~, \quad \vec p_\bot = (p^1, p^2)~,
                \quad p^- = {m^2+p_\bot^2\over p^+},
\end{equation}
and
\begin{eqnarray}
        &&\{d^3p\} \equiv {dp^+d^2p_\bot\over 2(2\pi)^3},
        \quad \delta^3(\tilde p)=\delta(p^+)\delta^2(\vec p_\bot),
        \nonumber \\
        &&\Big|Q(p_1,\lambda_1) [q_b q_c](p_2,\lambda_2)\Big\rangle
        = b^\dagger_{\lambda_1}(p_1) a^\dagger_{\lambda_2}(p_2) |0\rangle,\\
        &&[a_{\lambda'}(p'),a^\dagger_\lambda(p)] =2(2\pi)^3~\delta^3(\tilde p'-\tilde
        p)\,\delta_{\lambda',\lambda},\,
        \non\\
        &&\{b_{\lambda'}(p'),b_{\lambda}^\dagger(p)\} =
        2(2\pi)^3~\delta^3(\tilde p'-\tilde p)~\delta_{\lambda'\lambda},
                \nonumber
\end{eqnarray}
with $\lambda_2=S_2=0$ for a scalar diquark and $\lambda_2=0,\pm1$ and $S_2=1$ for an axial vector diquark. 
The coefficient $C_{\alpha\beta\gamma}$ is a
normalized color factor and $F^{bc}$ is a normalized flavor
coefficient, obeying the relation
 \be
 &&C_{\alpha'\beta'\gamma'} F^{b'c'}
 C_{\alpha\beta\gamma} F^{bc}
         \Big \la Q^{\alpha'}(p'_1,\lambda'_1)
                 [q_{b'}^{\beta'} q_{c'}^{\gamma'}](p'_2,\lambda'_2)
             \Big|Q^\alpha(p_1,\lambda_1) [q_a^\beta q_b^\gamma](p_2,\lambda_2) 
             \Big\rangle
 \non\\
&&=2^2(2\pi)^6~\delta^3(\tilde p'_1-\tilde p_1)
 \delta^3(\tilde p'_2-\tilde p_2)
 \delta_{\lambda'_1\lambda_1}\delta_{\lambda'_2\lambda_2}.
 \label{eq:norm}
 \en

The momenta can be defined 
in terms of the light-front internal momentum variables, 
$(x_i, \vec k_{i\bot})$ for $i=1,2$, 
\begin{eqnarray}
p^+_i=x_i P^{+}, 
\quad 
\sum_{i=1}^2 x_i=1, 
\quad
\vec  p_{i\bot}=x_i \vec P_\bot+\vec k_{i\bot}, 
\quad 
\sum_{i=1}^2 \vec k_{i\bot}=0.
\end{eqnarray}
The momentum-space wave-function $\Psi^{JJ_z}_{nL_KS_{[qq]} J_l}$ can be expressed
as
\be
\Psi^{JJ_z}_{nL_KS_{[qq]}J_l}(\tilde p_1,\tilde p_2,\lambda_1,\lambda_2)
&=&\sum_{s_1,s_2,L_z,J_{lz}} \langle \lambda_1|{\cal R}_M^\dagger(p_1^+,\vec p_{1\bot}, m_1)|s_1\rangle
         \langle \lambda_2|{\cal R}_M^\dagger(p_2^+,\vec p_{2\bot}, m_2)|s_2\rangle
\non\\ 
          && \la S_1 J_l; s_1 J_{lz}|S_1 J_l; J J_z\ra
                \la L_K S_{[qq]}; L_z s_2|L_k S_{[qq]};J_l J_{l z}\ra
\non\\
&&          
                  ~\phi_{n L_K L_z}(x_1,x_2,k_{1\bot},k_{2\bot}),
\label{eq: Psi}
\en
where $\phi_{n L_K L_z}(x_1,x_2,k_{1\bot},k_{2\bot})$
is the momentum distribution of the constituents in the
bound state,
$\la J' J''; m' m''|J' J'';J m\ra$ the
Clebsch-Gordan coefficients and $\langle
\lambda_i|{\cal R}_M^\dagger(p^+_1,\vec p_{1\bot}, m_i)|s_i\rangle$
the Melosh transform matrix element.

We normalize the state as
\begin{equation}
        \langle \B_Q(P',J',J'_z)|\B_Q(P,J,J_z)\rangle = 2(2\pi)^3 P^+
        \delta^3(\tilde P'- \tilde P)\delta_{J'J}\delta_{J'_z J_z}~,
\label{wavenor1}
\end{equation}
consequently, $\phi_{nLL_z}(x,p_\bot)$ satisfies the following orthonormal condition,
\begin{equation}
        \int {dx\,d^2p_{\bot}\over 2(2\pi)^3}~\phi^{\prime*}_{n'L^\prime L^\prime_z}(x,p_\bot)
                                                   \phi_{nLL_z}(x,p_\bot)
        =\delta_{n',n}~\delta_{L^\prime,L}~\delta_{L^\prime_z,L_z}.
\label{momnor}
\end{equation}

The wave function is defined as 
 \be
   \phi_{nLm}(\{x\},\{k_\bot\})
   &=&
   \sqrt{\frac{d k_{2z}}{d x_2}}
  ~\varphi_{nLm}
  \left(\frac{\vec k_1-\vec k_2}{2},\beta\right), 
\en
with
\be  
  \varphi_{n00}(\vec k,\beta)
  &=&\varphi_{ns}(\vec k,\beta),
\non\\  
  \varphi_{n1m}(\vec k,\beta)
  &=&k_{m} \varphi_{np}(\vec k,\beta)
  =-\varepsilon(k_1+k_2,m)\cdot k\varphi_{np}(\vec k,\beta),
  \non\\
  \varphi_{n2m}(\vec k,\beta)=(kk)_{m} \varphi_{nd}(\vec k,\beta)
  &=&\varepsilon_{\mu\nu}(k_1+k_2, m) k^\mu k^\nu \varphi_{nd}(\vec k,\beta).
  \label{eq: varphi}
 \en
where $k_m\equiv\vec\varepsilon(m)\cdot\vec k$ (or, explicitly
$k_{L_z=\pm1}\equiv\mp(k^x\pm i k^y)/\sqrt2$,
$k_{L_z=0}\equiv k^z$), 
and $\varphi_{ns}$ and $\varphi_{np}$ 
are $s$-wave and $p$-wave wave functions,
respectively. 
The kinematics are given by
 \be
 M_0^{(\prime) 2}&=&\sum_{i=1}^2\frac{m_i^{(\prime)2}+k^{(\prime)2}_{i\bot}}{x_i},\quad
 k^{(\prime)}_i
 =(\frac{m_i^{(\prime)2}+k^{(\prime)2}_{i\bot}}{x^{(\prime)}_i M^{(\prime)}_0},x^{(\prime)}_i M^{(\prime)}_0,\,\vec k^{(\prime)}_{i\bot})
 =(e^{(\prime)}_i-k^{(\prime)}_{iz},e^{(\prime)}_i+k^{(\prime)}_{iz},\vec k^{(\prime)}_{i\bot}),
 \non\\
 M^{(\prime)}_0&=&e^{(\prime)}_1+e^{((\prime)}_2,\quad
 e^{(\prime)}_i =\sqrt{m^{(\prime)2}_i+k^{(\prime)2}_{i\bot}+k^{(\prime)2}_{iz}}
 =\frac{x^{(\prime)}_i M^{(\prime)}_0}{2}+\frac{m_i^{(\prime)2}+k^{(\prime)2}_{i\bot}}{2 x^{(\prime)}_i M^{(\prime)}_0},
 \non\\
 k^{(\prime)}_{iz}&=&\frac{x^{(\prime)}_i M^{(\prime)}_0}{2}-\frac{m_i^{(\prime)2}+k^{(\prime)2}_{i\bot}}{2 x^{(\prime)}_i M^{(\prime)}_0},
 \qquad
 2M^{(\prime)}_0 (e^{(\prime)}_{1(2)}+m^{(\prime)}_{1(2)})=(M^{(\prime)}_0+m^{(\prime)}_{1(2)})^2-m_{2(1)}^{(\prime)2}.
 \label{eq: kinematics}
 \en
Under the constraint of $\sum_{i=1}^2 x_i=1$ and $\sum_{i=1}^2
\vec k_i=0$, one can easily obtain 
 \be
 \frac{d k_{2z}}{d x_2}=\frac{e_1 e_2 }{x_1 x_2 M_0}=\frac{d k_{1z}}{dx_1}.
 \en

For the heavy quark, the corresponding Melosh transform is
\cite{Jaus90,deAraujo:1999ugw},
 \be
        \la \lambda_1|{\cal R}^\dagger_M (p^+_1,\vec p_{1\bot},m_1)|s_1\ra
        &=&\frac{\bar
        u(p_1,\lambda_1) u_D(p_1,s_1)}{2 m_1}
\en 
with $u$ and $u_D$ the Dirac spinors in the light-front and instant
forms, respectively.
The Melosh transform for a scalar diquark is a trivial one, i.e.
\be
\la \lambda_2|{\cal R}^\dagger_M (p^+_2,\vec p_{2\bot},m_2)|s_2\ra
        &=&1,
\en
whereas the Melosh transform for an axial-vector diquark is more complicate
\be
\la \lambda_2|{\cal R}^\dagger_M (p^+_2,\vec p_{2\bot},m_2)|s_2\ra
        &=&-\varepsilon_{LF}^*(p_2,\lambda_2)\cdot \varepsilon_{I}(p_2,s_2),
\en
with $\varepsilon_{LF}$ and $\varepsilon_{I}$ the polarization vectors in light-front and instant forms, respectively.

It is convenient to use the covariant form for
$\Psi^{J J_z}_{nL_K S_{[qq]} J_l}$. We have
\be
\Psi^{1/2J_z}_{nL_K S_{[qq]} J_l}(\tilde p_1,\tilde p_2,\lambda_1,\lambda_2)
&=&  
      \frac{1}{\sqrt{(M_0+m_1)^2-m_2^2}}
        ~\bar u(p_1,\lambda_1)\Gamma_{L_K S_{[qq]} J_l} u(\bar P,J_z)
\non\\
&&
         ~\phi_{n L_K}(x_1,x_2,k_{1\bot},k_{2\bot}),  
\non\\
\Psi^{3/2 J_z}_{nL_K S_{[qq]} J_l}(\tilde p_1,\tilde p_2,\lambda_1,\lambda_2)
&=&  
      \frac{1}{\sqrt{(M_0+m_1)^2-m_2^2}}
        ~\bar u(p_1,\lambda_1)\Gamma^\mu_{L_K S_{[qq]} J_l} u_\mu(\bar P,J_z)
\non\\
&&
         ~\phi_{n L_K}(x_1,x_2,k_{1\bot},k_{2\bot}),              
\en
with $\Gamma^{(\mu)}_{L_K,n, S_{[qq]}}$ given in Eq.~(\ref{eq: Gamma}).
Note that we have 
\be
\phi_{n L_K}\equiv\sqrt{\frac{d k_{2z}}{d x_2}}\varphi_{n L_K},
\label{eq: phinLK}
\en
with $\varphi_{n L_K}$ given in Eq.~(\ref{eq:wavefn}).
The vertex functions $\Gamma_{s00}$, $\Gamma_{s11}$ and $\Gamma_{p01}$
are taken from \cite{Chua:2018lfa}, while $\Gamma^\mu$ are new and the derivations
can be found in Appendix \ref{appendix: vertex}.

\section{Vertex functions}\label{appendix: vertex}

\subsection{Some useful identities}

We collect some useful identities for the derivation of vertex functions.
Relations involving Melosh transform for spin-1/2 and spin-1 particles are given by 
\be
\la\lambda_1|{\cal R}_M^\dagger(x_1,k_{1\bot}, m_1)|s_1\rangle\bar u_D(k_1,s_1)
&=&
       \bar u(k_1,\lambda_1) \frac{u_D(k_1,s_1)\bar u_D(k_1,s_1)}{2 m_1}
      =\bar u(k_1,\lambda_1),       
\label{eq: meloshspin1/2}
\\
\la\lambda_2|{\cal R}_M^\dagger(x_2,k_{2\bot}, m_2)|s_2\rangle \varepsilon_I^*(k_2,s_2)
&=&
      - \varepsilon^*_{LF}(k_2,\lambda_2) \cdot \varepsilon_I(k_2,s_2) \varepsilon_I^*(k_2,s_2)
      = \varepsilon^*_{LF}(k_2,\lambda_2),
\label{eq: meloshspin1}      
\en
where the polarization vector in the light-front form $\varepsilon_{LF}$ has the following expression,
\begin{eqnarray}
        &&\varepsilon^\mu_{LF}(\pm 1) =
                \left[{2\over P^+} \vec \varepsilon_\bot (\pm 1) \cdot
                \vec P_\bot,\,0,\,\vec \varepsilon_\bot (\pm 1)\right],
                \quad \vec \varepsilon_\bot
                (\pm 1)=\mp(1,\pm i)/\sqrt{2}, \nonumber\\
        &&\varepsilon^\mu_{LF}(0)={1\over M_0}\left({-M_0^2+P_\bot^2\over
                P^+},P^+,P_\bot\right).   
                \label{polcom}
\end{eqnarray} 
Note that in the particle rest frame, the polarization vectors in the instant and light-front forms, $\varepsilon_I$ and $\varepsilon_{LF}$, are identical, and likewise the spinors in the two different forms, $u_D$ and $u$, are identical.

It is useful to expressed the relevant Clebsch-Gordan coefficients occurring in Eq.~(\ref{eq: Psi}) in compact forms:
\be
\la \frac{1}{2} 1; s_1 s_2|\frac{1}{2}1 ;\frac{1}{2} J_z\ra
&=&
      \frac{1}{\sqrt3}\chi^\dagger_{s_1}\vec\sigma\cdot
      \vec\varepsilon^*(k_1+k_2,s_2)\chi_{_{J_z}}
\non\\
&=&
       \frac{1}{\sqrt {3 [(M_0+m_1)^2-m_2^2]}}
\non\\
&&
       \times  \bar u_D(k_1,s_1)\gamma_5\not\!\varepsilon^*(k_1+k_2,s_2) u(k_1+k_2,J_z),
\label{eq: CG1/211/2}
\\
\la \frac{1}{2} 1; s_1 J_{lz}|\frac{1}{2}1 ;\frac{3}{2} J_z\ra
&=&
      -\chi^\dagger_{s_1} \varepsilon^{*\mu}(k_1+k_2,J_{lz})\chi_{\mu J_z}
\non\\
&=&
      -\frac{1}{\sqrt {(M_0+m_1)^2-m_2^2}}
\non\\
&&
      \times\bar u_D(k_1,s_1)\varepsilon^{*\mu}(k_1+k_2,J_{lz}) u_\mu(k_1+k_2,J_z), 
\label{eq: CG1/213/2}
\\
\la \frac{1}{2} 2; s_1 J_{lz}|\frac{1}{2} 2;\frac{3}{2} J_z\ra
&=&-\sqrt{\frac{2}{5}}\frac{1}{\sqrt {(M_0+m_1)^2-m_2^2)}}\varepsilon^*_{\mu\nu}(\bar P,J_{lz})
\non\\
&&\times
     \bar u_D(k_1,s_1)\gamma_5\gamma^\nu
     u^\mu(k_1+k_2,J_z),
\label{eq: CG1/223/2}     
\\
\la 1 1; s_2 s_4|11 ;2 J_{lz}\ra
&=&
      \varepsilon^\nu(\bar P, s_2)\varepsilon^\mu(\bar P, s_4)\varepsilon^*_{\mu\nu}(\bar P, J_{lz}),
\label{eq: CG112}                     
\en
where $u^\mu$ is the Rarita-Schwinger spinor in the light-front form
with
\be
u^\mu(k_1+k_2,J_z)
=
\la \frac{1}{2} 1;s_3 s_4|\frac{1}{2}1 ;\frac{3}{2} J_z\ra \varepsilon^\mu_{LF}(k_1+k2,s_4) u(k_1+k_2, s_3),
\label{eq: umu}
\en
and $\varepsilon_{\rho\sigma}(\bar P, m)$ is the spin-2 polarization vector defined as
\be
\varepsilon_{\mu\nu}(\bar P,s)
\equiv\la 11;m'm''|11; 2 s\ra \varepsilon_\mu(\bar P,m')\varepsilon_\nu(\bar P,m'').
\label{eq: epspin2}
\en
Note the Rarita-Schwinger spinor satisfies the following relations: (see, for example, \cite{Moroi:1995fs})
\be
\bar P^\mu u_\mu(\bar P, J_z)=0,
\quad
\gamma^\mu u_\mu(\bar P, J_z)=0,
\quad
\not\!\bar P u_\mu(\bar P)=M_0 u_\mu(\bar P, J_z).
\label{eq: umu relations}
\en
It is useful to note that we have
\be
u^+(\bar P,J_z)=\sqrt{\frac{2}{3}}\delta_{J_z S_z}\varepsilon^{+}_{LF}(\bar P, 0)\,u (\bar P,S_z)
=\sqrt{\frac{2}{3}}\delta_{J_z S_z}\frac{\bar P^+}{M_0}\,u (\bar P,S_z),
\label{eq: u+}
\en
which can be easily proved by using Eqs. (\ref{polcom}) and (\ref{eq: umu}), 
in particular, $u^+(\bar P,\pm3/2)=0$.

The following relations of the polarization vectors will be proved to be useful,
\be
\varepsilon^\mu(k_2,s_2)
&=& 
       \varepsilon^\mu(\bar P, s_2)
       -\frac{M_0k_2^\mu+m_2 \bar P^\mu}{m_2 M_0} \frac{\varepsilon(\bar P, s_2)\cdot k_2}{e_2+m_2},
\label{eq: ep2}
\\
\varepsilon^*_\mu(\bar P, m)\varepsilon_\nu(\bar P, m)
&=&
       -G_{\mu\nu},
\label{eq: epsum}       
\\
\varepsilon^*_{\mu\nu}(\bar P, m)\varepsilon_{\rho\sigma}(\bar P, m)  
&=&
       \frac{1}{2} G_{\mu\rho} G_{\nu\sigma}+\frac{1}{2} G_{\mu\sigma} G_{\nu\rho}-\frac{1}{3} G_{\mu\nu} G_{\rho\sigma},
\label{eq: epspin2sum}
\en
where $G_{\mu\nu}$ is defined as
\be
G_{\mu\nu}(\bar P)\equiv g_{\mu\nu}-\frac{\bar P_\mu \bar P_\nu}{M^2_0}.
\en

The relations in Eqs.
(\ref{eq: CG1/211/2}), 
(\ref{eq: CG1/213/2}) and (\ref{eq: CG112}) can be
easily proved by using 
Eq.~(\ref{eq: epspin2}) and Eq. (\ref{eq: kinematics}).
Explicitly, we use
\be
u_D(k_1,s)
&=&\frac{k_1\cdot\gamma+m_1}{\sqrt{e_1+m_1}}
\Bigg(
\begin{array}{c}
\chi_s\\
0
\end{array}
\Bigg)
=\frac{1}{\sqrt{e_1+m_1}}
\Bigg(
\begin{array}{c}
(e_1+m_1)\chi_s\\
\vec\sigma\cdot\vec p\chi_s
\end{array}
\Bigg),
\non\\
u(k_1+k_2,\lambda)
&=&\frac{(k_1+k_2)\cdot\gamma+M_0}{\sqrt{2 M_0}}\gamma^+\gamma^0
\Bigg(
\begin{array}{c}
\chi_\lambda\\
0
\end{array}
\Bigg)
=\sqrt{2 M_0}
\Bigg(
\begin{array}{c}
\chi_\lambda\\
0\\
\end{array}
\Bigg),
\non\\
u^\mu(k_1+k_2,\lambda)
&=&\frac{(k_1+k_2)\cdot\gamma+M_0}{\sqrt{2 M_0}}\gamma^+\gamma^0
\Bigg(
\begin{array}{c}
\chi^\mu_\lambda\\
0
\end{array}
\Bigg)
=\sqrt{2 M_0}
\Bigg(
\begin{array}{c}
\chi^\mu_\lambda\\
0\\
\end{array}
\Bigg),
\non\\
\chi^\mu_{\lambda}
&\equiv&\la \frac{1}{2} 1; m' m''| \frac{1}{2} 1;\frac{3}{2}, \lambda\ra \varepsilon^\mu(k_1+k_2, m'')\,\chi_{m'},
\label{eq: spinors}
\en 
the standard Dirac representation of $\gamma^\mu$, $\gamma_5$, 
$\varepsilon(k_1+k_2,s)=(0,\vec \varepsilon(s))$ with $ \vec \varepsilon (\pm 1)=\mp(1,\pm i,0)/\sqrt{2}$, $\vec \varepsilon(0)=(0,0,1)$.
One can see \cite{Chua:2018lfa} for the derivation of 
Eqs.
(\ref{eq: CG1/211/2}) and (\ref{eq: ep2}).
Eq. (\ref{eq: CG1/213/2}) can be easily proved using the above equation and the usual normalization of the polarization vector: $\varepsilon^*_\mu(\bar P,m)\varepsilon^\mu(\bar P,m')=-\delta_{m,m'}$.

To prove Eq.~(\ref{eq: CG1/223/2}), we need to use the following identity on Clebsch-Gordan coefficient:
\be
\la \frac{1}{2} 2; s_1 J_{lz}|\frac{1}{2} 2;\frac{3}{2} J_z\ra
&=&
\sqrt{\frac{6}{5}}\la \frac{1}{2} 1;s_1 s_2|\frac{1}{2}1 ;\frac{1}{2} s_3\ra
\la 1 1; s_2 s_4|11 ;2 J_{lz}\ra
\la \frac{1}{2} 1;s_3 s_4|\frac{1}{2}1 ;\frac{3}{2} J_z\ra,
\label{eq: CGid1}
\en
and use Eqs.~(\ref{eq: CG1/211/2}), (\ref{eq: CG112}), (\ref{eq: umu}) and (\ref{eq: epsum}),
\be
\la \frac{1}{2} 2; s_1 J_{lz}|\frac{1}{2} 2;\frac{3}{2} J_z\ra
&=&
\sqrt{\frac{6}{5}}
\la \frac{1}{2} 1;s_1 s_2|\frac{1}{2}1 ;\frac{1}{2} s_3\ra
\la 1 1; s_2 s_4|11 ;2 J_{lz}\ra
\la \frac{1}{2} 1;s_3 s_4|\frac{1}{2}1 ;\frac{3}{2} J_z\ra
\non\\
&=&
-\sqrt{\frac{2}{5}}
 \frac{1}{\sqrt {(M_0+m_1)^2-m_2^2}}\varepsilon^*_{\mu\nu}(J_{lz})
       \bar u_D(k_1,s_1)\gamma_5\gamma^\nu u^\mu(k_1+k_2,J_z).
\en

Using Eq.~(\ref{eq: epspin2}) and the following identity of the Clebsch-Gordan coefficients,
\be
\la 1 1; a,b| 1 1; 2 m\ra \la 1 1; c,d| 1 1; 2 m\ra
=\frac{1}{2} \delta_{a,c}\delta_{b,d}+\frac{1}{2}\delta_{a,d}\delta_{b,c}
-\frac{1}{3} (-)^{a+c} \delta_{a,-b}\delta_{c,-d},
\label{eq: CGid2}
\en
and noting that $\varepsilon^*(\bar P, m)=(-)^m \varepsilon(\bar P, -m)$, we have
\be
\varepsilon^*_{\mu\nu}(\bar P, m)\varepsilon_{\rho\sigma}(\bar P, m)
&=&
       \frac{1}{2}\varepsilon^*_\mu(\bar P, m)\varepsilon_\rho(\bar P, m) 
       \varepsilon^*_\nu(\bar P, m')\varepsilon_\sigma(\bar P, m') 
\non\\
&&           
       +\frac{1}{2}\varepsilon^*_\mu(\bar P, m)\varepsilon_\sigma(\bar P, m) 
       \varepsilon^*_\nu(\bar P, m')\varepsilon_\rho(\bar P, m')
\non\\
&& 
     -\frac{1}{3}
     \varepsilon^*_\mu(\bar P, m)\varepsilon_\nu(\bar P, m) 
     \varepsilon_\rho(\bar P, m')\varepsilon^*_\sigma(\bar P, m'),
\en     
which leads to Eq.~(\ref{eq: epspin2sum}) after performing the polarization sum, Eq.~(\ref{eq: epsum}).

Note that using Eqs. (\ref{eq: ep2}), (\ref{eq: meloshspin1}) and (\ref{eq: epsum}), we have
\be
\la \lambda_2|{\cal R}^\dagger_M (x_2,k_{2\bot},m_2)|s_2\ra \varepsilon^{*\nu}(\bar P,s_2)
&=&-\varepsilon^*_{LF}(k_2,\lambda_2)\cdot\varepsilon(k_2,s_2) \varepsilon^{*\nu}(\bar P,s_2)
\non\\
&=&\varepsilon_{LF}^{*\nu}(k_2,\lambda_2)
-\frac{M_0 k^\nu_2+m_2 \bar P^\nu}{(\bar P\cdot k_2+m_2 M_0)}\frac{\varepsilon_{LF}^*(k_2,\lambda_2)\cdot \bar P}{M_0},
\label{eq: meloshspin1ep2}
\en
which will be useful in obtaining vertex functions.

\subsection{$\Gamma^\mu$ for states with configuration
     $(L_K, S_{[qq]}^P, J_l^P,J^P)=( 1, 0^+,1^-, \frac{3}{2}^-)$} 
 
From Eq. (\ref{eq: Psi}) the corresponding momentum-space wave-function $\Psi^{JJ_z}_{n L_K S_{[qq]} J_l}$ is given by
\be
\Psi^{\frac{3}{2}J_z}_{np01}(\tilde p_1,\tilde p_2,\lambda_1,\lambda_2)
&=& \langle \lambda_1|{\cal R}_M^\dagger(p_1^+,\vec p_{1\bot}, m_1)|s_1\rangle
\non\\ 
          && \la \frac{1}{2} 1; s_1 J_{lz}|\frac{1}{2} 1; \frac{3}{2} J_z\ra
                \la 1 0; L_z 0|1 0;1 J_{l z}\ra
\non\\
&&          
                  ~\phi_{n 1 L_z}(x_1,x_2,k_{1\bot},k_{2\bot}).
\en
It can be expressed
as
\be                  
\Psi^{\frac{3}{2}J_z}_{np01}(\tilde p_1,\tilde p_2,\lambda_1,\lambda_2)
&=&\frac{1}{\sqrt{(M_0+m_1)^2-m_2^2}}
        ~\bar u(p_1,\lambda_1)\Gamma^{\mu}_{p01} u_\mu(\bar P,J_z)
\non\\
&&         
        \phi_{np}(x_1,x_2,k_{1\bot},k_{2\bot}),
\en
with
\begin{eqnarray}
\Gamma^\mu_{p01}=-\frac{1}{2}(p_1-p_2)^\mu,
\end{eqnarray}
where Eqs. (\ref{eq: meloshspin1/2}), (\ref{eq: CG1/213/2}), (\ref{eq: varphi}), (\ref{eq: epsum}) and (\ref{eq: umu relations}) have been used.

\subsection{$\Gamma^\mu$ for states with configuration
     $(L_K, S_{[qq]}^P, J_l^P,J^P)=( 0, 1^+,1^+, \frac{3}{2}^+)$}     

From Eq. (\ref{eq: Psi}) the corresponding momentum-space wave-function $\Psi^{JJ_z}_{nL_KS_{[qq]}J_l}$ is given by
\be
\Psi^{3/2J_z}_{ns11}(\tilde p_1,\tilde p_2,\lambda_1,\lambda_2)
&=& \langle \lambda_1|{\cal R}_M^\dagger(p_1^+,\vec p_{1\bot}, m_1)|s_1\rangle
         \langle \lambda_2|{\cal R}_M^\dagger(p_2^+,\vec p_{2\bot}, m_2)|s_2\rangle
\non\\ 
          && \la \frac{1}{2} 1; s_1 J_{lz}|\frac{1}{2} 1; \frac{3}{2} J_z\ra
                \la 0 1; 0 s_2|0 1;1 J_{l z}\ra
\non\\
&&          
                  ~\phi_{n 0 0}(x_1,x_2,k_{1\bot},k_{2\bot}).
\en
It can be expressed
as
\be
\Psi^{3/2J_z}_{ns11}(\tilde p_1,\tilde p_2,\lambda_1,\lambda_2)
&=&\frac{1}{\sqrt{(M_0+m_1)^2-m_2^2}}
        ~\bar u(p_1,\lambda_1)\Gamma^\mu_{s11} u_\mu(\bar P,J_z)
\non\\
&&         
        \phi_{n s}(x_1,x_2,k_{1\bot},k_{2\bot}),
\en
with
\begin{eqnarray}
\Gamma^\mu_{s11}=-
\bigg(\varepsilon_{LF}^{*\mu}(p_2,\lambda_2)
        -\frac{p^\mu_2}{\bar P\cdot p_2+m_2 M_0}
        \varepsilon_{LF}^*(p_2,\lambda_2)\cdot \bar P\bigg),
\end{eqnarray}
where Eqs. (\ref{eq: meloshspin1/2}), 
(\ref{eq: CG1/213/2}), (\ref{eq: meloshspin1ep2}) and (\ref{eq: umu relations}) have been used.

\subsection{$\Gamma^\mu$ for states with configuration      
     $(L_K, S_{[qq]}^P, J_l^P,J^P)=(1, 1^+,2^-, \frac{3}{2}^-)$}

From Eq. (\ref{eq: Psi}) the corresponding momentum-space wave-function $\Psi^{JJ_z}_{nL_KS_{[qq]}J_l}$ 
is given by
\be
\Psi^{3/2J_z}_{np12}(\tilde p_1,\tilde p_2,\lambda_1,\lambda_2)
&=& \langle \lambda_1|{\cal R}_M^\dagger(p_1^+,\vec p_{1\bot}, m_1)|s_1\rangle
         \langle \lambda_2|{\cal R}_M^\dagger(p_2^+,\vec p_{2\bot}, m_2)|s_2\rangle
\non\\ 
          && \la \frac{1}{2} 2; s_1 J_{lz}|\frac{1}{2} 2; \frac{3}{2} J_z\ra
                \la 1 1; L_z s_2|1 1;2 J_{l z}\ra
\non\\
&&          
                  ~\phi_{n 1 L_z}(x_1,x_2,k_{1\bot},k_{2\bot}).
\en
It can be expressed
as
\be
\Psi^{3/2J_z}_{np12}(\tilde p_1,\tilde p_2,\lambda_1,\lambda_2)
&=&\frac{1}{\sqrt{(M_0+m_1)^2-m_2^2}}
        ~\bar u(p_1,\lambda_1)\Gamma^\mu_{p12} u_\mu(\bar P,J_z)
\non\\
&&         
        \phi_{n p}(x_1,x_2,k_{1\bot},k_{2\bot}),
\label{eq: Psi3/2p12}
\en
with
\begin{eqnarray}
\Gamma^\mu_{p12}
&=&
-\frac{1}{2\sqrt{10}}\gamma_5
\bigg[
\bigg(\varepsilon^{*\mu}(p_2,\lambda_2)+(p_1-p_2)^\mu \frac{\varepsilon^{*}_{LF}(p_2,\lambda_2)\cdot\bar P}{\bar P\cdot p_2+M_0m_2}
\bigg)
\non\\
&&
\qquad\qquad
\times\bigg(
\not\! p_1-\not\! p_2
- \frac{m_1^2-m_2^2}{M_0}
\bigg)
\non\\
&&\quad
+(p_1-p_2)^\mu
\bigg(
\not\! \varepsilon^*_{LF}(p_2,\lambda_2) 
-\frac{\varepsilon^*_{LF}(p_2,\lambda_2)\cdot\bar P}{M_0}
\bigg)
\bigg].
\label{eq: Gammamup12}
\end{eqnarray}
The above result can be obtained by using 
Eqs.~(\ref{eq: meloshspin1/2}), 
(\ref{eq: meloshspin1}), 
(\ref{eq: CG1/223/2}),
(\ref{eq: varphi}),
(\ref{eq: ep2}), (\ref{eq: epspin2}) and (\ref{eq: epspin2sum}).
In fact, from Eqs.~(\ref{eq: meloshspin1}), 
(\ref{eq: ep2}) and (\ref{eq: epspin2}), we have
\be
&&\la 1 1; L_z s_2|1 1;2 J_{l z}\ra \varepsilon_\sigma(\bar P,L_z) 
\la \lambda_2|{\cal R}^\dagger_M (x_2,k_{2\bot},m_2)|s_2\ra
\non\\
&&=-\la 1 1; L_z s_2|1 1;2 J_{l z}\ra\varepsilon^*_{LF}(k_2,\lambda_2)\cdot\varepsilon(k_2,s_2)\varepsilon_\sigma(\bar P,L_z)
\non\\
&&=
-\bigg(\varepsilon^{*\rho}_{LF}(k_2,\lambda_2)
+\frac{\varepsilon^*_{LF}(k_2,\lambda_2)\cdot \bar P}{M_0 (e_2+m_2)}\frac{(k_1-k_2)^\rho}{2}\bigg)
\varepsilon_{\rho\sigma}(\bar P, J_{lz})
\en

Putting them together we obtain
\be
\Gamma^\mu_{p12} u_\mu
&=&-\sqrt{\frac{2}{5}}\bigg(\frac{k_1-k_2}{2}\bigg)^\sigma \gamma_5\gamma^\nu
\non\\
&&\bigg(
\varepsilon^{*\rho}_{LF}(k_2,\lambda_2)
+\frac{\varepsilon^{*}_{LF}(k_2,\lambda_2)\cdot\bar P}{M_0(e_2+m_2)} \bigg(\frac{k_1-k_2}{2}\bigg)^\rho
\bigg)
\non\\
&&
\bigg(
\frac{1}{2} G_{\mu\rho} G_{\nu\sigma}+\frac{1}{2} G_{\mu\sigma} G_{\nu\rho}-\frac{1}{3} G_{\mu\nu} G_{\rho\sigma}
\bigg) u^\mu
\non\\
&=&-\frac{1}{\sqrt{10}}\gamma_5
\non\\
&&\bigg(
\varepsilon^{*\rho}_{LF}(k_2,\lambda_2)\bigg(\frac{k_1-k_2}{2}\bigg)^\sigma
+\frac{\varepsilon^{*}_{LF}(k_2,\lambda_2)\cdot\bar P}{M_0(e_2+m_2)} 
\bigg(\frac{k_1-k_2}{2}\bigg)^\sigma\bigg(\frac{k_1-k_2}{2}\bigg)^\rho
\bigg)
\non\\
&&
\bigg[
g_{\mu\rho} \bigg(\gamma_{\sigma}-\frac{\bar P_\sigma}{M_0}\bigg)
+g_{\mu\sigma} \bigg(\gamma_{\rho}-\frac{\bar P_\rho}{M_0}\bigg)
\bigg] u^\mu.
\en
It can be further simplified into the following expression:
\be
\Gamma^\mu_{p12} u_\mu
&=&
-\frac{1}{2\sqrt{10}}\gamma_5
\bigg[
\bigg(\varepsilon^{*\mu}(k_2,\lambda_2)+(k_1-k_2)^\mu \frac{\varepsilon^{*}_{LF}(k_2,\lambda_2)\cdot\bar P}{M_0(e_2+m_2)}
\bigg)
\non\\
&&
\qquad\times\bigg(
\not\! k_1-\not\! k_2
- \frac{m_1^2-m_2^2}{M_0}
\bigg)
\non\\
&&+(k_1-k_2)^\mu
\bigg(
\not\! \varepsilon^*_{LF}(k_2,\lambda_2) 
-\frac{\varepsilon^*_{LF}(k_2,\lambda_2)\cdot\bar P}{M_0}
\bigg)
\bigg]
 u_\mu.
\en
Boosting $k_i\to p_i$ in the LF boost we obtain Eqs.~(\ref{eq: Psi3/2p12}) and (\ref{eq: Gammamup12}).

\section{Kinematics}~\label{appendix: kinematics}

The decay rates and asymmetries read \cite{Cheng:1996cs}
 \be
 \Gamma[\B_b\to \B_c(1/2) P]&=&\frac{p_c}{8\pi}\left[\frac{(M+M')^2-m_P^2}
 {M^2}|A|^2+\frac{(M-M')^2-m_P^2}{M^2}|B|^2\right],
 \non\\
 \Gamma[\B_b\to \B_c(1/2) V(A)]&=&\frac{p_c}{4\pi}\frac{E'+M'}{M}\left[2(|S^{(\prime)}|^2+|P^{(\prime)}_2|^2)
 +\frac{E_{V(A)}^2}{m_{V(A)}^2}(|S^{(\prime)}+D^{(\prime)}|^2+|P^{(\prime)}_1|^2)\right],
 \non\\
 \label{eq: C1}
 \en
\be
\alpha[\B_b\to \B_c(1/2) P]&=&-\frac{2\kappa {\rm Re}(A^* B)}{|A|^2+\kappa^2|B|^2},
\non\\
\alpha[\B_b\to \B_c(1/2) V(A)]&=&
\frac{4 m_{V(A)}^2{\rm Re}(S^{(\prime)*} P_2)+2 E_V^2 {\rm Re}(S^{(\prime)}+D^{(\prime)})^* P^{(\prime)}_1}
{2m_{V(A)}^2(|S^{(\prime)}|^2+|P^{(\prime)}_2|^2)+E_{V(A)}^2(|S^{(\prime)}+D^{(\prime)}|^2+|P^{(\prime)}_1|^2)},
\en 
with $\kappa\equiv p_c/(E'+M')$ and $p_c$ is the momentum in the center of mass frame
 \be
 S^{(\prime)}&=&-A^{(\prime)}_1,
 \qquad 
 P^{(\prime)}_1=-\frac{p_c}{E_{V(A)}}\left(\frac{M+M'}{E'+M'} B^{(\prime)}_1+M B^{(\prime)}_2\right),
 \non\\
 P^{(\prime)}_2&=&\frac{p_c}{E'+M'} B^{(\prime)}_1,
 \qquad
 D^{(\prime)}=-\frac{p_c^2}{E_{V(A)}(E'+M')}(A^{(\prime)}_1-M A^{(\prime)}_2).
 \en

The $\B_b\to\B_c(3/2) P$ decay rates and asymmetries are given by
 \be
 \Gamma[\B_b\to \B_c(3/2) P]&=&\frac{p_c^3}{12\pi}\left[\frac{(M+M')^2-m_P^2}
 {M^{\prime 2}}|C|^2+\frac{(M-M')^2-m_P^2}{M^{\prime 2}}|D|^2\right]
 \label{eq: C4}
 \en
and 
\be
\alpha[\B_b\to \B_c(3/2) P]&=&-\frac{2\kappa {\rm Re}(A^* B)}{|A|^2+\kappa^2|B|^2},
\en 
respectively.
The decay rates and asymmetries of $\B_b\to\B_c(3/2) V$ decays read~\cite{Cheng:1996cs,Korner:1992wi,Ebert:2006rp}
\be
\Gamma&=&\frac{p_c}{16 \pi M^2}\sum_{\lambda_{V}}
(|h_{\lambda_{V}+1/2,\lambda_{V};1/2}|^2+|h_{-(\lambda_{V}+1/2),-\lambda_{V};-1/2}|^2)
\non\\
&=&\frac{p_c}{32 \pi M^2}\sum_{\lambda_{V}}
(|h^{PV}_{\lambda_{V}+1/2,\lambda_{V};1/2}|^2+|h^{PC}_{\lambda_{V}+1/2,\lambda_{V};1/2}|^2)
\non\\
\alpha&=&
\frac{ \sum_{\lambda_{V}}|h_{\lambda_{V}+1/2,\lambda_{V};1/2}|^2-|h_{-(\lambda_{V}+1/2),-\lambda_{V};-1/2}|^2}
{\sum_{\lambda_{V}} |h_{\lambda_{V}+1/2,\lambda_{V};1/2}|^2+|h_{-(\lambda_{V}+1/2),-\lambda_{V};-1/2}|^2}
\non\\
&=& 
\frac{\sum_{\lambda_{V}} 2 h^{PV}_{\lambda_{V}+1/2,\lambda_{V};1/2} \, h^{PC}_{(\lambda_{V}+1/2),\lambda_{V};1/2}}
{\sum_{\lambda_{V}} |h^{PV}_{\lambda_{V}+1/2,\lambda_{V};1/2}|^2+|h^{PC}_{\lambda_{V}+1/2,\lambda_{V};1/2}|^2}
\en
where $h_{\lambda',\lambda_V;\lambda}\equiv\la \B_c(\lambda') V(\lambda_V)|H_W|\B_b(\lambda)\ra$ is the helicity amplitudes
and
\be
h^{PV}_{\lambda',\lambda_V;1/2}\equiv h_{\lambda',\lambda_V,1/2}\pm h_{-\lambda',-\lambda_V,-1/2},
\en
with
\be
h^{PV(PC)}_{3/2,1;1/2}&=&\mp\sqrt{2Q_\pm} E_1^{PV(PC)},
\non\\
h^{PV(PC)}_{-1/2,-1;1/2}&=&\mp\sqrt{\frac{2}{3}}\sqrt{Q_\pm} \bigg[E_1^{PV(PC)}-\frac{Q_\mp}{M'} E_2^{PV(PC)}\bigg],
\non\\
h^{PV(PC)}_{1/2,0;1/2}&=&\mp\frac{\sqrt{Q_\pm} }{2\sqrt3 M' m_V}
\bigg[2(M^2-M^{\prime 2}-m_V^2)E_1^{PV(PC)}\pm 2Q_\mp (M\pm M') E_2^{PV(PC)}
\non\\
&&
\qquad+Q_+ Q_-E_3^{PV(PC)}\bigg],
\en
$E^{PV}_i\equiv C_i$, $E^{PC}_i\equiv D_i$ and $Q_\pm\equiv(M\pm M')^2-m_V^2$.
The formulas for the decay rates and asymmetries of $\B_b\to\B_c(3/2) V$ decays can be obtained by using the above formulas, but with $\lambda_V$, $m_V$, $C_i$ and $D_i$ and replaced by $\lambda_A$, $m_A$, $C'_i$ and $D'_i$, respectively. 

\newcommand{\bi}{\bibitem}

\end{document}